\documentclass{aastex62}
\usepackage{graphicx}
\usepackage[space]{grffile}
\usepackage{lineno}

\newcommand{\gama}{$\gamma$}

\newcommand{\fermi}{\textsl{Fermi}}

\shorttitle{$\gamma$-ray variability of blazars}
\shortauthors{Bhatta \& Dhital}

\begin{document}

\title{The Nature of $\gamma$-ray Variability in Blazars}

\correspondingauthor{Gopal Bhatta}
\email{gopal@oa.uj.edu.pl}

\author{Gopal Bhatta}
\affiliation{Astronomical Observatory of the Jagiellonian University \\
ul. Orla 171 \\
 30-244 Krak\'ow, Poland}

\author{Niraj Dhital}
\affiliation{Institute of Nuclear Physics PAN\\
Radzikowskiego 152 \\
31-342, Krak\'ow, Poland}



\begin{abstract}

We present an in-depth and systematic variability study of a sample of 20 powerful blazars, including 12 BL Lacs and 8 flat spectrum radio quasars, applying various analysis tools such as flux distribution, symmetry analysis, and time series analysis on the decade-long Fermi/LAT observations. The results show that blazars with steeper \gama-ray spectral indexes are found to be more variable; and the \gama-ray flux distribution closely resembles lognormal probability distribution function. The statistical variability properties of the sources as studied by power spectral density analysis are consistent with \emph{flicker noise} ($P(\nu)\propto1/\nu$) -- an indication of long-memory processes at work. Statistical analysis of the distribution of flux rise and decay rates in the light curves of the sources, aimed at distinguishing between particle acceleration and energy dissipation timescales, counter-intuitively suggests that both kinds of rates follow a similar distribution and the derived mean variability timescales are in the order of a few weeks. The corresponding emission region size is used to constrain location of \gama-ray production sites in the sources to be a few parsecs. Additionally, using Lomb-Scargle periodogram and weighted wavelet z-transform methods and extensive Monte Carlo simulations, we detected year timescale quasi-periodic oscillations in the sources S5 0716+714, Mrk 421, ON +325, PKS 1424-418 and PKS 2155-304. The detection significance was computed taking proper account of the red-noise and other artifacts inherent in the observations. We explain the results in the light of current blazar models with relativistic shocks propagating down the jet viewed close to the line of sight.

\end{abstract}

\keywords{accretion, accretion disks --- radiation mechanisms: non-thermal, \gama-ray --- galaxies: active --- BL Lac objects, flat spectrum radio quasars --- galaxies: jets, method: time series}

\section{Introduction \label{sec:intro}}
Active galactic nuclei (AGN) are the most luminous sources (L$\sim10^{47}$ erg/s) with supermassive black holes lurking at their centers \citep[see][for M87 galaxy black hole]{2019ApJ...875L...6E}. The sources are  powered by accretion on to the supermassive black holes. A small fraction of AGN ($\sim$10\%) profusely emit in radio frequency and thereby are known as radio-loud sources. A sub-population of radio-loud AGN that eject relativistic jets towards us are known as blazars.  The sources are known to possess extreme properties such as high luminosity, rapid flux and polarization variability.  Also, blazars are the sources of the abundant nonthermal emission that is Doppler boosted due to the relativistic effects, which makes them appear highly variable over a wide range of spatial and temporal frequencies.  Besides, the objects could be sources of extra-solar neutrinos \citep[see][]{2018Sci...361.1378I,2018Sci...361..147I}. The spectral energy distribution (SED) of the broadband continuum emission from the sources is usually characterized by two distinct spectral peaks. The low energy peak, which usually lies between the radio and the X-ray energies, is attributed to the synchrotron emission from the relativistic particles, whereas the high energy peak, usually observed between the UV and the $\gamma$-ray energies, is believed to originate from inverse Compton scattering of low energy photons \citep[for recent blazar overview, see][and references therein]{Bottcher2019}. However, there is no common agreement on the source of these low energy seed photons. Of the two widely discussed models, the synchrotron self-Compton (SSC) model (e.g., \citealt{Maraschi1992,Mastichiadis2002}),  assumes that the same population of the electrons emitting synchrotron photons up-scatters these photons to higher energies; whereas, the external Compton (EC) model assumes that the softer seed photons are provided by various regions of AGN, such as accretion disk (AD; \citealt{Dermer1993}), broad-line region (BLR; \citealt{Sikora1994}), and dusty torus (DT; \citealt{Blazejowski2000}).

Based on the presence of the emission lines over the continuum in their SEDs, blazars are grouped into two sub-classes: more luminous flat-spectrum radio quasars (FSRQ) which show emission lines over the continuum, and less powerful BL Lacertae (BL Lac) sources which show weak or no such lines. In case of FSRQs, the synchrotron peak is in the lower frequency, and the most plausible process responsible for the high energy emission is believed to be EC as opposed to SSC \citep[][]{Ghisellini1998}. This is because the sources are known to have abundant seed photons from AD, BLR and DT  \citep{Ghisellini2011}.  In case of BL Lac objects, the synchrotron peak lies in the optical or X-ray regions.  These constitute an extreme class of sources featuring high energy emission from a few tens of keV to TeV energies that results from the combination of the synchrotron and IC processes.  Absence of strong circumnuclear photon fields and relatively low accretion rates could be the possible reasons behind the apparent low luminosity for such sources. Another scheme for blazar classification is based on the frequency of the synchrotron peak ($\nu_\mathrm{s}$), following which blazars are either  high synchrotron peaked blazars  (HSP; $\nu_s > 10^{15}$ Hz), intermediate synchrotron  peaked blazars (ISP; $10^{14}<\nu_s <10^{15} $ Hz), and  low synchrotron peaked blazars (LSP; $\nu_s < 10^{14}$ Hz) (\citealt[][]{Abdo2010};  see \citealt{Fan2016} for similar recent classification scheme).  In the  blazar sequence, a scheme to unify diverse appearance of the sources, bolometric luminosity is found to decrease along with the $\gamma$-ray emission in the direction from FSRQ to HSP but the peak frequencies move towards higher energies \citep{Fossati1998,Ghisellini2017}.  Also, while  synchrotron and $\gamma$-ray emission are comparable in HSP sources, FSRQs are mostly $\gamma$-ray (or Compton) dominant.

Blazars exhibit variability across the electromagnetic spectrum on diverse timescales that span a few minutes to a few decades. For the reason, multi-frequency variability studies could be one of the most relevant tools that can offer important insights into the physical conditions prevailing the innermost regions of blazar jets, including the nature of the dominant particle acceleration and energy dissipation mechanism, magnetic field geometry, jet content, etc. There have been numerous attempts to model the phenomenon by relating the sources of the variability to a wide range of possible physical processes occurring either in the accretion disk and/or in the jet; the various scenarios include emission sites at the accretion disk revolving around the supermassive black hole, various magnetohydrodynamic instabilities in the disk and the jets, shocks traveling down the turbulent jets, and relativistic effects due to jet orientation \citep[e.g.][and the references therein]{Camenzind92,Wagner1995,bhatta13, Marscher14}. However, the exact details of the underlying processes are still under debate.  In such context, study of $\gamma$-ray variability of  blazars provides us with an important tool to probe into  jet dynamics, and associated particle acceleration and energy dissipation mechanisms resulting high energy emission.

\begin{figure}[ht!]
\plotone{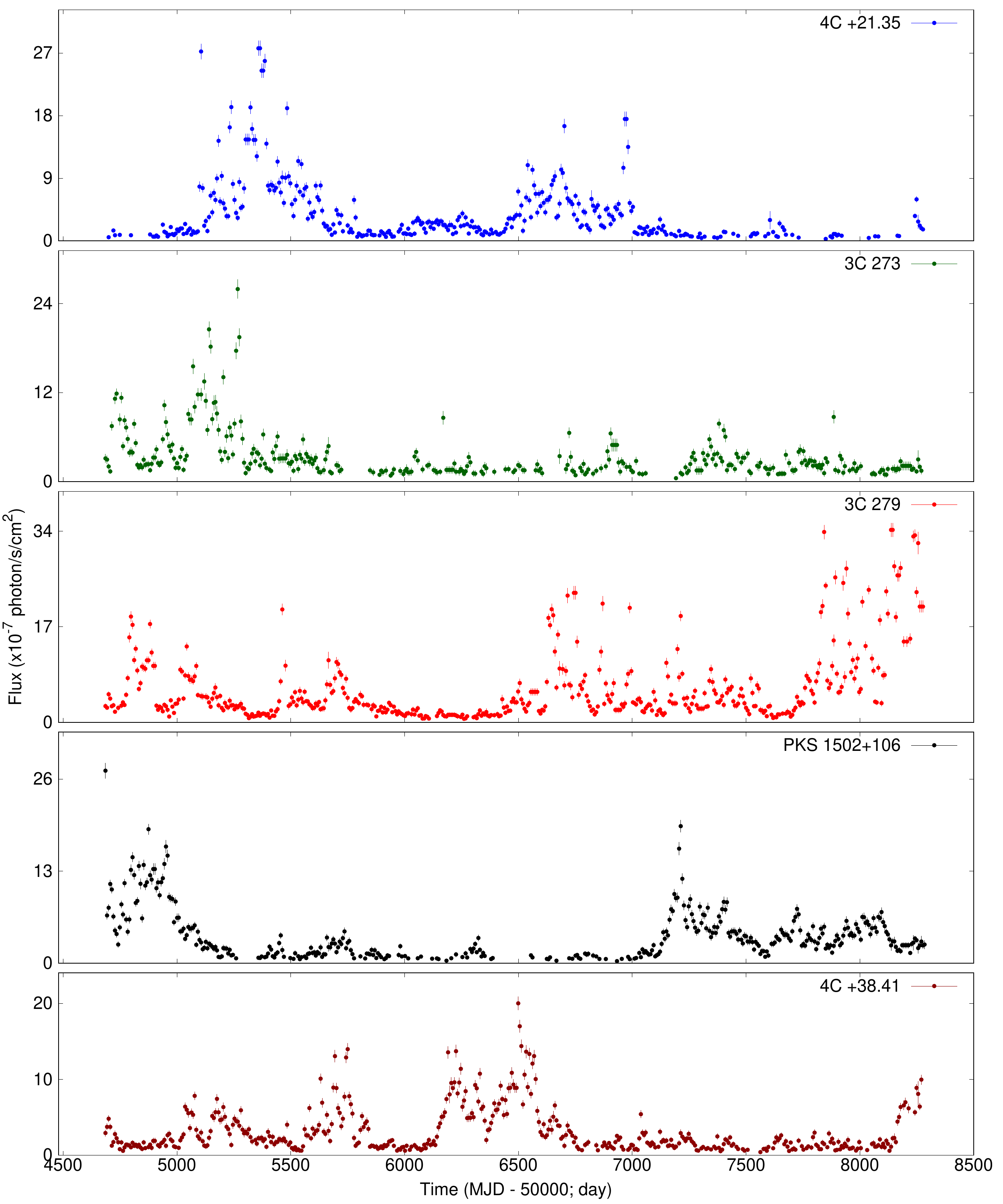}
\label{Fig:1}
\caption{Weekly binned Fermi/LAT light curves of 5 well studied blazars. The light curves for the rest of the sources in the sample as listed in Table \ref{table:1} are presented in Appendix \ref{apndx1}.}
\end{figure}

 In blazars, the flux modulations due to disk processes could easily be swamped by the Doppler boosted emission from jets. Nonetheless, the signatures of the disk modulations should, in principle, propagate along the jet through disk-jet coupling mechanisms such that the traces  of characteristic timescales related to disk processes could be revealed through robust time series analysis.  Such timescales then can be linked to the various processes in the jet as well as the accretion disk such as dynamical, thermal, and viscous processes \citep{Czerny2006}. For example, for blazars with typical masses between $\sim 10^{8}-10^{9} M_{\odot}$ the dynamical, thermal and viscous time scales are in the order of a few hours to a few years. Besides, several AGN models predict quasi-periodic oscillations (QPO) in the flux with the characteristic timescales ranging from a few hours to a few years. For instance, in the scenario of {\em magnetic flux paradigm} \citep[see][]{Sikora2013}, magnetic field at accretion disk threads black hole in launching the jets in AGN, and consequently it gives rise to various magnetohydrodynamical instabilities at the disk-magnetosphere interface. These instabilities in turn can produce QPOs which subsequently could propagate along modulating the jet emission, and could be observed in the multi-frequency observations, including \gama-ray light curves. In the similar context,  highly polarized optical flare discovered by \citet{Bhatta2015} might be a signature of dominance of magnetic field near blazar cores, so called magnetically arrested disk (MAD) scenario\citep[see][]{Narayan2003}. In observation,  detection of QPOs in various kinds of AGNs, including both radio-loud and radio-quiet, on various timescales has been reported in several works \citep[see][and the reference therein]{Bhatta2019,Bhatta2018d, Bhatta2017, bhatta16c}. In addition, QPOs have been observed to naturally develop in numerical studies involving simulations of the parsec scale relativistic jets \citep[e.g.][]{McKinney12}. 

 The state-of-the-art telescopes and detectors have enabled us to obtain a fair comprehension of these fascinating sources. In spite of the efforts to understand them, the details of processes including the nature of accretion processes, disk-jet connection and the role the magnetic field in launching the jets are still elusive. In such context, the current work is primarily motivated to the characterization of the statistical properties of   \gama-ray variability in blazars. The sources form a dominant group of sources that prominently shine in the $\gamma$-ray band: the recent fourth Fermi Large Area Telescope source catalog (4FGL) contains about 60\% of the \gama-detected sources as the blazar class \citep{Fermi-LAT2019}. Therefore, study of \gama-ray emission from blazar can compliment similar studies on the origin and propagation of high energy emission in the Universe \citep[see][]{2019Galax...7...28R,Madejski2016}. 

In this work, we carry out systematic in-depth analysis of 20 blazars utilizing decade long Fermi/LAT observations.  In Section \ref{sec:2}, the sample of the blazars and its physical properties are listed in Table \ref{table:1}. In addition, data processing method for \fermi/LAT instrument is outlined.  In Section \ref{sec:3}, several approaches to the analysis adopting various methods including fractional variability, flux distribution, PSD and QPO are introduced,  and also the results of the analyses on the \gama-ray light curves are presented. Then discussion on the results along with their possible implications on the nature of \gama-ray emission from the sources are presented in Section \ref{sec:4}, and  we summarize our  conclusions at the end in Section \ref{sec:5}. Additionally, one table and several figures resulting from the analyses are placed in Appendices.

\section{Sample sources and Fermi/LAT data processing \label{sec:2}}

 Source Sample: We included most of the 
Fermi/Third Source Catalog (3FGL; \citealt{Acero2015}) sources for which there could be weekly flux with significant test statistic (TS) value.  The sources included in the study are generally \gama-ray bright (mostly TeV blazars) and consists of 12 BL Lacs  and 8 FSRQs.  The source names, their 3FGL  catalog name, source classification, RA, Dec and red-shift are presented in column 1, 2, 3, 4, 5 and 6, respectively, of Table \ref{table:1}.  Of the 20 sources, the source with the highest red-shift is PKS 1502+106 with $z=1.84$, and the closest one ($z=0.03$) is Mrk 421.

\begin{deluxetable*}{llllllll}
\tablecaption{The source sample of the 
Fermi/LAT blazars \label{table:1}}
\tablewidth{500pt}
\tabletypesize{\scriptsize}
\tablehead{
\colhead{Source name} & \colhead{3FGL name} & 
\colhead{Source class} & \colhead{R.A. (J2000)} & 
\colhead{Dec. (J2000)} & \colhead{Red-shift} & 
\colhead{FV (\%)} & \colhead{$\beta \pm \Delta \beta$}\\
} 
\colnumbers
\startdata
	3C 66A 	&	3FGL J0222.6+4301 	&	BL Lac 	&	 $02^h22^m41.6^s$ 	&	 $+43^d02^m35.5^s$ 	&	0.444	&	58.43	$\pm$1.78 	&	0.90	$\pm$0.17	\\			
	AO 0235+164 	&	3FGL J0238.6+1636	&	BL Lac 	&	 $02^h 38^m38.9^s$ 	&	 $+16^d 36^m 59^s$ 	&	0.94	&	95.53	$\pm$1.12 	&	1.40	$\pm$0.19	\\
	PKS 0454-234 	&	3FGLJ0457.0-2324	&	BL Lac 	&	 $04^h 57^m03.2^s$ 	&	 $-23^d 24^m 52^s$ 	&	1.003	&	68.25	$\pm$1.06	&	1.10	$\pm$0.09 	\\
	S5 0716+714 	&	3FGL J0721.9+7120 	&	BL Lac 	&	$07^h21^m53.4^s$ 	&	 $+71^d20^m36^s$ 	&	0.3	&	62.20	$\pm$1.05 	&	1.00	$\pm$0.15 	\\
	Mrk 421 	&	3FGLJ1104.4+3812 	&	BL Lac 	&	 $11^h04^m273^s$ 	&	 $+38^d12^m32^s$ 	&	0.03	&	43.65$\pm$1.45	&	1.00	$\pm$0.08	\\
	 TON 0599 	&	3FGL J1159.5+2914 	&	BL Lac 	&	$11^h59^m31.8^s$ 	&	 $+29^d14^m44^s$ 	&	0.7247	&	111.69$\pm$0.88	&	1.30	$\pm$0.15 	\\
	ON +325 	&	3FGL J1217.8+3007 	&	BL Lac 	&	 $12^h17^m52.1^s$ 	&	 $+30^d07^m01^s$ 	&	0.131	&	43.78	$\pm$4.60 	&	0.80	$\pm$0.14	\\
	W Comae 	&	3FGL J1221.4+2814	&	BL Lac 	&	 $12^h 21^m31.7^s$ 	&	 $+28^d 13^m 59^s$ 	&	0.102	&	24.70	$\pm$8.87	&	1.10	$\pm$0.09	\\
	4C +21.35 	&	3FGLJ1224.9+2122 	&	FSRQ 	&	 $12^h24^m54.4^s$ 	&	 $+21^d22^m46^s$ 	&	0.432	&	114.91$\pm$0.59 	&	1.10$\pm$0.12	\\
	3C 273 	&	3FGL J1229.1+0202 	&	FSRQ 	&	 $12^h29^m06.6997^s$ 	&	 $+02^d03^m08.598^s$ 	&	0.158	&	94.66$\pm$0.98 	&	1.20$\pm$0.17 	\\
	3C 279 	&	3FGL J1256.1-0547 	&	FSRQ 	&	 $12^h56^m11.1665^s$ 	&	 $-05^d47^m21.523^s$ 	&	0.536	&	104.29$\pm$	0.46	&	1.10$\pm$0.16	\\
	PKS 1424-418 	&	3FGLJ1427.9-4206 	&	 FSRQ	&	 $14^h27^m56.3^s$ 	&	 $-42^d06^m19^s$ 	&	1.522	&	70.44$\pm$0.69	&	 1.5$\pm$0.13	\\
	PKS 1502+106 	&	3FGLJ1504.4+1029	&	FSRQ 	&	 $15^h 04^m25^s.0$ 	&	 $+10^d 29^m 39^s$ 	&	1.84	&	90.11	$\pm$0.70	&	1.3$\pm$0.10 	\\
	4C+38.41 	&	3FGL J1635.2+3809 	&	FSRQ 	&	 $16^h35^m15.5^s$ 	&	 $+38^d08^m04^s$ 	&	1.813	&	92.99	$\pm$0.72 	&	1.2$\pm$0.15 	\\
	Mrk 501 	&	3FGL J1653.9+3945 	&	 BL Lac 	&	 $16^h53^m52.2167^s$ 	&	 $+39^d45^m36.609^s$ 	&	0.0334	&	33.47$\pm$	3.76	&	 1.10$\pm$10	\\
	1ES 1959+65 	&	3FGL J2000.0+6509 	&	BL Lac 	&	 $19^h59^m59.8521^s$ 	&	 $+65^d08^m54.652^s$ 	&	0.048	&	49.55$\pm$2.84	&	1.10	$\pm$0.14	\\
	PKS 2155-304 	&	3FGL J2158.8-3013 	&	BL Lac 	&	 $21^h58^m52.0651^s$ 	&	 $-30^d13^m32.118^s$ 	&	0.116	&	45.93	$\pm$2.02	&	0.90$\pm$0.20 	\\
	BL Lac 	&	3FGL J2202.7+4217 	&	BL Lac 	&	$22^h02^m43.3^s$ 	&	 $+42^d16^m40^s$ 	&	0.068	&	64.10$\pm$1.05	&	1.0$\pm0.10$ 	\\
	CTA 102 	&	3FGL J2232.5+1143 	&	FSRQ 	&	 $22^h32^m36.4^s$ 	&	 $+11^d43^m51^s$ 	&	1.037	&	117.42	$\pm$0.37 	&	1.20$\pm$0.19	\\
	3C 454.3  	&	3FGL J2254.0+1608 	&	FSRQ 	&	 $22^h53^m57.7^s$ 	&	 $+16^d08^m54^s$ 	&	0.859	&$81.30\pm0.30$		& 1.30$\pm0.17$		\\
\enddata
\end{deluxetable*}

 The Large Area Telescope (LAT) onboard Fermi Gamma-ray Space Telescope (\fermi) is one of the most useful instruments in the study of the Universe in high energy. It is equipped with a large effective area ($> 8000\ cm^{2}$), wide field of view ($>$ 2 sr) and high angular resolution ($ < 3.5^{o}$  around 100 MeV and $< 0.15^{o}$  above 10 GeV). The instrument continuously scans the sky every 90 minutes across a wide  spectral energy range that spans 20 MeV to TeV energies  \citep{Atwood2009}. However, for most of the practical purposes, the data analysis  is limited between the range 100 MeV--300 GeV. This is because a large point spread function (PSF) below the range and low event statistics above the range might render unreliable results.  For this work, the \fermi/LAT observations from the period 2008-08-04 -- 2018-06-22 ($\sim$ 10 years)  and in the 100 MeV--300 GeV energy range were considered for the analysis. The \fermi/LAT observations of the sources were processed following  the standard procedures of the unbinned likelihood  analysis\footnote{\url{https://fermi.gsfc.nasa.gov/ssc/data/analysis/scitools/likelihood_tutorial.html}}. In particular,  the Fermi Science Tools were used that made the use of the \fermi/LAT catalog, Galactic diffuse emission model and isotropic model for point sources.\footnote{\url{https://fermi.gsfc.nasa.gov/ssc/data/analysis/scitools/}} As a first step, selections of the events were made using the Fermi tool \emph{ gtselect} which selected only the events in a circular region of interest (ROI) of $10^{\circ}$ radius centered around the source.   To minimize the contamination of $\gamma$-rays from the Earth limb,  zenith angle was limited to  $ <$ 90$^{\circ}$. Similarly, the \fermi\ tool \emph{gtmktime} was used to select the good time intervals (GTI) to ensure that the satellite was operating in the standard science mode so that only good  quality of the observations enter the final analysis. After creating an exposure map using \emph{gtexpmap} and \emph{gtltcube}, a source model file was created  using the Python application make3FGLxml.py\footnote{\url{https://fermi.gsfc.nasa.gov/ssc/data/analysis/user/make3FGLxml.py}}. Subsequently, the diffuse source response  was calculated using the Galactic and extra-galactic models of the diffuse $\gamma$-ray emission, namely, \emph{ gll\_ iem v06.fit} and \emph{ iso\_P8R2 SOURCE V6 v06.txt}. To generate the light curves the data were binned in weekly bins and  the task \emph{ gtlike} was run to carry out maximum-likelihood analysis \citep{Mattox1996}. As one of the input parameters to the likelihood analysis, spectral index of the source model was frozen to the average index value from the 3FGL catalog.  With the set of the parameters given in the input source models, the task attempts to  maximize the probability, that the models represent the observations, by fitting all the sources within the ROI, and consequently computes  significance of the $\gamma$-ray events from the source. The maximum-likelihood test statistic, measuring significance of a detection, is given as
 TS = 2 $\times$ (logL$_{1}$ - logL$_{0})$, where L$_{1}$ and L$_{0}$ represent the likelihood of the data given the model with and without a point source at the position, respectively. Then the significance of a source detection can be expressed by $\sim \sqrt{TS} \sigma$ \citep{Abdo10}. In the current work, to ensure a robust analysis, only the observations  with TS value $>$ 10 (equivalently $\gtrsim 3 \sigma$) were included \citep[see also][]{Bhatta2019,Bhatta2017}.

\section{Analysis \label{sec:3}}
In order to constrain the statistical variability properties of the blazars, the \gama-ray light curves of the sample sources were intensively studied applying various analysis methods including fractional variability, flux distribution, RMS-flux relation, and symmetry analysis. Moreover, time series analysis in the form of power spectral density analysis, Lomb-Scargle periodogram, and weighted wavelet z-transform were carried out along with an extensive Monte Carlo (MC) simulations. The description of methods and the corresponding results of the analyses are presented below.

\begin{figure*}[t!]  
\gridline{
		\fig{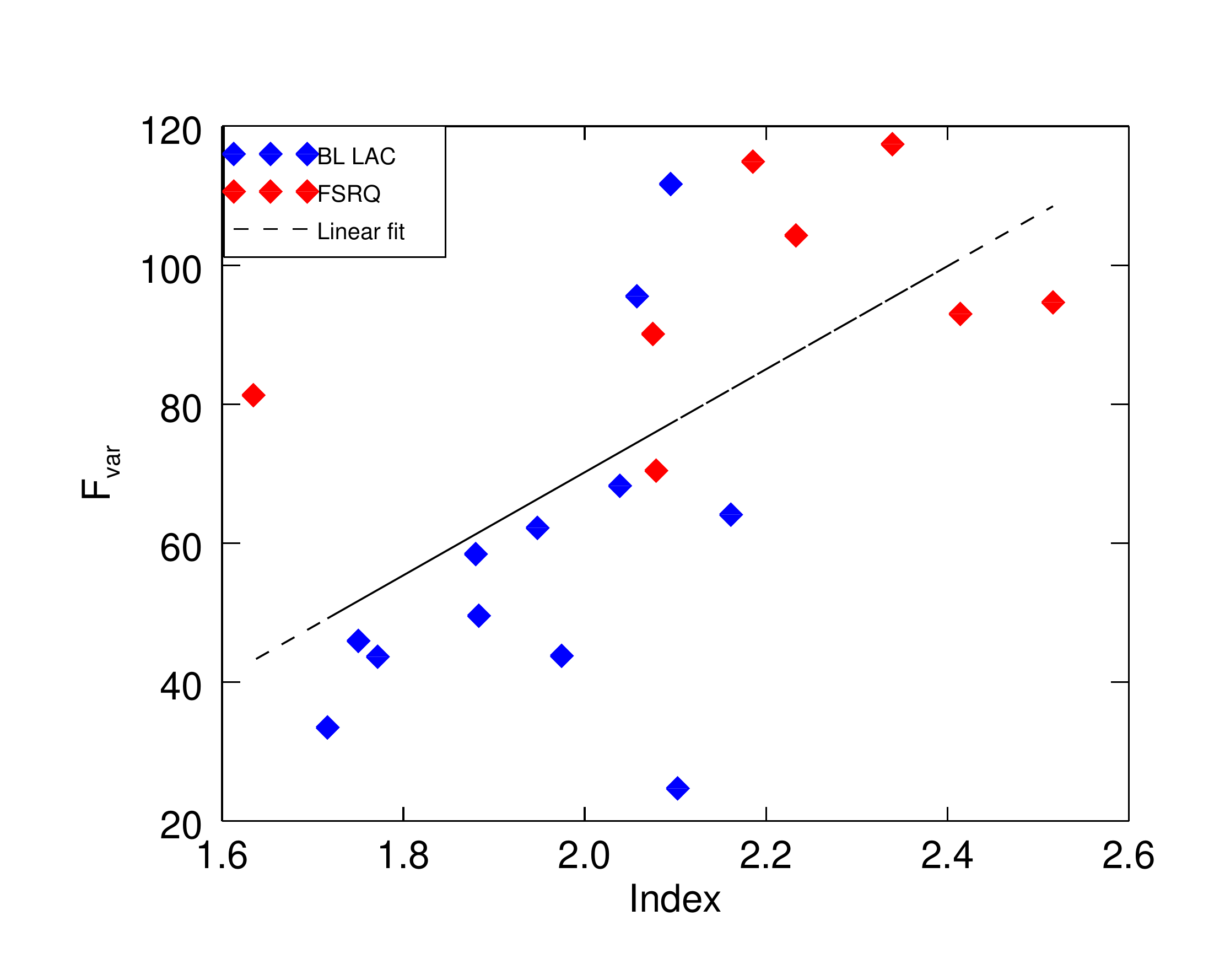}{0.5\textwidth}{}\hspace{-0.7cm}
    \fig{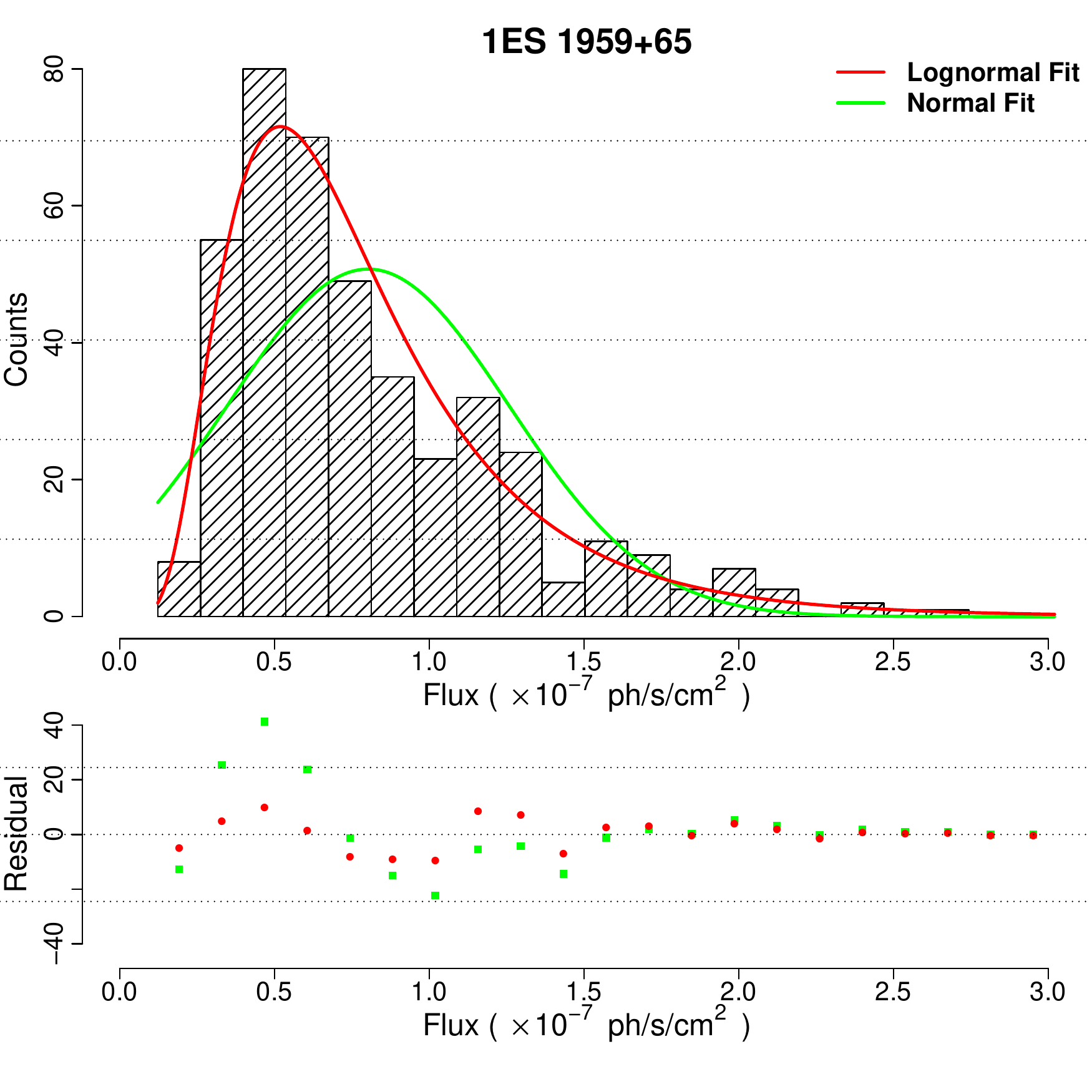}{0.4\textwidth}{}\hspace{-0.7cm}
}
\caption{\textit{Left:} Fractional variability plotted against corresponding Fermi/LAT band spectral indexes of the sources. BL Lacs and FSRQs are shown in blue and red symbols, respectively, and the black line shows the best linear fit to the data. \textit{Right}:  \gama-ray flux distribution (hatched black), and normal (green) and lognormal (red) PDF fitting to the histogram of the source 1ES 1959+65. Similar plots for other sources are presented in Appendix \ref{apndx3}.}
\label{fig:2}
\end{figure*}

\subsection{Fractional variability}
The decade-long light curves (e.g. see Figure  \ref{Fig:1}), distinctly revealing the variable nature of the sources,  imply that the sources might have undergone dramatically violent episodes to result in the observed large flux variation over the period. The average variability during the entire period can be quantified by estimating their fractional variability (FV; a measure of normalized excess variance) given as
\begin{equation}
F_{var}=\sqrt{\frac{S^{2}-\left \langle \sigma _{err}^{2} \right \rangle}{\left \langle F \right \rangle^{2}}} ,
\end{equation}
for which uncertainty can be expressed as
\begin{equation}
\centering
\sigma_{F_{var}}=\sqrt{ F_{var}^{2}+\sqrt{ \frac{2}{N}\frac{\left \langle \sigma _{err}^{2} \right \rangle^{2}}{\left \langle F \right \rangle^{4}}+  \frac{4}{N}\frac{\left \langle \sigma _{err}^{2} \right \rangle }{\left \langle F \right \rangle^{2}} F _{var}^{2}}} - F_{var}
\end{equation}
 (\citealt{Vaughan2003}, see also \citealt{Bhatta2018a}). The resulting FVs for the sample sources are listed in the 7th column of Table \ref{table:1}. The analysis shows that the blazar light curves display remarkable variability in the $\gamma$-ray band with the mean (of the sources in the sample)  FV of 73.37\% --  the mean FV of BL Lacs is 58.44 \% with  standard deviation of 24.83 \% and that of the FSRQs is 95.76\% with a standard deviation of 16.03 \%.  Of the sample sources, the most variable source is FSRQ CTA 102 (z= 1.037) with FV $\sim$117\%. Similarly, the next most variable sources are FSRQ 4C+21.35 and BL Lac TON 0599 with $\rm{FV}\sim115$ and $\sim$ 111\%, respectively. Whereas the least variable source turns out to be BL Lac W Comae with just FV $\sim$ 25\% followed by another BL Lac  Mrk 501 with FV $\sim$ 33\%. Although it appears that in general FSRQs sources are more variable than BL Lac sources in the sample, for a stronger conclusion, the analysis should be carried out on larger sample.

 As an attempt to characterize variability properties of the sources, we also studied correlation between the FV and the spectral indexes of the sources taken from the 3FGL catalog. In particular, the Pearson linear correlation coefficient between the parameters turned out to be 0.61 with a  p-value of 0.004.
The plot between FV and the spectral index is shown in the left panel of Figure \ref{fig:2}.  In the figure, BL Lacs and FSRQs are shown in blue and red color, respectively, and the best linear fit, with coefficient of determination (R$^2$)=0.3701 and p-value=0.004, is represented by the black line. The fit results in a steep slope of $\sim 70$ signifying high sensitiveness of variability on the spectral index. Although the sample is small, it suggests a hint of positive correlation between the quantities and encourages to carry out similar future studies involving large number of sources. Among the sources, two outliers are easily distinguished visually: FSRQ 3C 454.3 with flatter spectral index and high variability and BL Lac W Comae with steeper index and lowest FV.

\subsection{Flux distribution: Lognormality and RMS-flux relation}
 A study of the long term flux distribution of blazars can hold some of the important clues to the origin and nature of their variability. In particular, a statistical probe of probability density function (PDF) of gamma-ray flux can provide important insights  on the nature of high energy emission processes, and thereby help constrain the underlying processes that drive observed variability in  blazars. With such a goal, we studied \gama-ray flux distribution of the blazar samples by constructing the histograms to ascertain PDF of the distribution, which can be approximated  by the model fit to the distribution of the fluxes  from the long-term light curves.  We mainly attempted fitting normal and lognormal PDFs.

A normal distribution is defined by,
\begin{equation}
f_\mathrm{normal}\left(x\right) = \exp(-\frac{(x - \mu)^{2}}{2\sigma^{2}}) \mathrm{\quad ,}
\end{equation}
where $\mu$ and $\sigma$ are the mean and the standard deviation of the normal distribution, respectively, expressed in the unit of flux, i. e., counts/sec/cm$^2$. Similarly  lognormal distribution is defined by,
\begin{equation}
f_\mathrm{lognormal}(x) = \frac{1}{x s \sqrt{2\pi}} \exp({{-\frac{\left(\ln x\  -\  m\right)^{2}}{2 s^{2}}}}) \mathrm{\quad ,}
\end{equation} 
where $m$ and $s$  are the mean location and the scale parameters of the distribution, respectively; and $m$ is expressed in the unit of natural log of flux.

First, we performed curve fitting to the flux histograms of the sample sources using the above two PDFs by employing weighted least-square (WLS) method on the binned data, and the resulted fit statistics for both lognormal and normal PDFs are listed in Appendix \ref{apndx2}. The mean of the flux, scale  and reduced $\chi^2$ for the lognormal fitting are listed in column 2, 3, and 4, respectively. Similarly,  the mean, standard deviation and reduced $\chi^2$ for the normal fitting are presented in column 5, 6, and 7, respectively.  Based on the reduced $\chi^2$ i.e.,  $\chi^2$/dof, we find that for most of the sources lognormal PDF fits better than normal PDF. However,  in the sources AO 0235+164, ON+325, BL Lac and 3C 454.3  the reduced $\chi^2$ for both of the PDFs are comparable and normal distribution fits provide slightly better representation. We also studied the flux distribution using more robust maximum likelihood estimation (MLE) method, and implemented the PDF fitting using the software package \emph{fitdistrplus}\footnote{\url{https://cran.r-project.org/web/packages/fitdistrplus/index.html}} \citep{Delignette2015} available in the MASS library of the public domain R statistical software system. The package attempts to fit PDFs to the unbinned  flux distribution using MLE method, as opposed to the PDF fitting on the binned histogram in WLS method. The resulting fit statistics along with the log-likelihood (LL), Bayesian information criterion (BIC) and Akaike information criterion (AIC) quantities for the sample sources are presented in Table \ref{table:2}.  The smaller AIC and BIC  values for the log-normal PDF suggest that, compared to the normal PDF, it is more preferable representation of blazar flux distribution. The fitting of histogram of the source 1ES 1959+65 using normal and lognormal distributions and the corresponding residuals are shown in the right panel of Figure \ref{fig:2}, and similar plots for the rest of the sources in the sample are presented in Appendix \ref{apndx3}. From the figures, it is evident that the observed flux distribution is asymmetric with a heavy tail.

\begin{deluxetable*}{llllll|lllll}
\tablecaption{ Lognormal and normal distribution fit statistics for the \gama-ray flux distribution of the Fermi/LAT sources using maximum likelihood estimation method. Similar table using weighted least-square method is presented in Appendix \ref{apndx2} \label{table:2}}
\tablewidth{500pt}
\tabletypesize{\scriptsize}
\tablehead{
\colhead{} & 
\multicolumn{3}{c}{Lognormal fit} & \multicolumn{3}{c}{Normal fit} \\
\hline
\colhead{Source name} &
\colhead{$m$} &\colhead{$s$} &\colhead{$LL$} & 
\colhead{AIC}  &  \colhead{$BIC$} &
\colhead{$\mu$} & \colhead{$\sigma$} &\colhead{$LL$} &
\colhead{AIC}  &  \colhead{$BIC$} 
} 
\colnumbers
\startdata
3C 66A  & -0.03 $\pm$ 0.02  & 0.52 $\pm$ 0.02 &  -369  &743  &  751  & 1.12 $\pm$ 0.03  & 0.70 $\pm$ 0.02 &  -526  &1057  &  1065 \\
  AO 0235+164  & 0.43 $\pm$ 0.05  & 0.78 $\pm$ 0.03 &  -439  &882  &  889  & 2.16 $\pm$ 0.13  & 2.09 $\pm$ 0.09 &  -588  &1181  &  1188 \\
  PKS 0454-234  & 0.72 $\pm$ 0.03  & 0.71 $\pm$ 0.02 &  -848  &1700  &  1708  & 2.61 $\pm$ 0.08  & 1.82 $\pm$ 0.06 &  -952  &1909  &  1917 \\
  S5 0716+714  & 0.60 $\pm$ 0.03  & 0.68 $\pm$ 0.02 &  -799  &1602  &  1610  & 2.25 $\pm$ 0.06  & 1.43 $\pm$ 0.05 &  -870  &1745  &  1753 \\
  Mrk 421  & 0.65 $\pm$ 0.02  & 0.41 $\pm$ 0.01 &  -603  &1210  &  1218  & 2.09 $\pm$ 0.04  & 0.96 $\pm$ 0.03 &  -699  &1403  &  1411 \\
  TON 0599  & 0.35 $\pm$ 0.04  & 0.77 $\pm$ 0.03 &  -536  &1077  &  1084  & 2.02 $\pm$ 0.12  & 2.28 $\pm$ 0.09 &  -796  &1597  &  1605 \\
  ON +325  & -0.32 $\pm$ 0.03  & 0.64 $\pm$ 0.02 &  -289  &581  &  590  & 0.86 $\pm$ 0.02  & 0.51 $\pm$ 0.02 &  -331  &666  &  674 \\
  W Comae  & -0.49 $\pm$ 0.03  & 0.40 $\pm$ 0.02 &   -3  & 11  &   18  & 0.66 $\pm$ 0.02  & 0.29 $\pm$ 0.01 &  -35  & 74  &   80 \\
  4C +21.35  & 0.91 $\pm$ 0.05  & 0.97 $\pm$ 0.04 &  -854  &1712  &  1720  & 4.02 $\pm$ 0.24  & 4.63 $\pm$ 0.17 &  -1101  &2206  &  2214 \\
  3C 273  & 0.96 $\pm$ 0.04  & 0.69 $\pm$ 0.03 &  -727  &1459  &  1466  & 3.45 $\pm$ 0.17  & 3.31 $\pm$ 0.12 &  -950  &1904  &  1911 \\
  3C 279  & 1.39 $\pm$ 0.04  & 0.87 $\pm$ 0.03 &  -1342  &2688  &  2696  & 6.02 $\pm$ 0.28  & 6.30 $\pm$ 0.20 &  -1637  &3277  &  3286 \\
  PKS 1424-418  & 1.52 $\pm$ 0.03  & 0.68 $\pm$ 0.02 &  -1203  &2410  &  2418  & 5.69 $\pm$ 0.19  & 4.04 $\pm$ 0.13 &  -1332  &2667  &  2676 \\
  PKS 1502+106  & 1.01 $\pm$ 0.04  & 0.87 $\pm$ 0.03 &  -878  &1760  &  1768  & 3.93 $\pm$ 0.18  & 3.57 $\pm$ 0.13 &  -1033  &2071  &  2078 \\
  4C +38.41  & 0.81 $\pm$ 0.04  & 0.81 $\pm$ 0.03 &  -937  &1877  &  1886  & 3.20 $\pm$ 0.14  & 3.01 $\pm$ 0.10 &  -1164  &2332  &  2340 \\
  Mrk 501  & -0.63 $\pm$ 0.02  & 0.45 $\pm$ 0.01 &    7  &-10  &   -1  & 0.59 $\pm$ 0.01  & 0.25 $\pm$ 0.01 &  -20  & 44  &   52 \\
  1ES 1959+65  & -0.37 $\pm$ 0.03  & 0.54 $\pm$ 0.02 &  -182  &369  &  377  & 0.80 $\pm$ 0.02  & 0.46 $\pm$ 0.02 &  -266  &536  &  544 \\
  PKS 2155-304  & 0.08 $\pm$ 0.02  & 0.47 $\pm$ 0.01 &  -383  &769  &  778  & 1.21 $\pm$ 0.03  & 0.61 $\pm$ 0.02 &  -467  &939  &  947 \\
  BL Lac  & 1.08 $\pm$ 0.03  & 0.66 $\pm$ 0.02 &  -991  &1987  &  1995  & 3.63 $\pm$ 0.11  & 2.38 $\pm$ 0.08 &  -1085  &2175  &  2183 \\
  CTA 102  & 1.58 $\pm$ 0.05  & 1.08 $\pm$ 0.04 &  -1309  &2623  &  2631  & 8.66 $\pm$ 0.49  & 10.18 $\pm$ 0.35 &  -1589  &3182  &  3190 \\
  3C 454.3  & 2.25 $\pm$ 0.05  & 1.00 $\pm$ 0.03 &  -1694  &3392  &  3401  & 14.11 $\pm$ 0.53  & 11.49 $\pm$ 0.38 &  -1784  &3571  &  3580 \\
\enddata
\tablecomments{For the normal fit $\mu$ and $\sigma$ are presented in the unit of  flux in 10$^{-7}$ $\times$ counts/sec/cm$^2$, whereas for the lognormal fit $m$ is in the unit of natural log of flux.}
\end{deluxetable*}

\begin{figure*}[t!]  
\begin{center}  
\plottwo{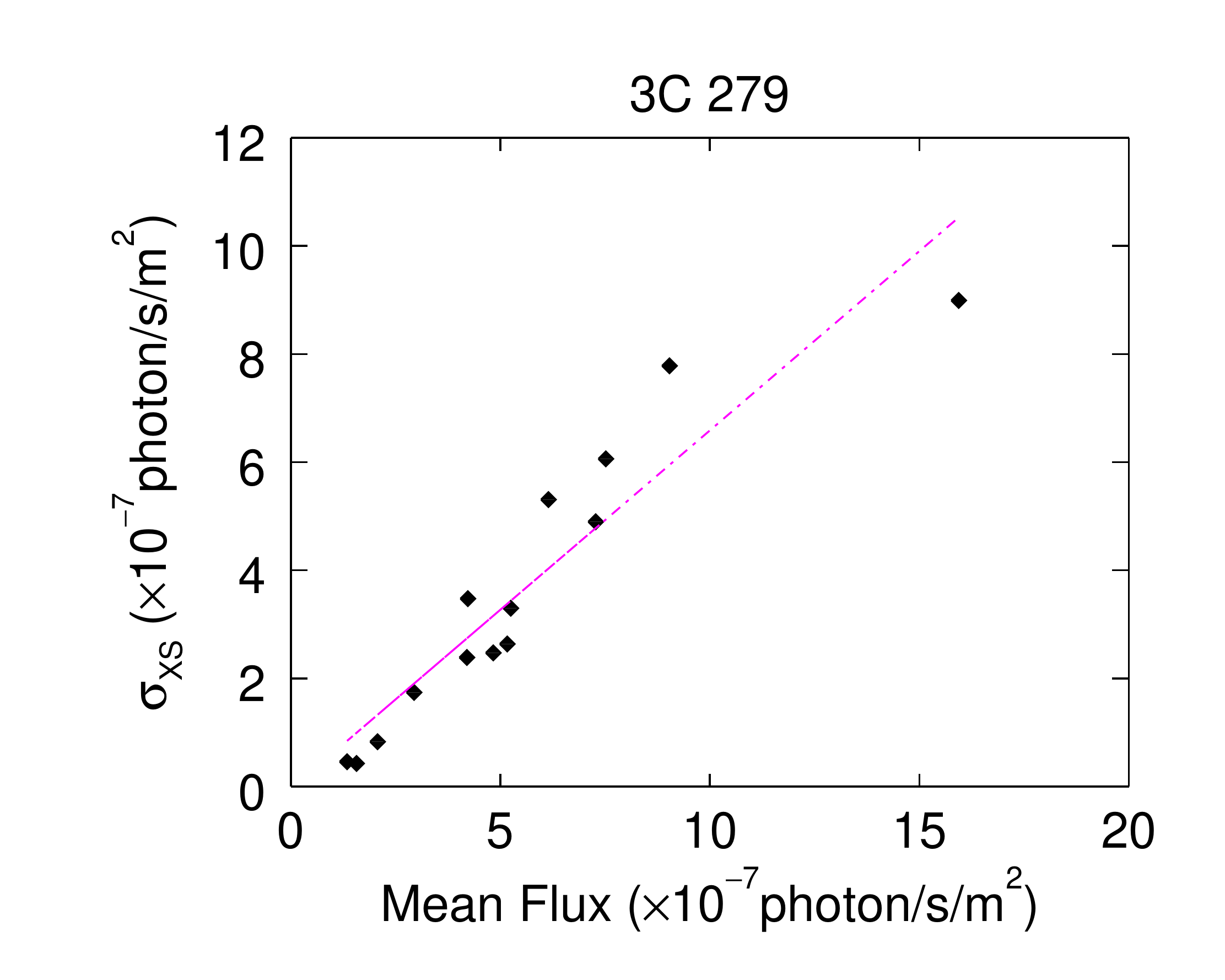}{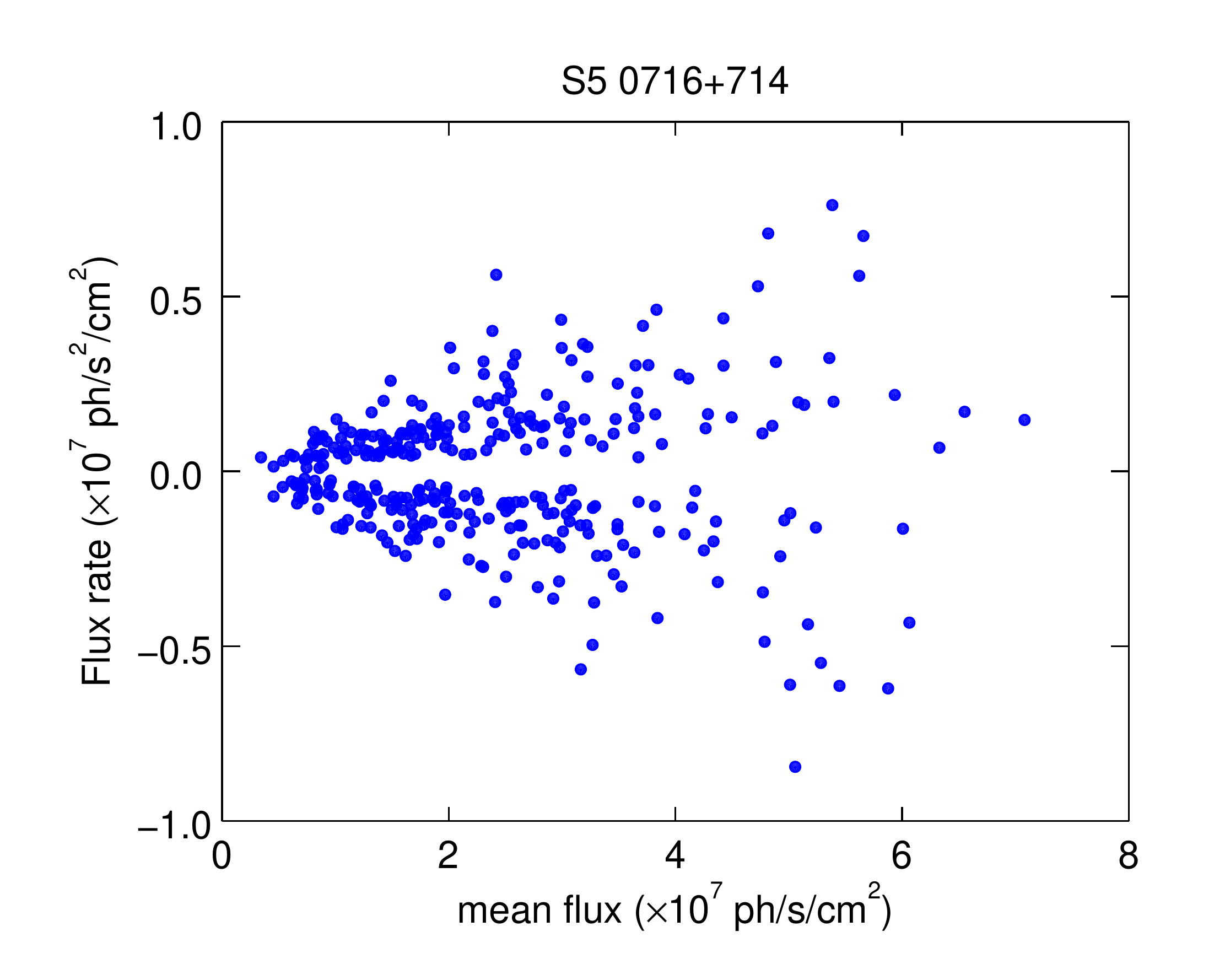}
\caption{\textit{Left:} RMS-flux relation in the FSRQ 3C 279. The magenta line represents the linear fit to the observations.  Similar figures for other sources are presented in Appendix \ref{apndx4}. \textit{Right:} Distribution of rates of flux changes over the mean of the consecutive fluxes of the BL Lac S5 0716+714.}
\label{fig:3}
\end{center}
\end{figure*}

We further tried to characterize the flux distribution and variability properties by investigating  the correlation between the flux states and source activity as represented by  root mean square (RMS), commonly known as \emph{RMS-flux relation}. For the purpose, the light curves were divided into N segments of equal lengths such that each segment contains at least 20 observations. This is to ensure that we can conduct a meaningful statistical analysis. For each segment of the light curve, the Poisson noise corrected excess variance is given as $\sigma _{XS}^{2}=S^{2}-\bar{\sigma}_{err} ^{2}$; where $S^{2}$ represents the sample variance and $\bar{\sigma}_{err} ^{2}$ is the mean square of measurement error given by $\bar{\sigma}_{err} ^{2}=1/n\sum_{i}^{n}\sigma_{err,i} ^{2}$. From the light curve of the source 3C 279, the RMS values for each segment are plotted against the corresponding mean flux values in left panel Figure \ref{fig:3}. The magenta line on the figure represents the linear fit to the observations and also serves as a visual guide to the trend that appears on the RMS-flux plane. Similar figures for rest of the sample sources are presented in Appendix \ref{apndx3}.  We see that a linear trend distinctly appears in  most of the sources.  The slope parameters from the linear fit are listed in the 2nd column of Table \ref{tab:table3}. The mean slope of the sources is 0.47; BL Lac ON +325 has the flattest slope of 0.03 whereas another BL Lac  TON 0599 shows the steepest slope 0.82. It is seen that, in general, BL Lacs show flatter average slope 0.43 in comparison to the steeper average slope 0.56 for FSRQs. To further quantify the correlation between the RMS and flux in the sources, Spearmann rank correlation coefficients are estimated along with the corresponding $p$-values that represent the two-sided significance of its deviation from zero, which are presented in the 3rd and 4th column of the table, respectively. As the $p$-values indicate, the linear correlation between the flux and RMS seems to be a dominant trend in most of the sample sources. It is important to note that, except for the source ON +325, all the sources in the sample display a strong linear RMS-flux relation.  Similar results were reported in the work of \cite{Kushwaha2017} studying Fermi/LAT observations of 4 AGN.

\subsection{ Symmetry Analysis: Flux rise and decay profiles}
To investigate into the nature of particle acceleration that results in the flux rise and energy dissipation mechanism causing flux decay in a source light curve, the distribution of positive and negative  flux rates are studied by considering the two consecutive fluxes in the light curve. Such a statistical analysis, in principle, should reveal the inherent difference between the nature of acceleration and cooling mechanisms.  To carry out the analysis, flux rates between two fluxes consecutive in times were estimated as $\text{rate} ={\Delta F}/{\Delta t}$. This provides a simple measure of how swiftly the fluxes rise or decay in the weekly time bin.  A comparison between rise and decay rates (positive and negative rates, respectively) in all of the sources in the sample are presented in Table \ref{tab:4}.  Here for each of the sources the ${\Delta F}$s are normalized by the mean flux of the entire light curve so that the flux change rates can be expressed in percent per day and thereby conveniently compared with the values for the other sources in the sample.  We find that the positive and negative flux change rates for each of the sources in the sample, as  listed in 2nd and 3rd column of the table, are very similar.  As an  illustration, the flux rise and decay rates against the mean flux for the source S5 0716+714 are plotted  in the right panel of Figure \ref{fig:3}.   It is interesting to note that a similar conclusion was inferred in \citet{Abdo10} in the similar studies of a large number of sources in the Fermi/LAT observations. Furthermore, in order to see how similar/dissimilar are  the distribution of decay rates  from the rise rates, Kolmogorov-Smirnov (K-S) test was performed.  The K-S statistic (D), listed in the 4th column of Table \ref{tab:4}, specifies the maximum deviation between the cumulative distributions of the two samples. The p-value corresponding to the K-S statistic can be used to infer whether both samples can be considered as being drawn from the same parent population. As indicated by the p-values in the 5th column of Table \ref{tab:4}, none of the sources have the flux rise and decay rates that are significantly different from each other.  

Moreover, any two consecutive fluxes and their mean can be used to compute  rise/decay timescales as given by 
\begin{equation}
\frac{1}{\tau_\pm}=\pm \frac{\Delta F}{\Delta t}\frac{1}{\left \langle F \right \rangle},
\end{equation}
where $\Delta t=t_i-t_{i+1},\ \Delta F=F_i-F_{i+1}$ and $\left \langle F \right \rangle=(F_i+F_{i+1})/2$. Note that this timescale can be taken as a measure for the flux-doubling timescale.
The average of such timescales along with the corresponding 1$\sigma$ are listed in the 6th column of Table \ref{tab:4}. We find  that in almost all sources the flux rise and decay timescales are very similar, and are these are in the order of a few weeks.

\begin{table}
\begin{center}
\caption{Relation between RMS and the mean flux of  \gama-ray  light curves of the Fermi/LAT blazars \label{tab:table3}}
\begin{tabular}{lccl}
\hline
\hline
Source	&	slope	&	Spearman's r. c. ($\rho$)	&	$p$-value	\\
\colnumbers\\
\hline
3C 66A 	&	0.70	&	0.73	&	0.006	\\
AO 0235+164 	&	0.59	&	0.94	&	$< 0.001$	\\
PKS 0454-234 	&	0.47	&	0.73	&	0.007 	\\
S5 0716+714 	&	0.39	&	0.93	&	0.006	\\
Mrk 421 	&	0.65	&	0.78	&	 0.001	\\
 TON 0599 	&	0.82	&	0.94	&	$< 0.001$		\\
ON +325 	&	0.03	&	0.04	&	0.09	\\
W Comae 	&	0.22	&	0.80	&	0.02	\\
4C +21.35 	&	0.61	&	0.86	&	$< 0.001$		\\
3C 273 	&	0.77	&	0.90	&	 $< 0.001$		\\
3C 279 	&	0.66	&	0.96	&	 $< 0.001$		\\
PKS 1424-418 	&	0.25	&	0.71	&	0.004	\\
PKS 1502+106 	&	0.46	&	0.96	&	$< 0.001$		\\
4C+38.41 	&	0.48	&	0.96	&	 	$< 0.001$		\\
Mrk 501 	&	0.39	&	0.68	&	0.007	\\
1ES 1959+65 	&	0.32	&	0.85	&	 0.002	\\
PKS 2155-304 	&	0.39	&	0.68	&	0.007	\\
BL Lac 	&	0.35	&	0.78	&	$< 0.001$		\\
CTA 102 	&	0.52	&	0.95	&	 $< 0.001$		\\
3C 454.3  	&	0.42	&	0.90	&	 $< 0.001$		\\
\hline
\end{tabular}
\end{center}
\end{table}

\begin{table}
\begin{center}
\caption{Symmetry analysis of  \gama-ray  fluxes of the source light curves\label{tab:4}}
\begin{tabular}{lccccc}
\hline
\hline
source & rise rate (\%\ d$^{-1}$)  & decay  rate (\%\ d$^{-1}$)& D& $p$-value & ave. timescale (d)\\
\colnumbers\\
\hline
3C 66A	&	7.28	$\pm$ 	6.36	&	7.50	$\pm$ 	7.23	&	0.12	&	0.33	&	17.51	$\pm$ 	7.72	\\
AO 0235+164	&	8.36	$\pm$ 	6.51	&	7.85	$\pm$ 	6.64	&	0.11	&	0.70	&	21.98	$\pm$ 	21.67	\\
PKS 0454-234	&	6.04	$\pm$ 	5.61	&	5.76	$\pm$ 	4.55	&	0.08	&	0.63	&	22.19	$\pm$ 	11.90	\\
S5 716+714	&	7.06	$\pm$ 	5.92	&	7.11	$\pm$ 	5.96	&	0.06	&	0.93	&	20.10	$\pm$ 	13.02	\\
Mrk 421	&	5.00	$\pm$ 	3.16	&	5.00	$\pm$ 	3.01	&	0.10	&	0.51	&	24.94	$\pm$ 	12.22	\\
TON 0599	&	7.39	$\pm$ 	7.31	&	7.04	$\pm$ 	7.15	&	0.10	&	0.55	&	20.61	$\pm$ 	20.27	\\
ON +325	&	10.62	$\pm$ 	6.71	&	11.27	$\pm$ 	7.78	&	0.09	&	0.71	&	11.84	$\pm$ 	5.93	\\
W Comae	&	5.82	$\pm$ 	3.78	&	6.29	$\pm$ 	5.05	&	0.12	&	0.92	&	31.72	$\pm$ 	29.80	\\
4C +21.35	&	8.45	$\pm$ 	6.48	&	7.49	$\pm$ 	6.50	&	0.15	&	0.13	&	21.74	$\pm$ 	21.52	\\
3C 273	&	8.82	$\pm$ 	9.67	&	7.52	$\pm$ 	6.21	&	0.11	&	0.52	&	21.98	$\pm$ 	17.22	\\
3C 279	&	7.77	$\pm$ 	6.77	&	7.34	$\pm$ 	8.83	&	0.09	&	0.51	&	21.55	$\pm$ 	14.92	\\
PKS 1424-418	&	3.77	$\pm$ 	2.88	&	4.07	$\pm$ 	3.25	&	0.11	&	0.25	&	32.22	$\pm$ 	19.14	\\
PKS 1502+106	&	4.98	$\pm$ 	4.05	&	5.66	$\pm$ 	5.82	&	0.08	&	0.84	&	28.14	$\pm$ 	27.89	\\
4C +38.41	&	6.52	$\pm$ 	6.49	&	6.44	$\pm$ 	6.17	&	0.06	&	0.93	&	22.34	$\pm$ 	15.19	\\
Mrk 501	&	7.52	$\pm$ 	3.65	&	7.48	$\pm$ 	3.77	&	0.06	&	0.97	&	16.21	$\pm$ 	7.78	\\
1ES 1959+65	&	7.23	$\pm$ 	4.26	&	7.73	$\pm$ 	4.74	&	0.11	&	0.70	&	17.29	$\pm$ 	9.74	\\
PKS 2155-304	&	6.13	$\pm$ 	3.96	&	6.77	$\pm$ 	4.11	&	0.11	&	0.28	&	19.18	$\pm$ 	9.02	\\
BL Lac	&	6.40	$\pm$ 	4.96	&	6.28	$\pm$ 	5.01	&	0.08	&	0.69	&	20.44	$\pm$ 	11.96	\\
CTA 102	&	6.30	$\pm$ 	6.19	&	6.55	$\pm$ 	6.13	&	0.10	&	0.42	&	23.59	$\pm$ 	20.87	\\
3C 454.3	&	4.99	$\pm$ 	4.63	&	4.87	$\pm$ 	4.97	&	0.08	&	0.60	&	30.86	$\pm$ 	22.03	\\
\hline
\end{tabular}
\end{center}
\end{table}

\subsubsection{ Time Series Analysis}

\subsubsection*{ Power Spectral Density \label{PSD}}

 Discrete Fourier periodogram (DFP) of a light curve of a variable source provides a measure for the variability power at a given temporal frequency (or, equivalently, timescale). Mathematically, it can be given by the square of  absolute value of discrete Fourier transform.  For a time series $x(t_{j})$ sampled at times $t_{j}$ with $j = 1,2,..,n$ and spanning a total duration observations, $T$, the DFP at a temporal frequency $\nu$ is expressed as
\begin{equation}
P\!\left(\nu \right)=\frac{1}{n} \, \left | \sum_{j=1}^{n}x\!\left( t_{j} \right ) \, e^{-i 2\pi \nu t_{j}} \right |^{2} \, .
\end{equation}
The periodograms are computed for $n/2$ frequencies that are evenly sampled in log-space between the minimum  $\nu_{min}=1/T$ and $\nu_{max}=1/2 \Delta t$, where $\Delta t$ is mean sampling step in the light curve.  Moreover, the periodogram can be normalized to express it in a convenient unit. In particular, if we normalize it with a factor $2 T/\left(n \bar{x} \right)^{2}$,  the unit  becomes (rms/mean)$^{2}$ /d. The main advantage of this normalization is that the total integrated power of periodogram is nearly equal to the variance of the light curve - a result following from the Parseval's theorem.  The distribution of DFP  over the temporal frequencies reveal variability power at the corresponding timescales, and thereby provide  information about the underlying variability structures and dominant timescales. Power spectral density (PSD)  is a mathematical function that best approximates the shape of a source periodogram.  In general, blazar periodograms have been found to be best approximated by power-law function of the form $P(\nu) \propto \nu^{-\beta}$ with spectral power index ${\beta}$. However, in reality true underlying PSD of a source light curve sampled at discrete times for a finite duration often gets distorted by the effects of sampling pattern as represented by window function. Therefore, it is important that any robust evaluation of  PSD of real astronomical observations should be able to carefully untangle the effects of  the window function on the observed PSD. 

 Power Spectrum Response method  \citep[PSRESP;][]{Uttley02} is  one of such methods frequently applied in the characterization of PSD of AGN periodogram \citep[see][and the references therein]{Bhatta2019,Bhatta2018b,bhatta16c}. The main merits of the method are that it  properly accounts a number of important issues relating to the blazar light curves such as  dominant red-noise,  discrete sampling, finite observation length and uneven sampling of the light curve.  Moreover, since the nature of distribution of periodograms of unevenly spaced light curves of power-law type PSD are not well understood, the distribution of a large number of simulated light curves, that posses similar statistical properties such as mean, standard deviation,  sampling pattern and observation duration, are utilized to compute a measure for the goodness of fit of a model PSD.  

As mentioned earlier, the sampling properties of the observed light curve can impose distorting effects on the true underlying PSD in many ways. In particular, variability power leakage from lower to higher frequencies owing to the limited observation period, commonly known as red noise leakage,  can  alter the true PSD shape by flattening the high frequency tail of the power-law function \citep{Papadakis1993}. During the spectral analysis of the blazar PSD, the effects of the red noise leak were corrected through extensive MC simulations. Particularly, to address this issue within the scheme of the PSRESP, light curves were simulated first 10 times longer than the total source observation duration, and then divided into segments of 10 data sets \citep{Isobe2015}. Similarly, in case of the light curve with the finite time resolution (or inadequate sampling rate), \textit{aliasing} can alter the shape of PSD also leading to the flattening of the high frequency tail of the true PSD. The effect in general can be avoided by sampling the periodogram up to the  Nyquist frequency \citep[see][]{Uttley02}. 

\begin{figure*}[t!]  
\begin{center}  
\plottwo{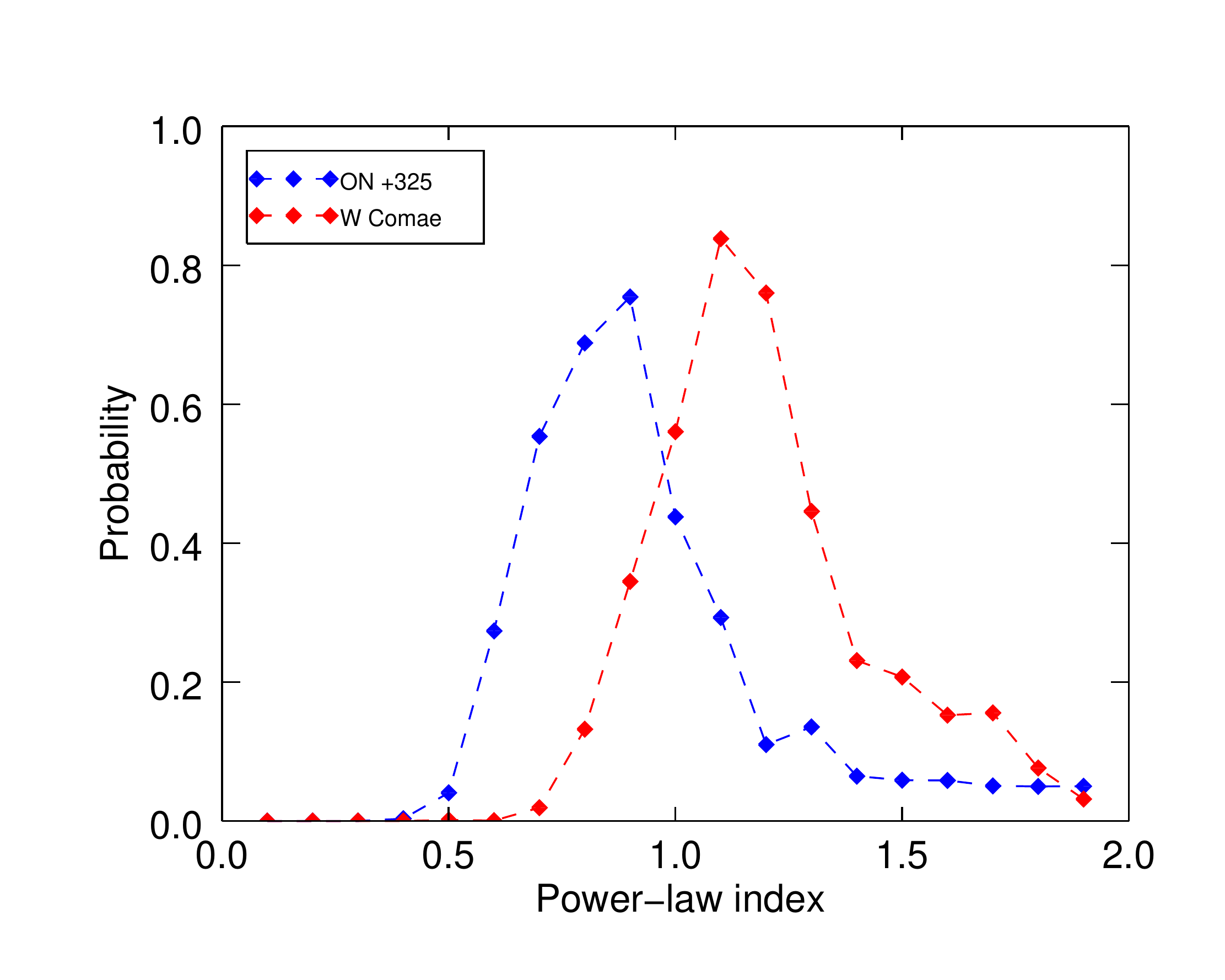}{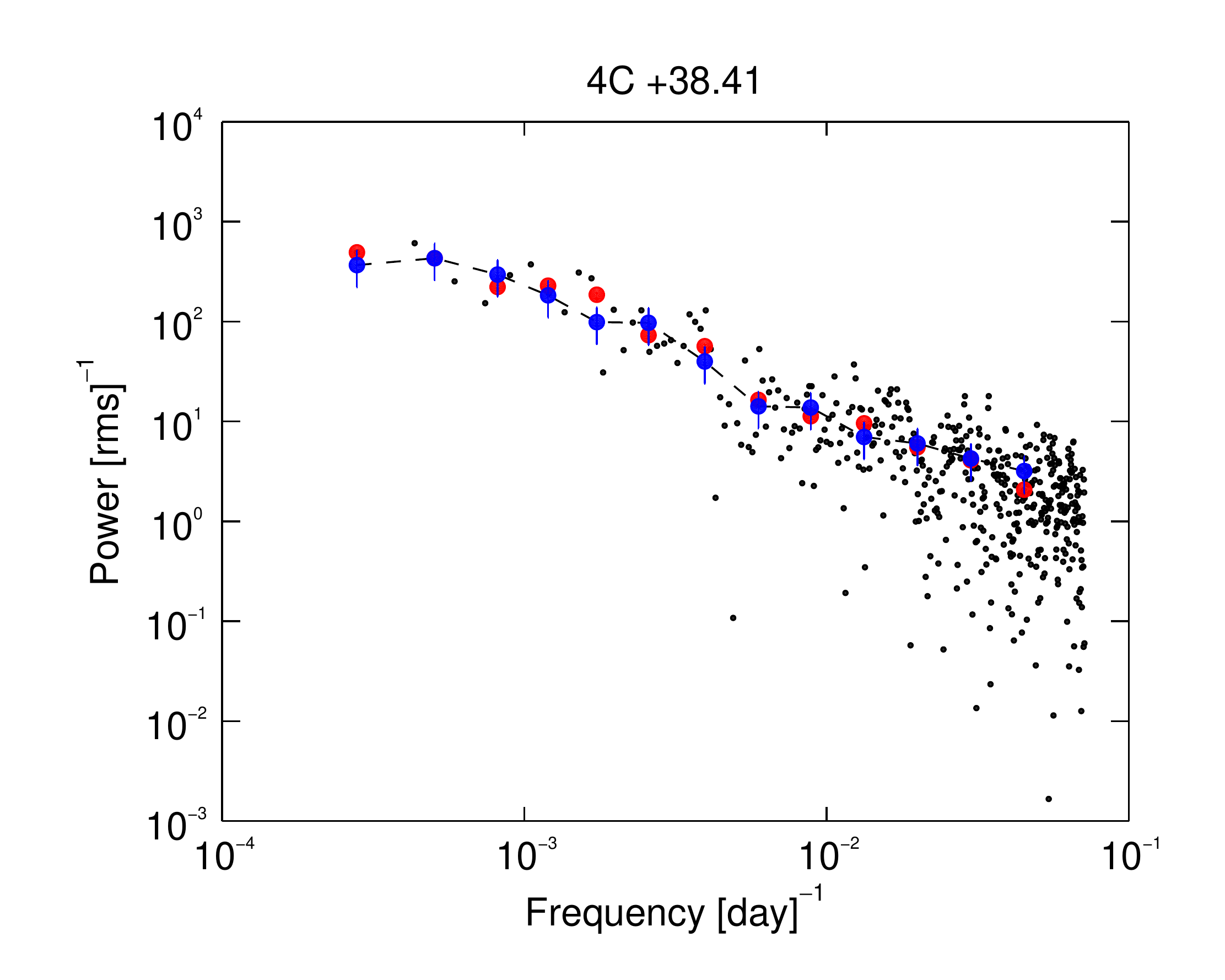}
\caption{\textit{Left:} Following PSRESP, distribution of probability, that given power-law PSD model best represents the source periodograms, is plotted over spectral indexes ranging from 0 to 2 for W Comae (red) and ON +325 (blue).  \textit{Right:} Discrete Fourier periodogram of the source 4C +38.41 (black), binned periodogram (red), and  best fit PSD (blue). Similar plots for other sources are presented in Appendix  \ref{apndx5}.}
\label{fig:4}
\end{center}
\end{figure*}

To implement PSRESP\footnote{We have described the PSRESP method and its implementation in detail in several of our previous works, \cite{Bhatta2016b,bhatta16c,Bhatta2017,Bhatta2019}}, 10000  light curves with a 7-day bin were simulated using single power law PSD model with spectral power index $\beta$ as given by $P(\nu) = \nu^{-\beta}+C$ \citep[see][]{TK95}. The  simulated light curves were assigned the same observational properties, e.g. mean, standard deviation, observational length and uneven sampling, and consequently periodograms for each simulated light curves were computed. The distribution of simulated periodograms in turn were utilized to estimate the best fit model PSDs for the source \gama-ray light curves. In the left panel of Figure \ref{fig:4}, the probability distribution over the spectral index are presented for the sources W Comae (red) and ON +325 (blue). The figure shows that the spectral indexes corresponding to the best-fit PSD are 1.10$\pm$0.09 (W Comae) and 0.84$\pm$0.14 (ON +325 ), where the half width at half maximum (HWHM) from the Gaussian fit to the observations are used to represent the uncertainties in the indexes. Following similar procedure, spectral indexes corresponding to the best-fit PSD model for the sample sources are listed in the 8th column of Table \ref{table:1}. In addition,  figures showing the DFP (black), binned periodogram (red) and the best-fit PSD model (blue) for the sample sources are presented  in Appendix \ref{apndx5}, whereas the plot showing the best-fit PSD for the source 4C+38.41 is presented in the right panel of Figure \ref{fig:4}.

It is found that the periodograms of  \gama-ray light curves of the 20  sources are consistent with a single power-law of the form $P(\nu) \propto \nu^{-\beta}$ where the slope index ranges between 0.8--1.5. The mean PSD index of all the sources in the sample turns out to be 1.13 with a standard deviation of 0.18. To compare between FSRQs and BL Lacs, mean index for BL Lacs is 1.05 with standard deviation 0.17; where as  the mean of FSRQs is 1.24 with a lesser spread of standard deviation 0.13. The results also show that the source with steepest index of 1.5 is PKS 1424-418 wheres the one with the flattest index tuns out to be ON +325. Our results are in close agreement with the recent results $1.15\pm0.10$ by \cite{Meyer2019}, and also largely in agreement with the work of \citet{Nakagawa13} using 4 year-long Fermi/LAT observations of 15 sources.  However, in a study of first 11 months of the Fermi survey involving several blazars,  \citet{Abdo10} reported steeper average slope indexes 1.5 for FSRQs and 1.7  for BL Lacs. The discrepancies can be ascribed to the difference in the method, sampling interval and total observation duration between the two works.

\begin{figure*}[t!]  
\begin{center}  
\plotone{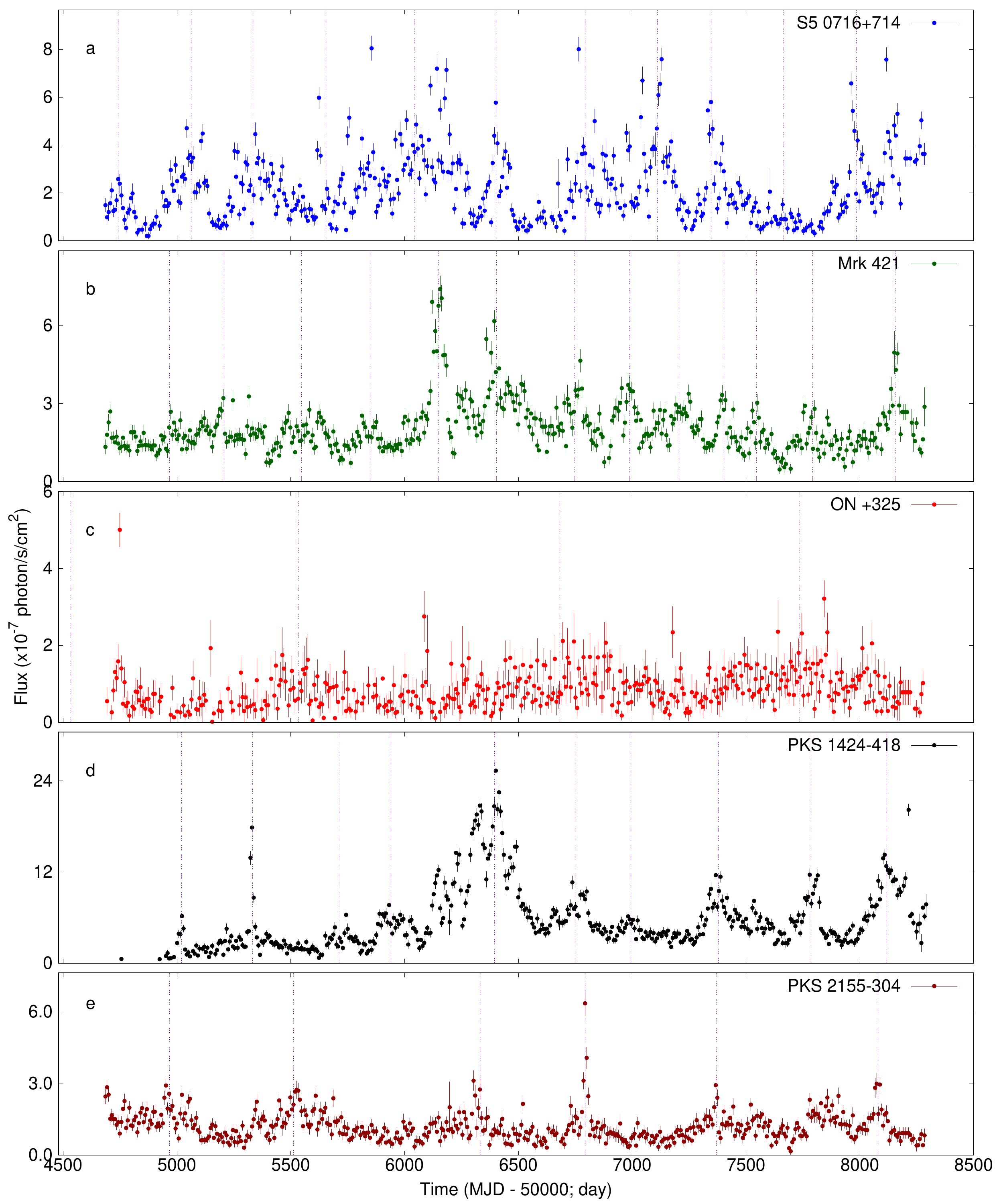}
\caption{Decade-long Fermi/LAT light curves of blazar displaying quasi-periodic oscillations. The vertical lines mark the tentative position of the peak or centroid  of the periodic flux modulation.}
\label{fig:5}
\end{center}
\end{figure*}

\subsection*{ Quasi-Periodic Oscillations }
As we saw above, the periodograms of source light curves can be largely characterized by a single power-law PSD.  But if we closely look at the periodogram structures, occasionally we find peaks at some frequencies suggesting possible presence of (quasi-) periodic signals in the observations. In fact, several sources are known to show QPOs in their light curves in different energy bands \citep[see][and the references therein]{Gupta2018,bhatta16c}. To cite a few cases, blazar OJ 287 is famous for showing characteristic double-peaks in its optical light curve that re-occur after every $\sim$12 years \citep[e. g.][]{Sillanpaa88}. In the \gama-ray energy band, the first case of year-scale QPO was observed in blazar PG 1153+113 which seemed to display a  $\sim$ 2-year periodic modulation in the Fermi/LAT observations \citep[see][]{Ackermann15}. Subsequently, a number of works have reported QPOs in the \gama-ray light curves of several blazars e. g. $\sim$ 230 d QPO in Mrk 501 \citep{Bhatta2019},   34.5 d in  PKS 2247-131 \citep{Zhou2018}),   $\sim$ 2 yr in PKS 0301-243 \citep{Zhang2017b} and  3.35 yr in  PKS 0426-380 in \citep{Zhang2017a}. In addition to these QPO studies focused on  individual sources, search for QPOs  in the \gama-ray light curves in a sample of \gama-ray bright sources has also been carried out in several works \citep[e.g., see][]{Benkhali2019,Sandrinelli2017,Sandrinelli2016,Sandrinelli14}. 

 The periodic \gama-ray flux modulations most likely originating in the blazar jets can, through long-memory processes, carry information about the violent processes occurring at the innermost regions of AGN. At a time when  the central engines of AGN  still can not be resolved by most of our current instruments, time series analysis carrying out studies of QPOs can serve as a probe to the nature of disk-jet connection and  jet ejection. With such a motivation, we analyzed the decade-long $\gamma$-ray observations of the sample source applying the Lomb-Scargle periodogram (LSP; \citealt{ Lomb76,Scargle82}), one of the most efficient methods of finding QPO signals in the data with irregularities and gaps. The method basically tries to least-square fit sine waves of the form $X_{f}(t)= A \cos\omega t +B \sin\omega t$ to the observations such that the periodogram is given according to
\begin{equation}
P=\frac{1}{2} \left\{ \frac{\left[ \sum_{i}x_{i} \cos\omega \left( t_{i}-\tau \right) \right]^{2}}{\sum_{i} \cos^{2}\omega \left (t_{i}-\tau \right) } + \frac{\left[ \sum_{i}x_{i} \sin\omega \left( t_{i}-\tau \right) \right]^{2}}{\sum_{i} \sin^{2}\omega \left( t_{i}-\tau \right)} \right\} \, ,
\label{modified}
\end{equation}
where $\tau$ is given by $\tan\left( 2\omega \tau \right )=\sum_{i} \sin\omega t_{i}/\sum_{i} \cos\omega t_{i} $\,.  The periodogram is evaluated for $N_{\nu}$ number of frequencies between the minimum, $\nu_{min} = 1/T$, and the maximum frequencies, $ \nu_{max}=1/(2 \Delta t)$.   The total number of frequencies can be empirically given as $N_{\nu}=n_{0} T \nu_{max} $,  where $n_{0}$ can be chosen in the range of $5-10$  \citep[see][]{VanderPlas2018}. A peak centered  at a temporal frequency may potentially suggest presence of periodic signal characteristic to the corresponding timescale. Unlike strictly periodic signals  which appear as sharp peaks in periodogram, QPO signals give rise to periodogram structures that are extended over to the frequencies nearby to a central characteristic  frequency.  In case of real astronomical observations that often show irregular sampling and gaps in the data, spurious peaks can arise due to a number of factors discussed in Section \ref{PSD}. More importantly, in blazar light curves, which are dominated by variability due to red-noise processes, high amplitude QPO features can arise especially  in the lower frequency region of the periodograms. Therefore, any significance estimation method should take into account  this behavior along with the other artifacts that are prevalent in finite duration time-series sampled at discrete and irregular time steps - in other words, artifacts introduced by window function.

\begin{figure*}
\gridline{\fig{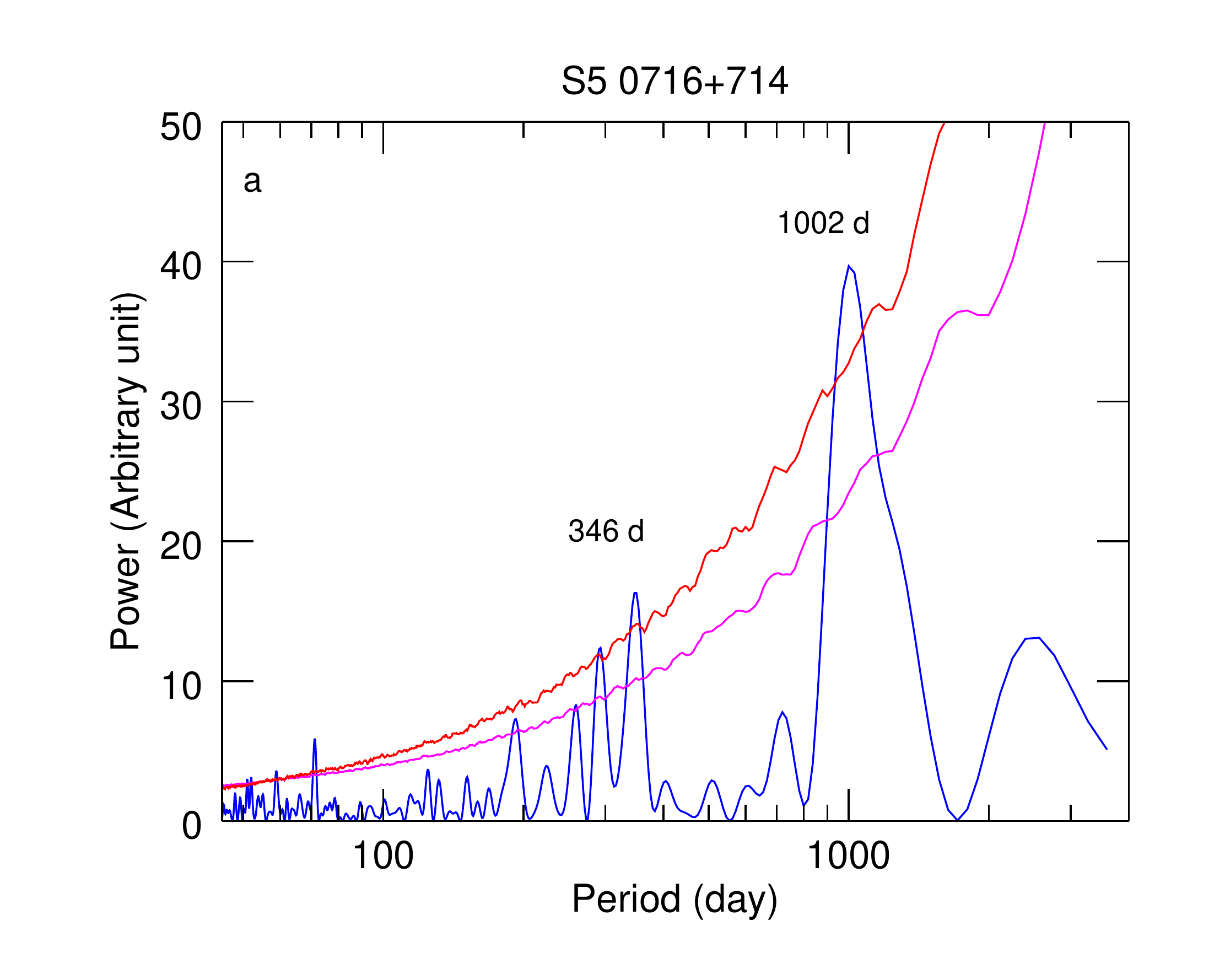}{0.5\textwidth}{}\hspace{-1cm}
  \fig{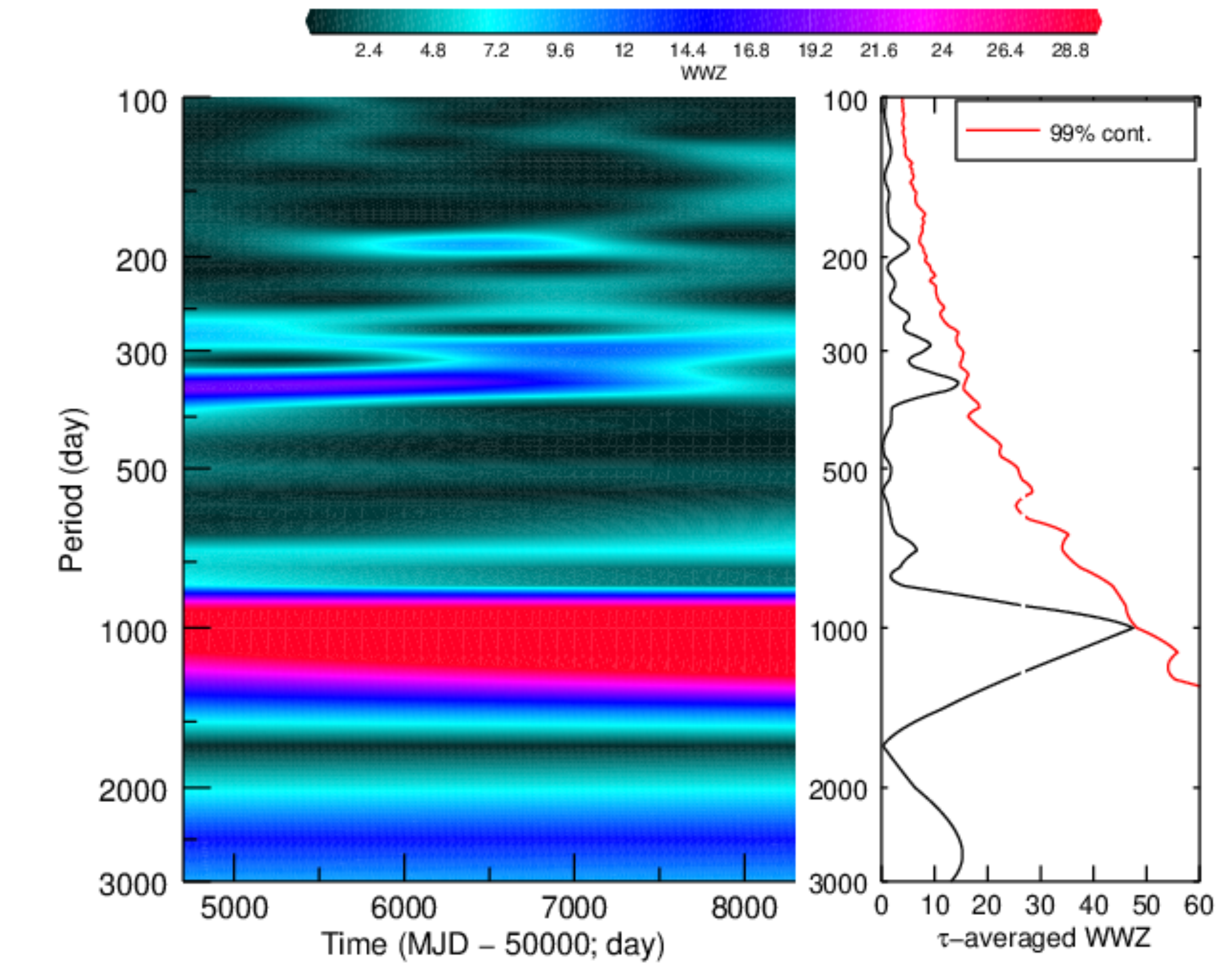}{0.45\textwidth}{}\hspace{-0.5cm}
		}
\vspace{-1.0cm}
\gridline{\fig{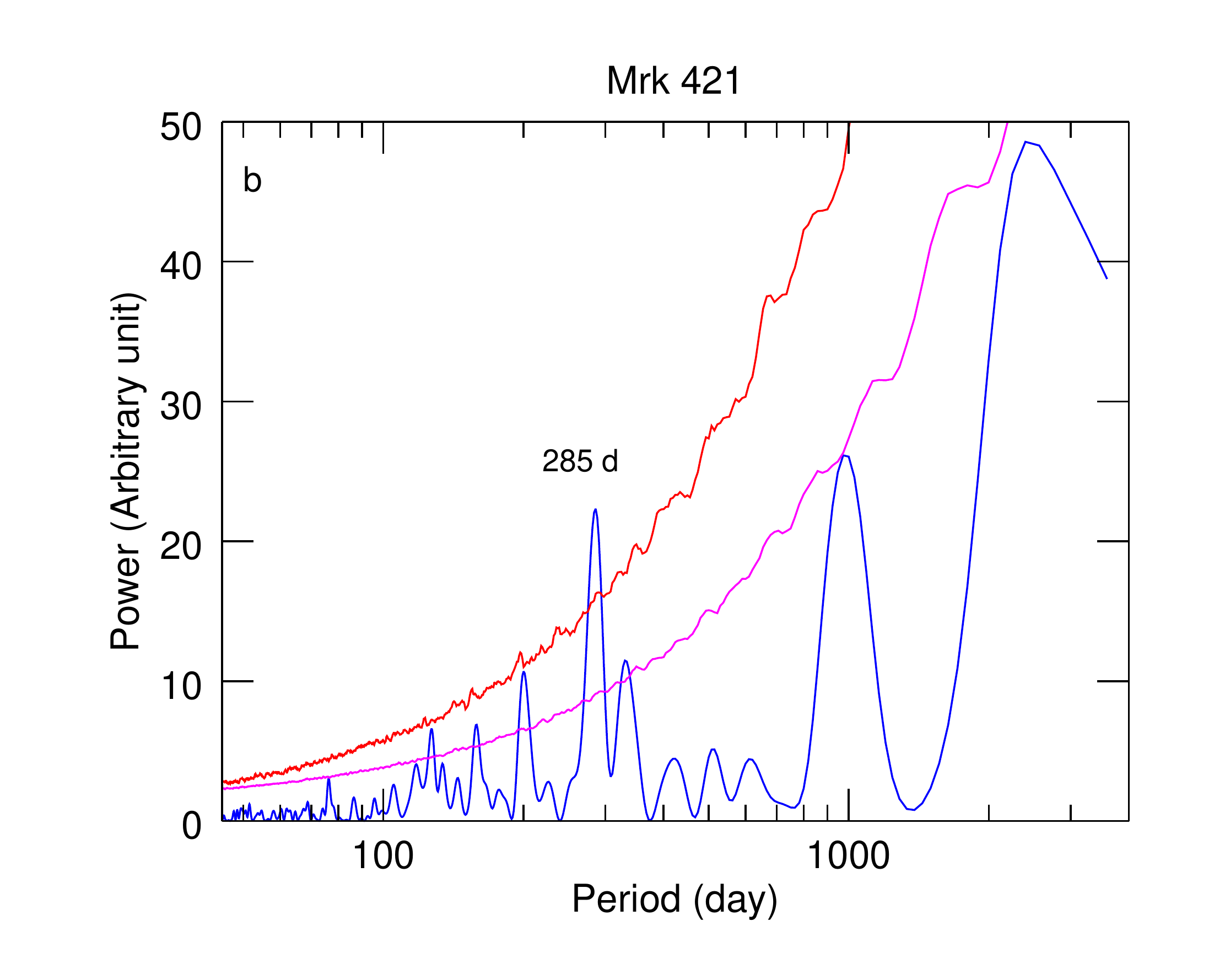}{0.5\textwidth}{}\hspace{-1cm}
          \fig{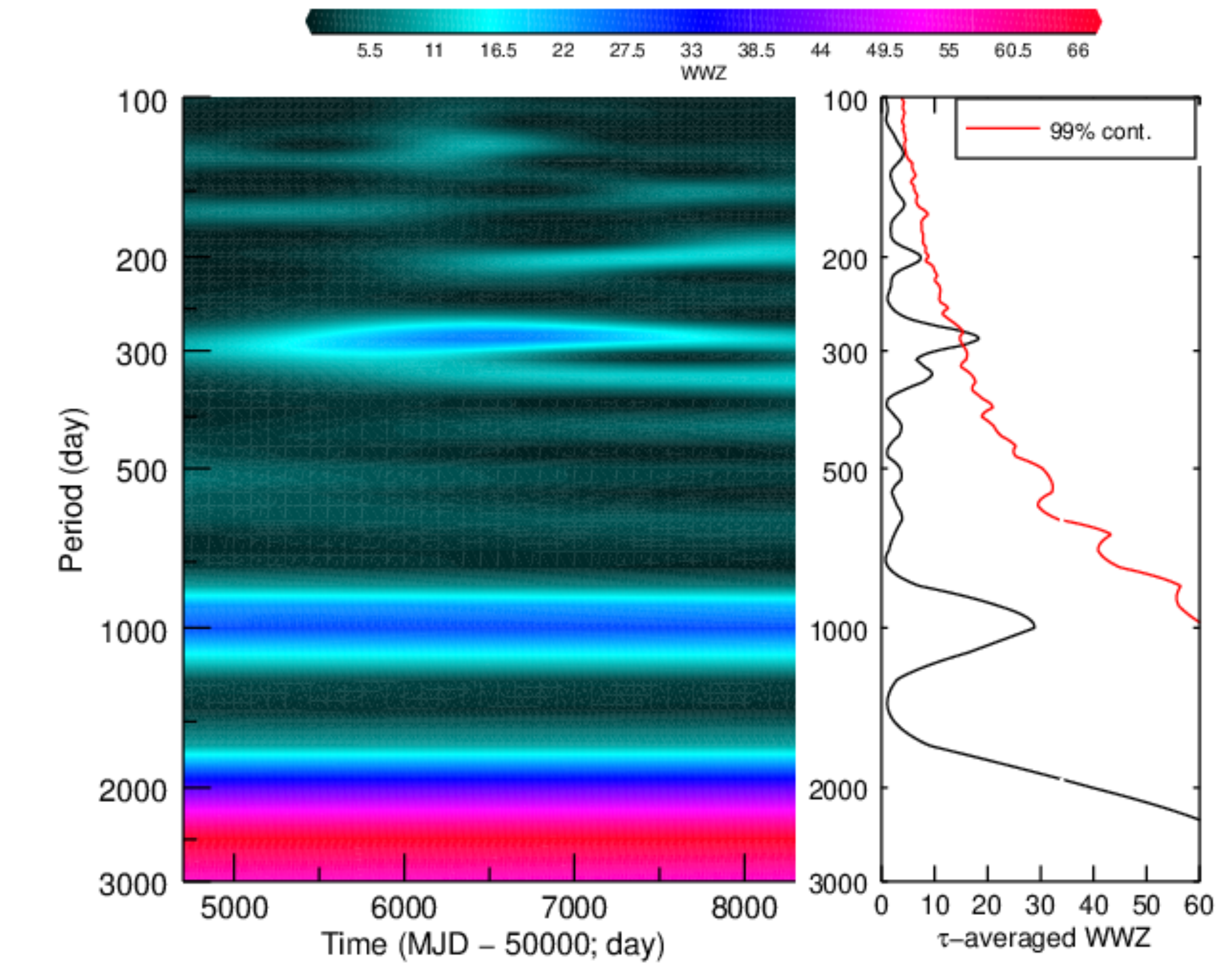}{0.45\textwidth}{}\hspace{-0.5cm}
		}
\vspace{-1.0cm}
          \gridline{\fig{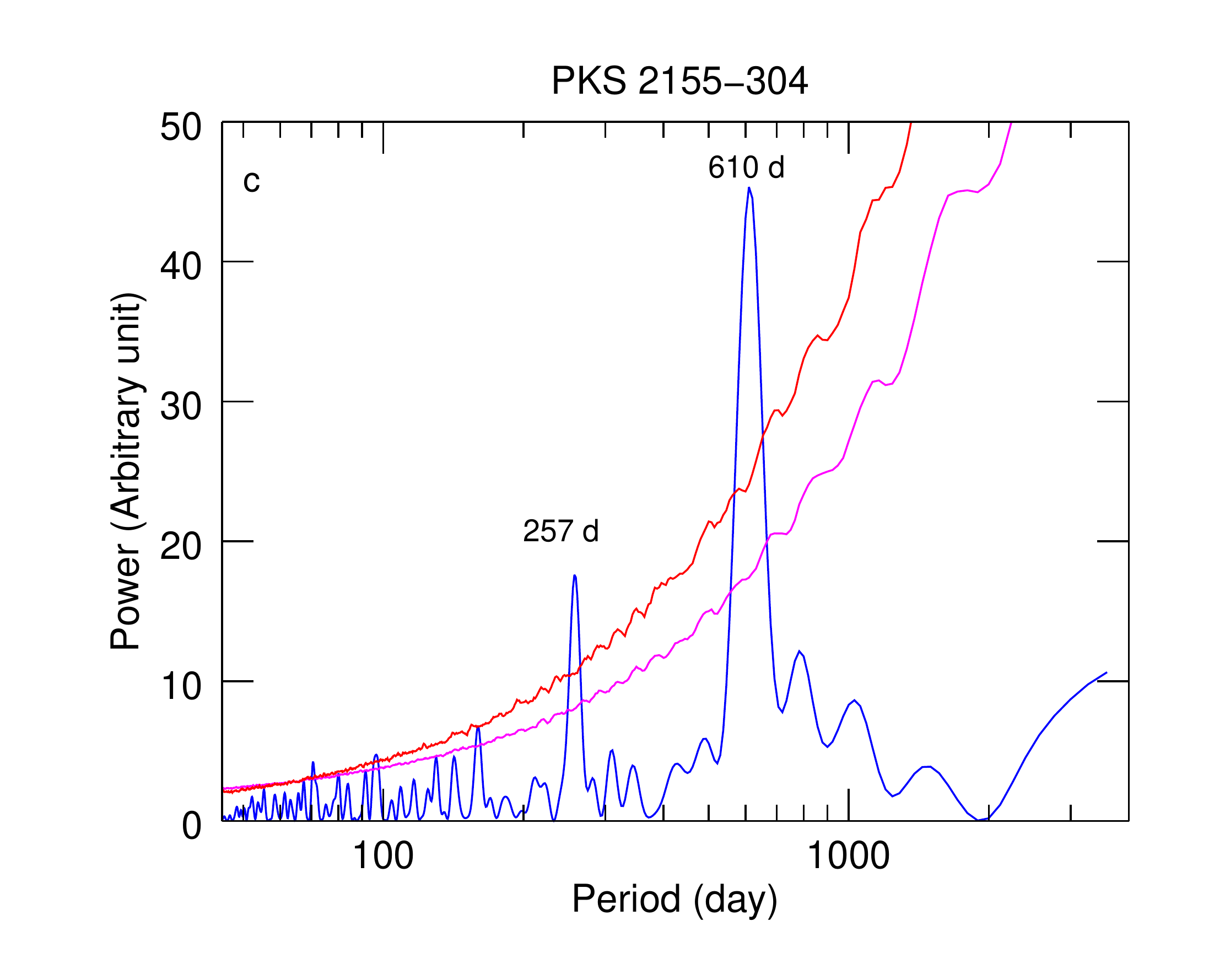}{0.5\textwidth}{}\hspace{-1cm}
			\fig{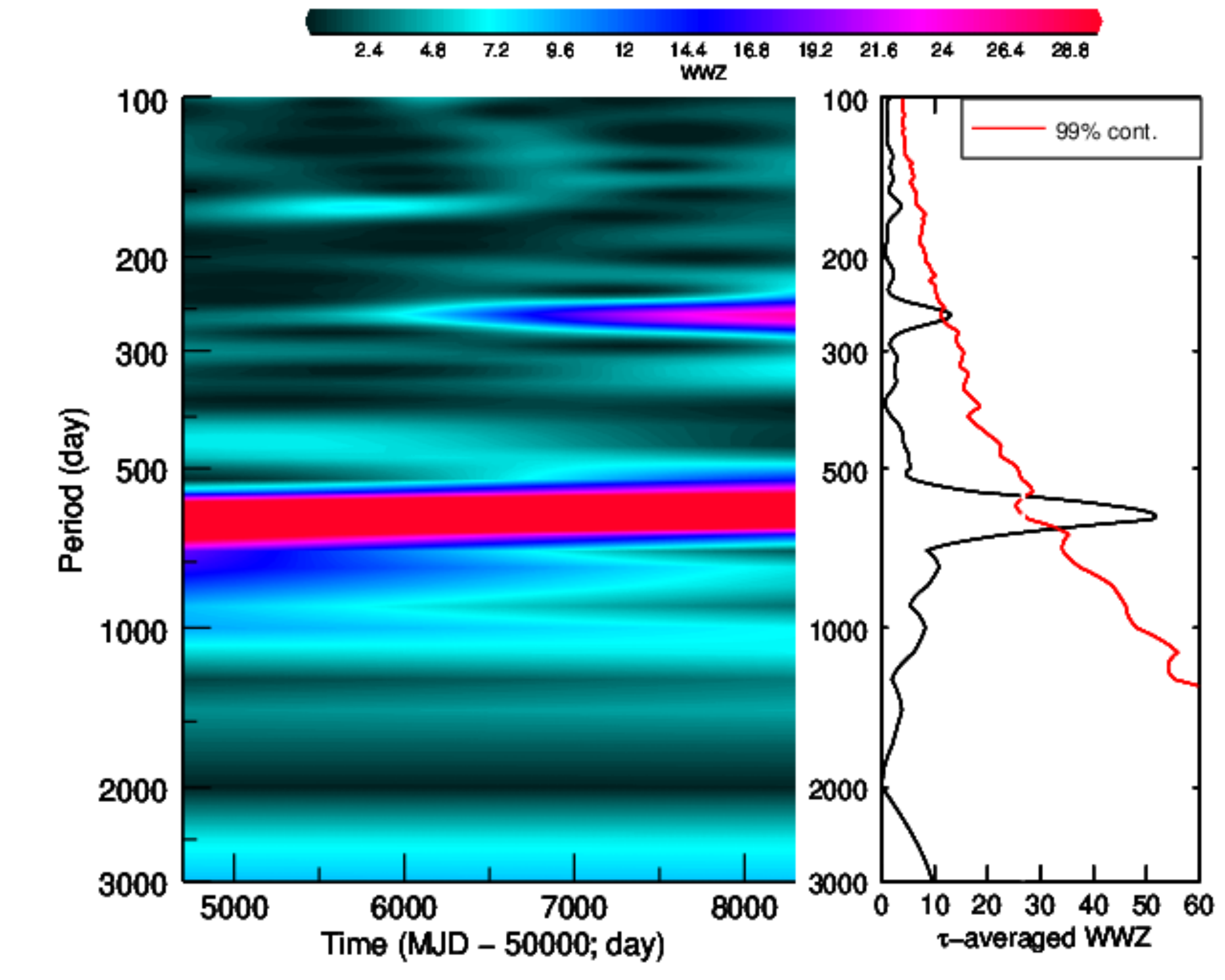}{0.45\textwidth}{}\hspace{-0.5cm}
                     } 
\caption{Detection of quasi-periodic oscillations in the \gama-ray light curves of blazars. The LS periodograms, 90\% and 99\% contours are shown in blue, magenta and red curves, respectively, in the left column panels. The right column panels show the WWZ power in colors, mean WWZ power at a given  period as the black curve, and 99\% significance contour as red curve. \label{fig:LSP1}}
\end{figure*}

\begin{figure*}
\gridline{\fig{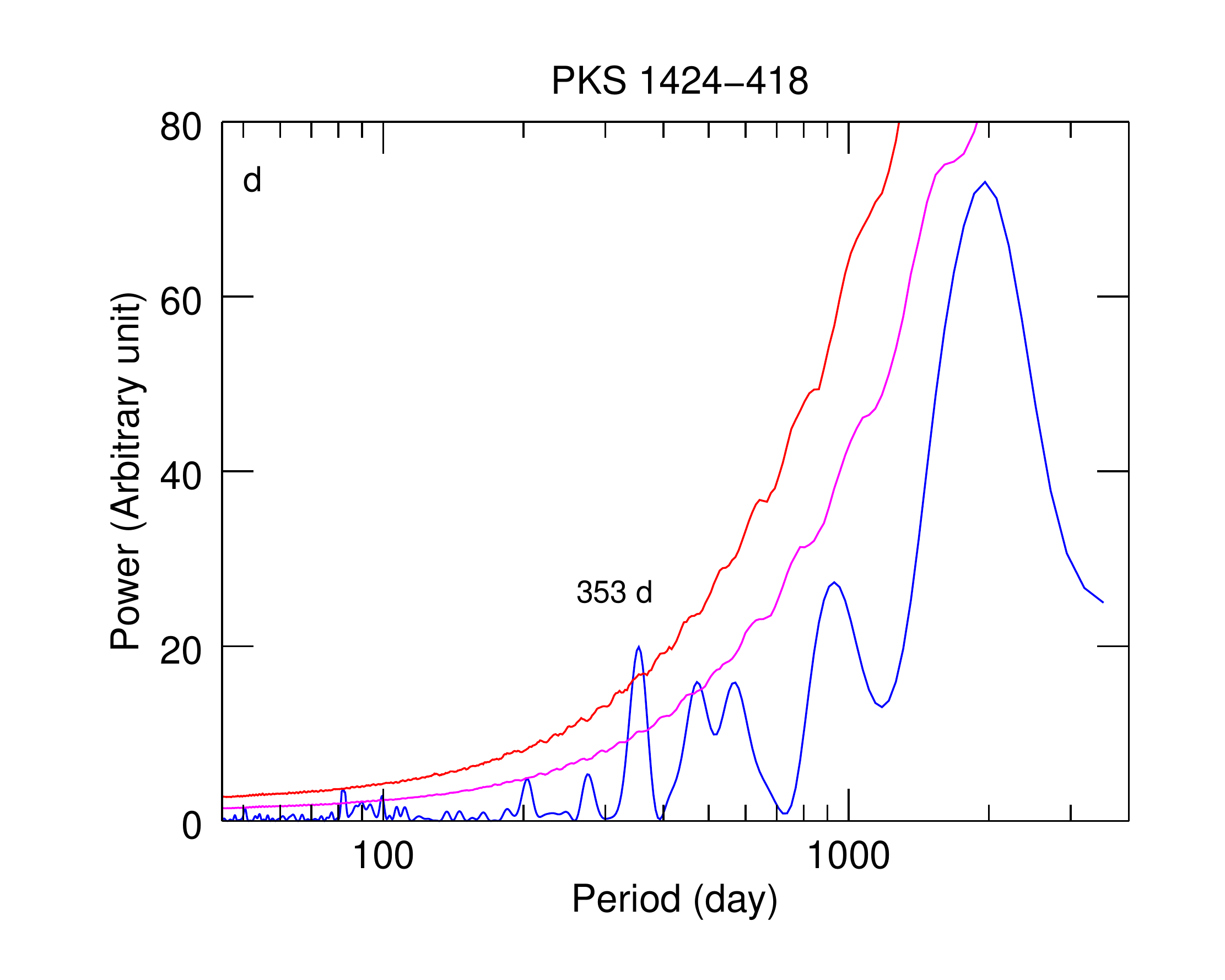}{0.5\textwidth}{}\hspace{-1cm}
          \fig{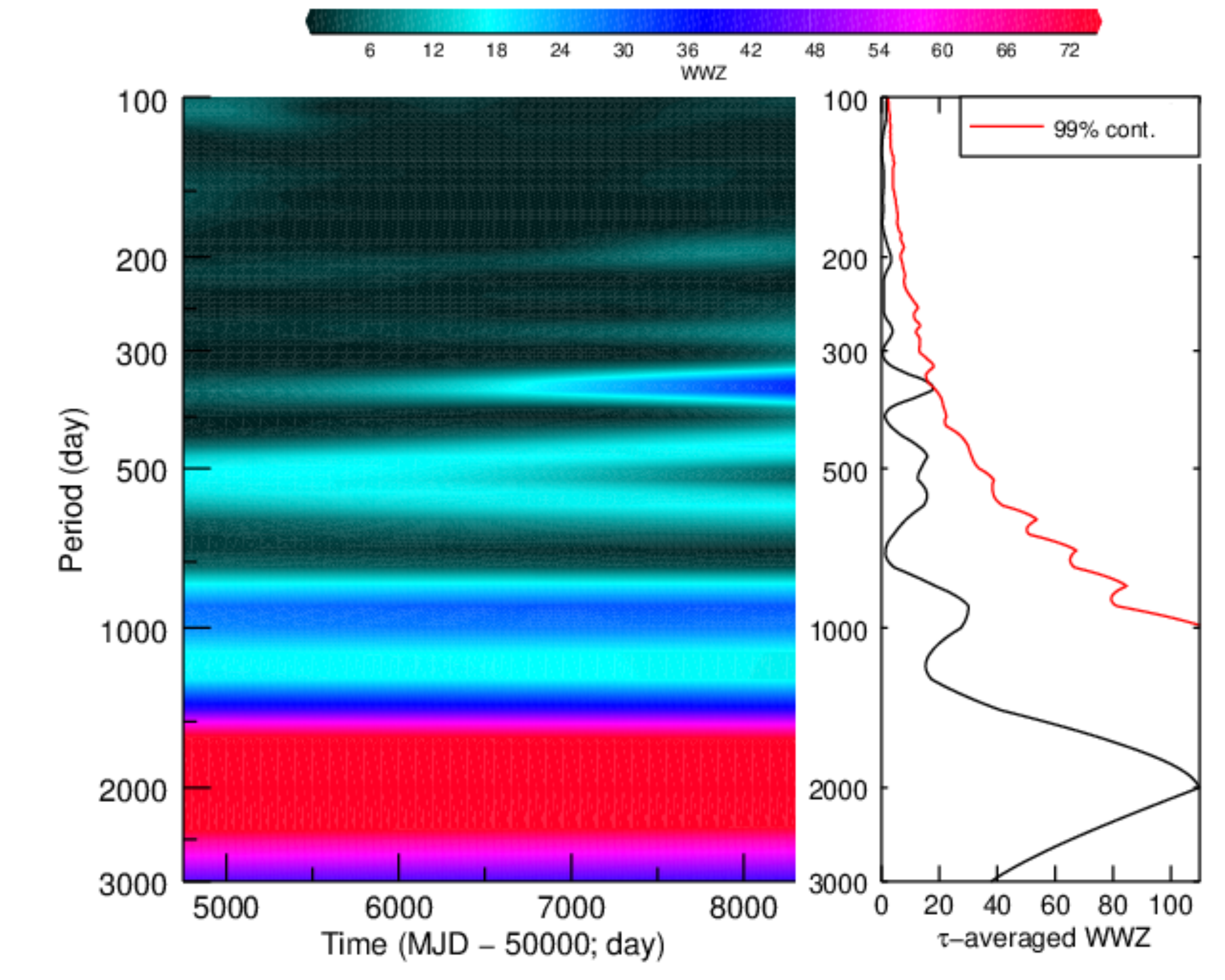}{0.45\textwidth}{}\hspace{-0.5cm}
		}
\vspace{-1.0cm}
          \gridline{\fig{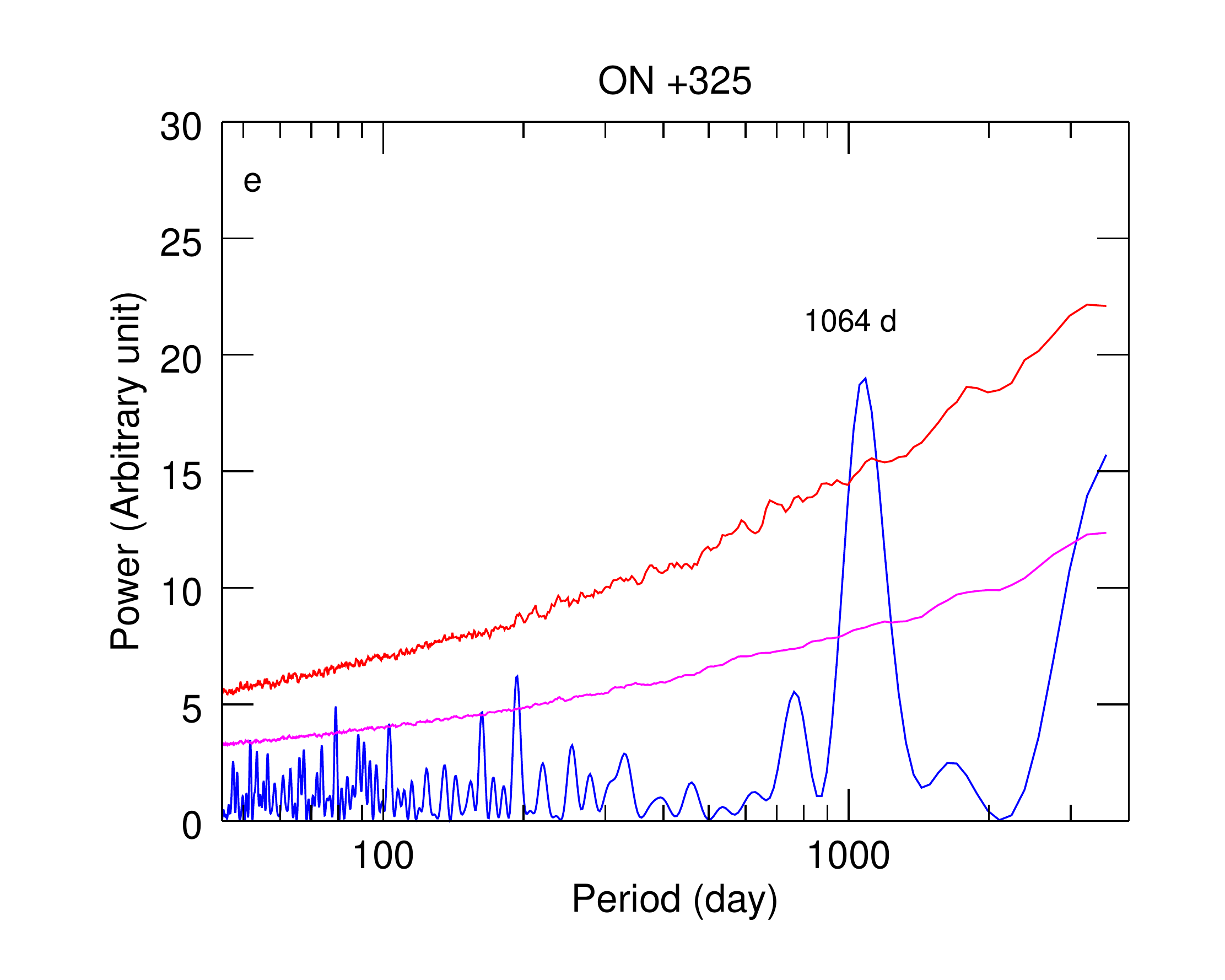}{0.5\textwidth}{}\hspace{-1cm}
			\fig{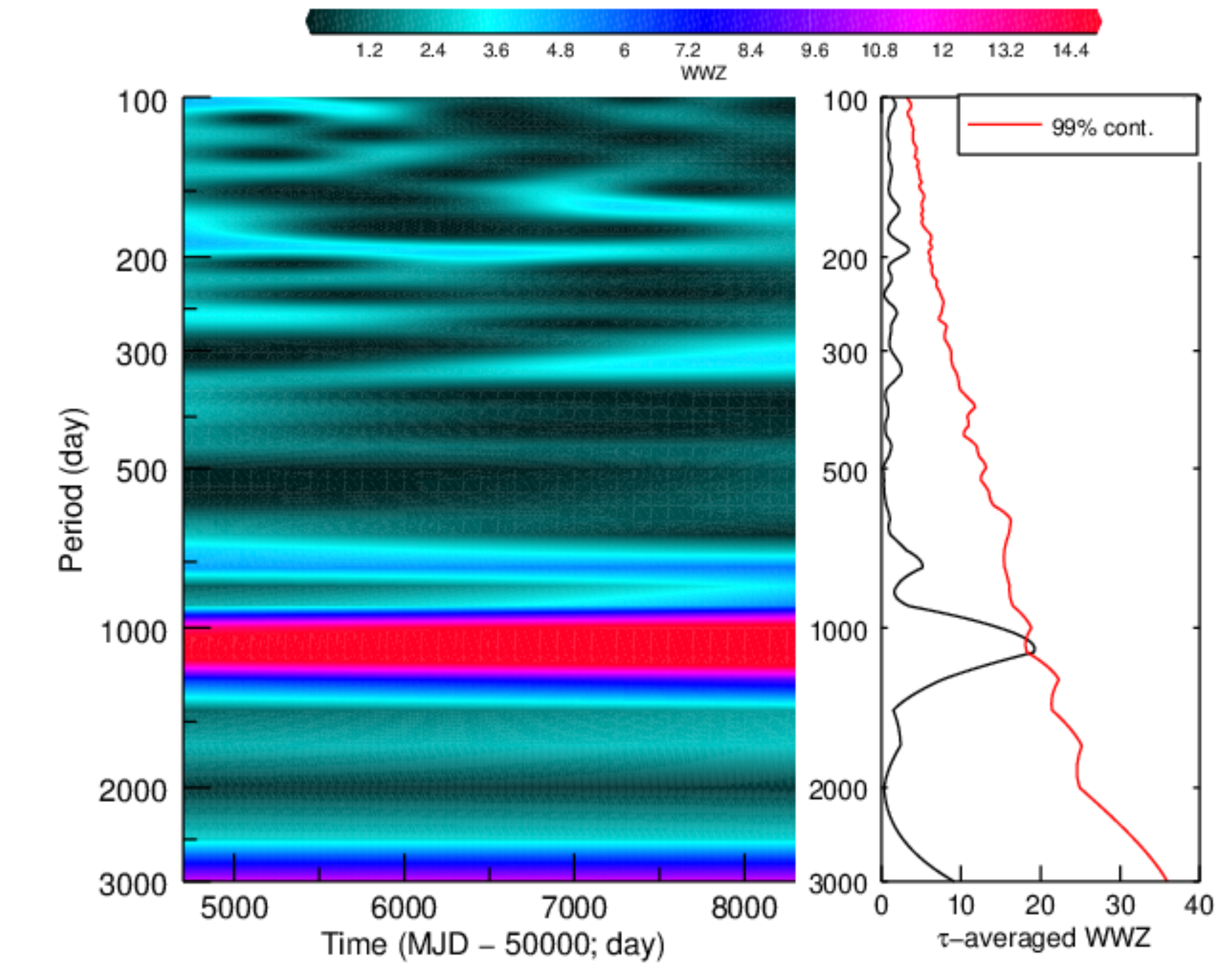}{0.45\textwidth}{}\hspace{-0.5cm}
                     }
                    \vspace{-1.0cm}
\gridline{	\fig{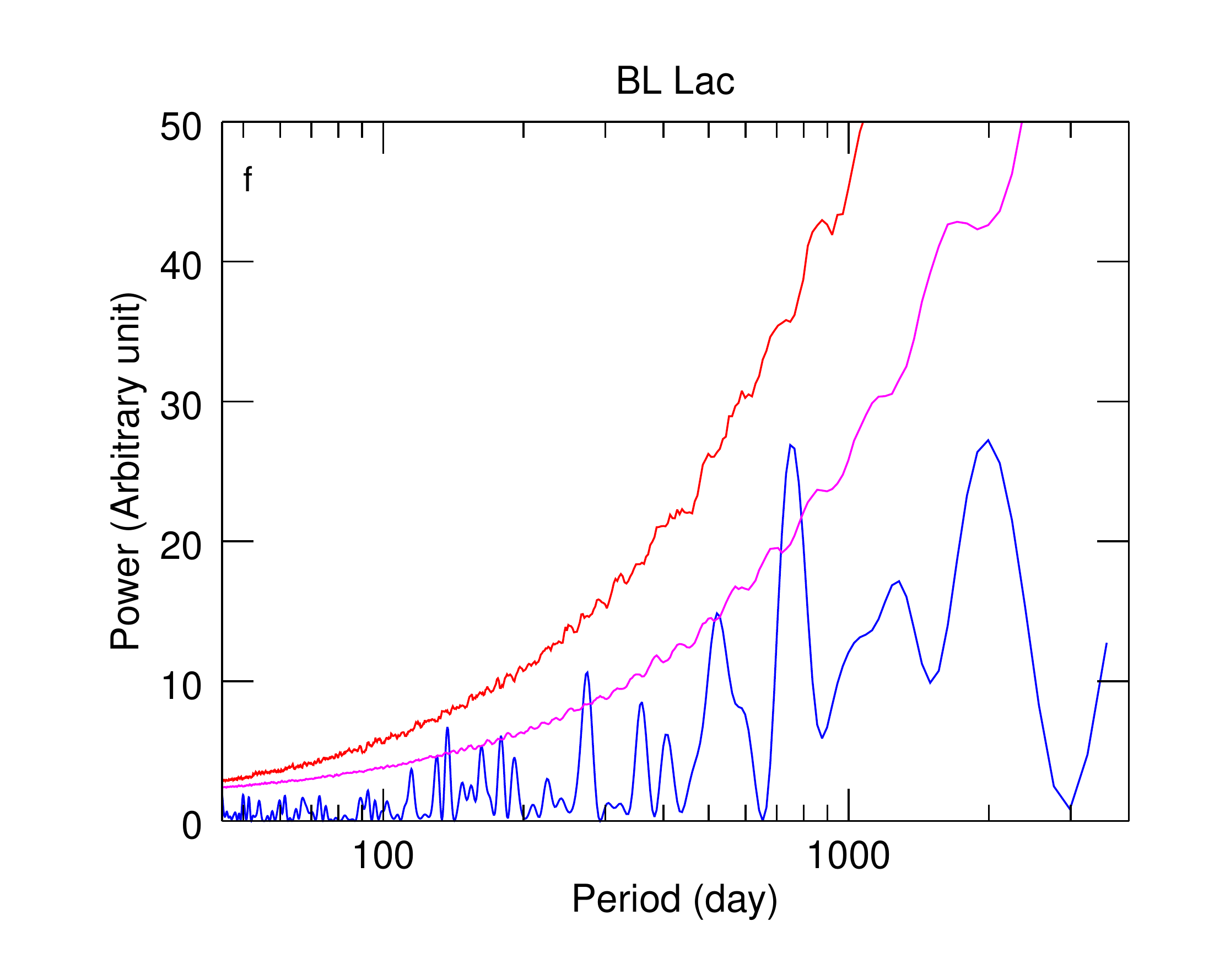}{0.5\textwidth}{}\hspace{-1cm}
                 \fig{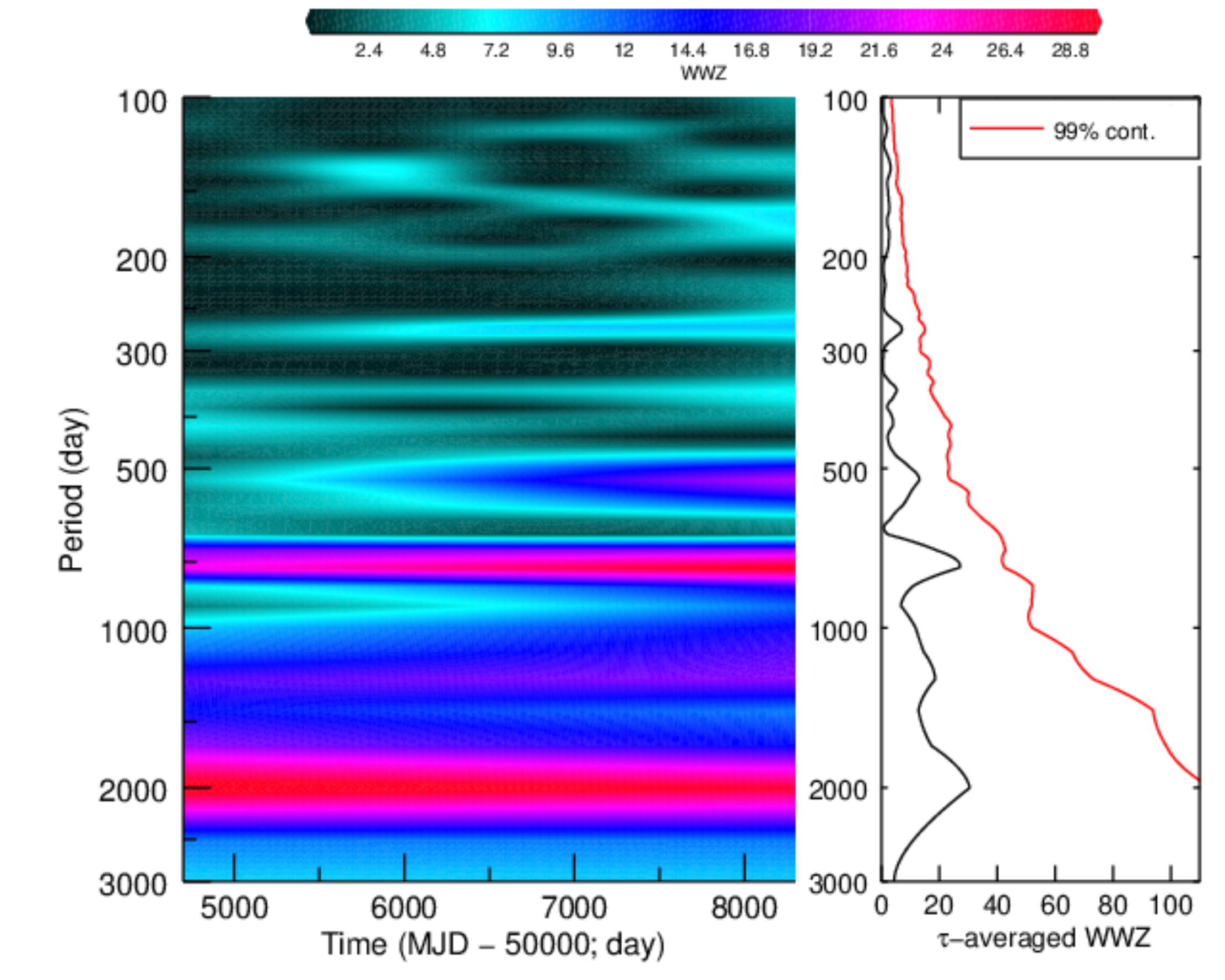}{0.5\textwidth}{}\hspace{-0.5cm}
		          }
\caption{Same as Figure \ref{fig:LSP1}.  \label{fig:LSP2}}
\end{figure*}

To further explore the transient (in frequency and amplitude) nature of the possible QPOs, we also performed weighted wavelet z-transform (WWZ), one of the robust wavelet methods \cite[see][for details]{Foster96}\footnote{We are skipping here some of the details including implementation of the method and the significance estimation, as we have extensive covered the topics in our multiple previous works \cite[see][]{Bhatta2019, Bhatta2017,bhatta16c}}. The decade-long \gama-ray observations of the blazars were analyzed to search for possible periodic flux modulations using both of the methods, i.e., the LSP and the WWZ methods. The analyses suggested presence of 
year-scale QPOs in some of the objects in the  source sample as listed in Table \ref{tab:QPO}. The LSP diagram for these sources are presented in the left column panels of Figure \ref{fig:LSP1} and \ref{fig:LSP2}; whereas the the corresponding WWZ diagrams are placed on the right column panels of the figures. Moreover, the significant periods resulted from the LSP and WWZ methods are listed in the 2nd and the 5th column of Table \ref{tab:QPO}, respectively. 

\begin{table}[!b]
\begin{center}
\caption{List of the blazars in the sample that show significant QPO in the \gama-ray light curves\label{tab:QPO}}
\begin{tabular}{lccc|ccc}
\hline
\multicolumn{3}{c}{LSP} & \multicolumn{3}{c}{WWZ} \\
\hline
source & period (d)  & local sig. ( \%)& global sig. (\%)& period (d)  & local sig. ( \%)& global sig. (\%)\\
\colnumbers\\
S5 716+714	&	346	$\pm$ 	23	&	99.97	&	99.96&	349	$\pm$ 	27&	99.982	&	99.980	\\
Mrk 421	&	285	$\pm$ 	27	&	99.99	&	99.97&287	$\pm$ 	32	&	99.997&	99.993	\\
PKS 2155-304	&	610	$\pm$ 	51	&	99.9994 &	99.99841	&617	$\pm$ 	53	&	99.995 &	99.9981\\
PKS 1424-418	&	353	$\pm$ 	21	&	99.98	&	99.95&349	$\pm$ 	24	&	99.985	&	99.981	\\
ON +325	&	1086	$\pm$ 	63	&	99.9986	&	99.9968	&1081	$\pm$ 	67	&	99.987	&	99.983\\

\hline
\end{tabular}
\end{center}
\end{table}

 The significance of the detected periodogram features were computed by employing  PSRESP method during which extensive MC simulation were performed.  In particular,  the spectral distribution of 10000 simulated light curves (simulated using their corresponding best-fit PSD models)  were employed to evaluate the local 90\% and 99\% significance contours \citep[see also][]{bhatta16c,Bhatta2017,Bhatta2018b, Bhatta2019}.  The estimation of local significance only makes the use of the simulated LSP distribution at the period of the detected QPO, whereas global significance, given that we do not have \textit{a priori} knowledge of the period of the detection, considers the simulated  spectral distribution at all the temporal frequencies considered \citep[see][for details]{Bell2011,Bhatta2017}. The resulting 90\% (magenta) and 99\% (red) significance contours in the LSP diagram are presented in the left column panels of Figure \ref{fig:LSP1} and \ref{fig:LSP2}, and the local and the global significance of dominant periods are listed in the 3rd and 4th columns, respectively, of Table \ref{tab:QPO}.  Similarly, the 99\% significance contours are shown in WWZ plots as red curves and the local and the global significance of dominant periods are listed in the 6th and 7th columns, respectively, of the table.  Moreover, a brief description of each sources showing possible QPO features are presented below. 

\begin{itemize}
\item S5 0716+714:  We detected  highly significant QPO at the period centered around 340 d. The tentative peaks of the periodic oscillation are marked in the source light curve with vertical lines at a separation of $\sim$ 340 d  as shown in panel a) of Figure \ref{fig:5}. The LSP (left panel) and WWZ (right panel) diagrams along with the corresponding significance contours are presented in panel a) of Figure \ref{fig:LSP1}. It is interesting to note that \citet{Prokhorov2017} in their analysis including the Fermi/LAT observations  from the year 2008  to 2016 also detected exactly the same periodicity  with a high significance (99\%) over power law. However, in their work, \citet{Sandrinelli2017} did not detect the QPO. It should be pointed out that, in addition to 346 d QPO, the 1002 d QPO -- possibly the 3rd harmonics -- also appears to be significant ($>$99\%) in both of the analyses.  But in such case the light curve would contain only 3 cycles, which would leave one indecisive  as to the signal being real. Interestingly, this period is close to that of the optical QPO detected by \citet{Raiteri2003} with possible interpretation that the \gama-ray QPO could be the counterpart of the optical one.

\item  Mrk 421: With significance greater than 99\%, we detected $\sim$280 d periodic flux modulations in the famous blazar  Mrk 421.  Both the LSP and WWZ diagrams of the source are presented in panel b) of Figure \ref{fig:LSP1}, and also the tentative peaks of the oscillation are marked with vertical lines drawn at an interval nearly equal to the period as shown in the panel b) of Figure \ref{fig:5}. The detection supports the previous claim by \cite{Li2016} who reported the exact same period in the \gama-ray band along with the similar one in 15 GHz radio observations. Similarly, \citet{Benitez2015} reported a similar period of 310 d in the multi-frequency (optical, hard X-ray and $\gamma$-ray) light curves of the source. However, in the analysis presented by \citet{Sandrinelli2017}, it was not found to be significant enough.

\item  PKS 2155-304: We detected $\sim$610 d periodic flux oscillations in the blazar  PKS 2155-304. The WWZ analysis reveals that over the year the period is gradually shifting towards slightly higher frequency.  We show the tentative peaks of the oscillation of the period which are marked with vertical lines in panel d) of Figure \ref{fig:5}.  The source LSP and WWZ spectral powers along with the significance contours are shown in panel c) of Figure \ref{fig:LSP1}. A number of previous works have also reported similar periods, e. g., 700d in the optical and $\gamma$-ray \citep[][]{Chevalier2019}, 620 d in the $\gamma$-ray  \citep{Benkhali2019},  635 d in the $\gamma$-ray \citep{Zhang2017},    640 d in the $\gamma$-ray \citep{Sandrinelli2016}, and  625 d in the lower energy band (300 MeV--1 GeV) of the Fermi/LAT light curve \citep{Sandrinelli14}. Also, it is intriguing to note that this timescale is nearly double the 317 d timescale reported by  \citet[][]{Zhang2014RAA}  in the light curves spanning 35 yrs using multiple methods such as epoch folding, the Jurkevich periodogram and discrete correlation function. The optical QPO was also reported in \citep{Sandrinelli14}.  In addition, we also detected $\sim260$ d QPO with a high significance ($>$ 99.99\% using both the methods), although the flux oscillations of the period are not visually clear in the light curve plot.

\item PKS 1424-418: We found 353 d periodic flux oscillations in the \gama-ray flux of the source blazar  PKS 1424-418 significant above 99\% over the power-law noise.  The possible QPO appearing in several cycles in the data is reported here for the first time. The LSP and the WWZ diagram along with the respective significance contours are shown in panel e) of Figure \ref{fig:LSP2}; also the tentative peaks of the periodic oscillations are shown with vertical lines in panel d) of Figure \ref{fig:5}.

\item  ON +325:  Both the LSP and WWZ analyses of the ON +325 light curve resulted in detection of a significant periodic oscillation with characteristic timescales of  $\sim$1070 d.  However, since only three cycles can be seen in the entire light curve, it is not clear if it is truly a segment of QPO oscillations. We drew the tentative peaks of the oscillations of the period which are marked with vertical lines in panel c) of Figure \ref{fig:5}, and the distribution of LSP and WWZ powers of the source are shown in panel e) of Figure \ref{fig:LSP2}.  It is noted that a $\sim 4.5$ year optical QPO was claimed in the source previously \citep{Fan2002A&A}.

\item  BL Lac: As shown in panel f) of Figure \ref{fig:LSP2}, the  structure of LSP and WWZ power distributions of the source appears to be rather complex. There does not seem to be one dominant sinusoidal component but there are a  number of possible  timescales, in particular $\sim$ 270, 520 and 750 d.  Note that the first two periods are close to the harmonic range.  However, these peaks are well below 99\% contour and therefore can not be considered significant. We note that the  680 d \gama-ray QPO  claimed in the work of \citep{Sandrinelli2017} is not visible in the analysis.

\item In addition, a $\sim$ 330 day period QPO in the TeV blazar Mrk 501 has been reported in \cite{Bhatta2019}.

\end{itemize}

\section{Discussion \label{sec:4}}
In this section, we present our interpretation and discussion on the results derived from the above analyses in the light of standard model of blazars, i.e., black hole powered central engine and the extended radio jets providing grounds for particle acceleration and energy dissipation events. 
\begin{itemize}
\item \gama-ray variability in blazars:\\
In the variability analysis, quantified measure of flux modulations as observed in the \gama-ray light curves of the blazar sources was provided by computing their fractional variabilities. The numerical values listed in the 7th column of Table \ref{table:1} suggest that blazar sources are distinctly characterized by their remarkable activity in the $\gamma$-ray band.
 The \gama-ray variable emission can be largely ascribed to the events occurring at the kilo-parsec scale radio jets aligned  within $\sim 5^{o}$ to the line of sight. These jets  are primarily fed with the energy that could be extracted from the fast spinning Kerr black hole in the presence of the magnetic field at the rotating accretion disk  \citep{Blandford2019,Blandford1982,Blandford1977}.  Shocks traveling down the jet can produce a power-law distribution of energetic electrons $N(\gamma)\propto \gamma ^{-p}$ such that the spectral index of the synchrotron emission can be related as $\alpha=(p-1)/2$. These synchrotron electrons responsible for non-thermal emission might be accelerated to the Lorentz factors as high as $\sim10^6$. Owing to the violent and energetic events prevalent in the jets, individual  radio knots appear to be moving with superluminal motion with apparent velocity up to $\sim$ 78c \citep{Jorstad2017}.  Observed \gama-ray variability could be intrinsically linked to the combined modulations  in a number of components such as distribution of high energy particles, seed photons and ambient magnetic field at the emission region. On the other hand, it could also be linked to extrinsic (e.g. projection) effects associated with a `plasma blob' that is moving down the jet with bulk Lorentz factors ($ \Gamma$)  as large as $\sim 50$.
 In BL Lacs, large Lorentz factors could be conceivable as viable explanation for the observed high-amplitude variability;  in FSRQs, however, it might pose problem because  the high energy emission is  most likely produced through  inverse Compton scattering of seed photons external to the jet such that large values of $ \Gamma$ enhances the energy density of these external photons in the comoving frame by $\sim \Gamma^2$.  As a result pair production  process becomes dominant which, in turn, should lead to the reduced \gama-ray emission.  But in BL Lacs, due to lack of circumnuclear seed photons and SSC being dominant process to result high energy emission, the above argument can not be applied \cite[see][]{Sbarrato2011}. Similarly, non-thermal emission from mini-jets which are further embedded in larger jets can also result in rapid \gama-ray variability \citep{2009MNRAS.395L..29G}. In hadronic models of blazar emission, $\gamma$-ray variability could arise owing to variability in synchrotron emission from extremely high ($E\sim 10^{19}$eV) energy  protons in highly magnetized (few tens of Gauss) compact regions of the jet with a moderate Doppler factor $\sim 15$ \citep{2000NewA....5..377A}.  

The observed linear correlation between fractional variability and \gama-ray spectral index suggests that the sources with steeper spectrum exhibit greater variability. Theoretically, one might expect such a relation in several cases. For instance, if the emission comes from a smaller volumes than lower-energy emission volumes, as in the radiative shock or turbulent extreme multi-zone models, the fluctuations have higher relative amplitude and shorter time-scales. Similarly, the observed flux could be more strongly dependent on the Doppler factor when the spectrum is steeper. In addition, a steeper spectrum indicates that the energies of the emitting particles are close to their upper limits, so that radiative losses are more severe and perhaps their acceleration is more sporadic, causing greater variability. The observed steep slope of the linear fit on FV-index plane strongly supports the last scenario.

\item Flux distribution:\\
The analysis of flux distribution of the Fermi/LAT light curves of the sample sources suggests that for most of the sources studied in this work, the best fit PDF closely follows lognormal distribution. Similar result is obtained by \citet{Shah2018} who used average monthly Fermi/LAT flux for 50 bright blazars.  The observed log-normal distribution of the blazar flux has been interpreted in terms of disk processes.  Accordingly, log-normal flux distribution could act as indicative of disk-jet connection in blazars.  The fluctuations  in the disk, contributing to flux variability, can take place at different radii and thereby be dictated by viscosity fluctuations in accordance with the local viscous timescales. In turn, these modulate the mass accretion rates at larger distances from black hole. Variable emission from accretion disks owing to variable accretion rate could be driven by uncorrelated fluctuations in the $\alpha$-parameter taking place at different radii of the disk \citep[see][]{Lyubarskii1997}.  The observed log-normal distribution of the blazar flux suggests multiplicative coupling of these perturbations at the disk, as opposed to additive coupling as in shot-noise-like perturbation  \citep{Arevalo2006}. The radiation being relativistically beamed,  \gama-variability in blazars could arise due to a combination of both  source intrinsic events such as instabilities at the disk and the jet, and source extrinsic  geometrical and projection effects.  Furthermore, the radiation by the up-scattered photons depend both on the population of the seed photons as well as high energy particles that contribute to the up-scattering. In such scenario, no single variable parameter can be considered as dominant to the variable emission, rather all possible contributing factors such as variable magnetic field and high energy particle density, seed photon density acted upon by the particle acceleration and diffusion processes could be coupled in a complex manner resulting in the log-normal distribution of the variable emission from the sources. 

On the other hand,  normal flux distribution can be interpreted as  integrated emission from individual shock or magnetic-reconnection events occurring stochastically in large-scale turbulent jets \citep[e.g][]{Xu2019,bhatta13}.  It is possible to interpret  both kinds of distributions  as being special cases of a more general class of skewed distribution, such as Pareto distributions, with variable degree of skewness.  In the context of relativistic jets, such distribution could be a natural consequence of emission from Poynting flux dominated jets that hosts mini-jets distributed isotropically within the emission region and that get ejected close to the line of sight with a high bulk Lorentz factor $\sim 50$.  In  such scenario,  the resulting flux distribution has been found to hold the RMS-flux relation \citep[see][]{Biteau2012}. Similarly,  in the acceleration due to shock scenario, a small perturbation in the acceleration timescale can result in variability in particle number density that is a linear combination of Gaussian and lognormal process.  Based on the relative weight associated with these processes, it  can in turn determine the dominant shape of the flux distribution \citep[see][and the references therein]{Sinha2018}. If the  variability in gamma-ray emission is dictated by such variability in the number density of the accelerated particles, then it is natural for the flux distribution to appear both as Gaussian and lognormal. Such a scenario, where both additive and multiplicative processes operate at various degrees along the extended jet, also looks plausible.

  In blazars,  although we infer the variability properties from the jet emission, the primary source of  variability could still be associated with fluctuations in the disk processes.  These fluctuations could then propagate through the relativistic  jets affecting the jet processes and get altered owing to the relativistic effects e.g.,  flux amplification and time dilation. In blazars, although the disk emission is often completely swamped by the non-thermal emission from jets, a careful and detailed study of flux distribution of blazars should be able to trace the origins of variability back to the disk, and thereby constrain the disk-jet connection.

\item Symmetry analysis:\\  We adopted a simple yet novel approach to investigate into the blazar emission regions. For the purpose,  a statistical analysis studying flux rise and decay in the \gama-ray light curves of a sample of sources was performed. The study aimed to unravel an intrinsic difference in the distribution of flux rise and decay rates which, if intrinsic to the source, should be associated with two inherently different mechanisms, e.g. particle acceleration and energy dissipative processes, respectively. However,  as revealed by the K-S test, we do not observe significant difference between rising and decaying profile of the flux distribution; which  we find surprising and counter-intuitive as  physical mechanisms driving the flux rise due to particle acceleration mechanisms such as shocks and magnetic reconnection should operate in different timescale from the cooling time scales due to emission processes, mainly considered to be inverse-Compton process. In such context, it is natural to expect characteristic difference between the flux rise and decay rates of long term \gama-ray light curves. Nevertheless these two different processes operating in various timescales could be blended over the extent of the jet such that the overall distribution takes the form which is not easily distinguishable.

It should be pointed out  that the method of symmetry analysis presented in this work differs from the one in which the rise and decay timescales are estimated by fitting an exponential curves to well resolved individual flares as in the works by \citet{Meyer2019}, \citet{Chatterjee12} and \citet{Abdo10}. In such case, the asymmetry in the flare could depend upon individual flares. But in the  approach adopted in this work the results rather provide a statistical measure of the average flux doubling timescales during a wide range of flux changes which includes both flaring and non-flaring (quiescent) states. This is reflected in the observed wide range of timescales corresponding to the diverse rates with large standard deviations presented as errors in the average timescales (see 6th column in Table \ref{tab:4}). It should be noted that, in spite of the different approaches to the analysis, the results of this work are in close agreement with that of these works which indicate no significant asymmetry between the rise and decay flux profiles. Interestingly, similar results were reported in the studies using long-term optical observation of the sources S5 0716+714 \citep[see][]{Li2017PASP}  and BL Lac \citep[see][]{Guo2016MNRAS}. 

 The timescale estimated using average flux rise/decay rates ($\tau$) can also shed light into one of the most important issues yet unresolved in blazar physics, namely the location of \gama-ray production site in reference to the central black hole. In literature, we mainly find two compelling arguments on the location of origin of the $\gamma$-ray emission relative to the central engine. Based on the observed rapid (a few minutes) $\gamma$-ray variability  \citep[e.g., see][]{Ackermann2016,Aharonian2007}, it is argued that the emission should originate at compact regions close to the central black hole ($\sim 20 r_g$), where bulk of the gravitational potential energy of the in-falling matter is released and processed into radiation energy. However,  to avoid an eventual depletion of the  $\gamma$ ray photons due pair production in a compact region, it requires a large Doppler factor, typically $\delta > 60$ to explain the observed $\gamma$-rays. On the other hand, most of the \gama-ray flare events have been found to coincide with the ejection of radio knots and rotation of polarization angle at the mm-VLBI cores which lie at a distance of few kilo-parsecs (kpc) from the central engine \citep[see e. g.,][]{Blinov2018,Jorstad2016,2016Galax...4...37M}. Also as \gama-ray flaring events are commonly observed to last a few weeks, it can be argued that \gama-ray emission is produced at the parsec scale distance away from the black hole. In such a context, the results obtained from the symmetry analysis can be used to estimate the size of the emission region where \gama-ray variability arise, and thereby obtain a lower limit for the distance between the black hole and the \gama-ray emission sites. If we let  $ r\sim \Gamma^2 \tau c$ for a typical  $\Gamma=15$ with mean $\tau=22$ d, we obtain $\sim$ 4 pc. This supports the idea that  \gama-ray emission could be predominantly produced along the jets on parsec-scale distances as opposed to regions within a few tens of gravitational radii. To reconcile both of the ideas, it can be suggested that the blazar variability as observed in the \gama-ray light curves could be a combination of the variable emission originating  at both the locations, i. e., the low-amplitude fast variability might chiefly originate at the innermost regions -- where conversion of gravitational potential energy of in-falling matter into high energy emission is most efficient -- and the \gama-ray flaring events, flux brightening at least by a factor of a few tens, that last about a few weeks could be located at a distance of a few pc.

\item Power spectral density:\\ 
 We find that the PSDs that best represent periodogram of  \gama - ray light curves of the 20 well-known sources are consistent with a single power-law of the functional form $P(\nu) \propto \nu^{-\beta}$ where the slope index ranges between 0.8--1.5.  In the given sample source light curves,  majority of the slope indexes  tend to center around 1.0.  Similar results were obtained by \citet{Sobolewska2014} in  their  PSD analysis of $\gamma$-ray light curves of 13 blazars, although the work followed different method of PSD estimation. This is interesting because 
$\beta$=1, often known as flicker noise, are exactly halfway between random walk ($\beta$=2) and white noise ($\beta$=0), and are prevalent in nature \citep[see][]{Press78}. Flicker noise diverges when integrated from a finite high frequency to lower frequencies - towards zero  frequency. But the divergence, being logarithmic, is so slow that the noise maintains its appearance over several orders of frequencies up to arbitrarily low values. Therefore, flicker noises are long-memory processes and therefore can appear coherent over several decades in timescale. In case of blazars, although our instrument primarily detects Doppler boosted emission from the jets, it can possess memory of the events occurring at the accretion disk, especially the disk modulations such as changes in the accretion rates, viscosity, magnetic field etc.,  that could be coupled with the jet processes such that the disk instabilities could drive the jet emission variability.  In other words, jet emission might ``remember'' disk processes, and this indicates strong disk-jet connection. 

In general, the power-law type PSD seen in most power spectra of blazar variability  can also be explained in the context of a turbulent flow behind propagating shock \citep[][]{Marscher1992} or a standing/reconfinement shock in blazar jets \citep{Marscher2008}.  If the emission from a single dominant turbulent cell get enhanced due to Doppler boosting, it contributes to the temporal frequency corresponding to the size or velocity of the cell. The stochastic nature of turbulence implies that cells of various sizes will be Doppler enhanced over time depending on their velocity and angle to the line of sight. Eventually,  this will result in a variability spectrum  over wide range of temporal frequencies that is consistent with the power-law noise seen in blazars \citep[see][]{Wiita2011}. In a similar context, the magnetic field at the accretion disk could be fairly magnetized owing to the material accreted over a considerable period of time. In such event, the magnetic field can extract the vast rotational energy by threading the black hole and channel into the jet as the bulk power of the relativistic jets. Moreover, as the radiation power is only 10\% of the total jet power, a significant contribution to the jet contents could be provided by poynting flux \citep{Ghisellini2014}, which then can facilitate the rampant magnetic reconnection events triggering stochastic particle acceleration and energy dissipation at various temporal and spatial scales. If the observed variable \gama-ray emission is  produced in such a scenario, the variability power spectrum should closely resemble  power-law shape.

\item Quasi-periodic oscillations: \\
We found the presence of year timescale QPOs in some of the sources with a high significance over the power-law PSDs.  The detected \gama-ray QPOs  can potentially offer profound insights on  the nature of high energy emission processes taking place in the sources. In particular, the studies can shed light on a number of current blazar issues such as  disk-jet connection, origin of relativistic jets from the central engine, and other extreme conditions near the fast rotating supermassive black holes. In principle, origin of QPOs can be conceived of mainly in three scenarios: supermassive binary black holes (SMBBH) system, accretion disk and jet instabilities.  Some of the possible explanation of the origins of QPOs  are discussed below. 

\begin{itemize}
\item SMBBH system:
In the context of SMBBH system, the observed timescales can be interpreted as the Keplerian periods of the secondary black hole around the central black hole as given by $T=2\pi a^{3/2} (G\ M)^{-1/2}$ with $M=M_p+M_s\sim 10^9 M_\odot$, where M$_s$ and  M$_p$ are the corresponding masses, respectively, and $a$ is the length of the semi-major axis of the elliptic orbit. Over the long course of merging galaxies, the dynamical friction present in the system can gradually smooth the elliptical orbits into circular orbits.  Then assuming a typical AGN total mass of  $M=M_p+M_s\sim 10^9 M_\odot$ and a mass ratio $M_s/M_p$ in the rage 0.1--0.01, the separation between the black holes can be estimated in the order of a few parsecs. Such a binary systems can undergo  orbital decay due to emission of low frequency (a few tens of nano-Hertz) gravitational waves (GW), which could be detected by future GW missions. For such a system,  the GW-driven orbital decay timescale can be estimated applying
 \begin{equation}
\tau_{insp}=3.05\times 10^{-6} \left ( \frac{M}{10^9M_\odot} \right )^{-3}\left ( \frac{a}{r_g} \right )^4 \rm years,
\end{equation}
\noindent \cite[see][]{Peters1964}, a few thousands years, rather short span of time relative to cosmic timescales.

\item Accretion disk:  Year time scale periodicity in blazars can be explained in the context of instabilities intrinsic to the accretion disk. To modulate flux periodically, a bright hotspot could be revolving around the central black hole with a Keplerian period, $\tau_k$, given by
\begin{equation}
\tau_k=0.36 \left ( \frac{M}{10^9M_\odot} \right )^{-1/2}\left ( \frac{a}{r_g} \right )^{3/2} \rm days,
\end{equation}
where $a$ is the length of the semi-major axis of the elliptic orbits.  Assuming circular orbits, for a typical black hole of mass of $10^9 M_\odot$ the radius of the Keplerian orbit for a year timescale can be estimated  to be a few tens of gravitational radius ($r_g$).    Similarly, in the case of globally perturbed thick accretion disks,  the disk can undergoes p-mode oscillations with a fundamental frequency that can be approximated as,  
   \begin{equation}
  f_{0}\approx 100\left ( \frac{r}{r_g} \right )^{-3/2}\left ( \frac{M}{10^8M\odot } \right )^{-1} \ \rm day^{-1}
 \end{equation}
\citep[see][and the reference therein]{An2013}.

\begin{figure*}[t!]  
\begin{center}  
\plotone{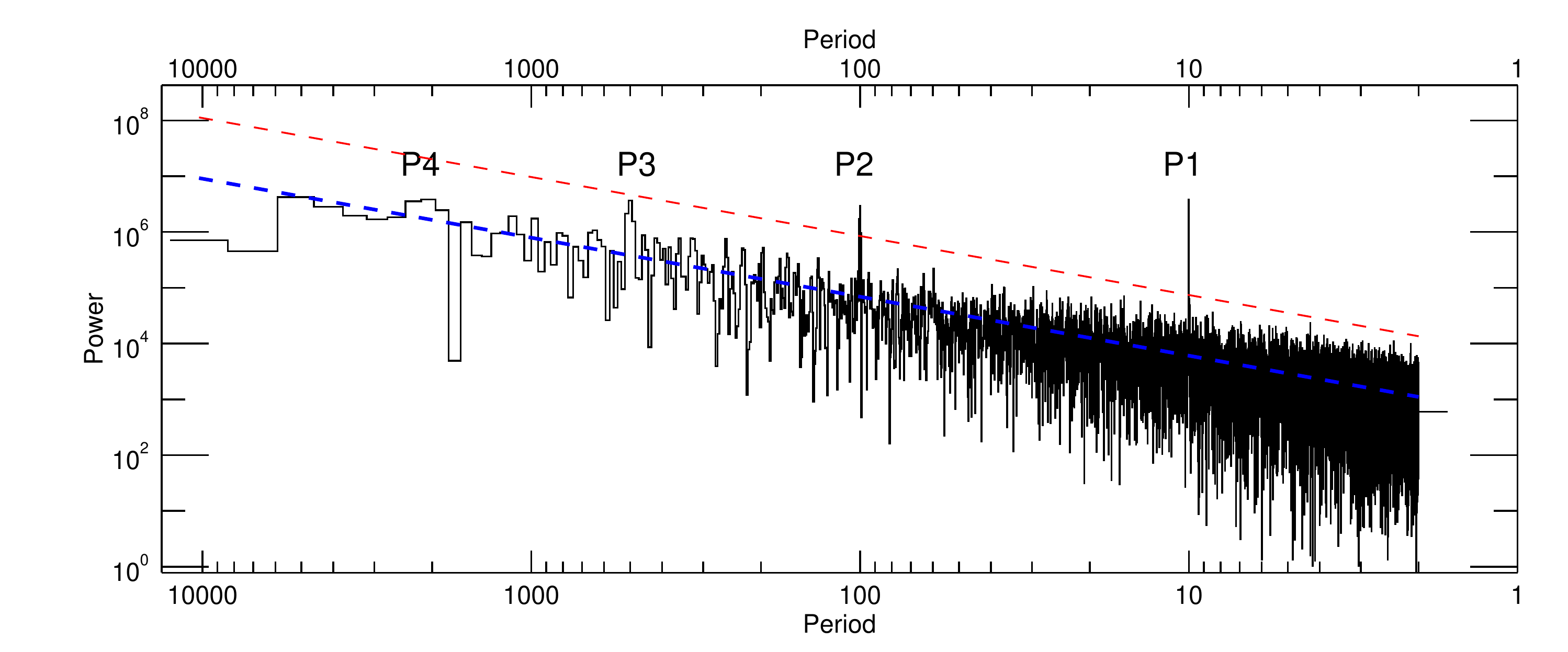}
\caption{ Discrete Fourier periodogram of evenly space light curve simulated applying  model PSD with spectral index of 1.5. Sinusoidal waves of periods 10, 100, 500 and 2000 in arbitrary time  unit  are superimposed on the pure power-law PSD. The blue line represents the log-linear fit to the periodogram and the red line shows the 99\% significance contour.}
\label{fig:8}
\end{center}
\end{figure*}

  To include the effects of strong gravitational field near a fast spinning  supermassive black hole, the frame dragging effect can warp  the inner part of the accretion disk. This might lead to the nodal precession of the tilted plane of the disk better known as the \emph{ Lense-Thirring precession}. The period of such precession can be expressed as 
 \begin{equation}
\tau_{LT}=0.18 \left ( \frac{1}{a_s} \right )\left ( \frac{M}{10^9M_\odot} \right )\left ( \frac{r}{r_g} \right )^{3} \rm days,
\end{equation}
\noindent  where $a_s$, $M$ and $r$ represent dimensionless spin parameter, mass of the black hole and the radial distance of the emission region from the black hole, respectively. For a ${10^9M_\odot}$ black  hole with a maximal spin ($a_s=0.9$),  a year timescale  would correspond to the inner part of accretion disk extending in the order of a few tens of $r_g$. In blazars, the precession of the disk also can lead to jet as precession thereby resulting in the periodic emission   \cite[e.g][]{Liska2018}.

\item Jets:   The observed quasi-periodic flux modulations can also be linked to relativistic motion of the emission regions along  helical path of the magnetized jets \cite[e.g.,][]{Camenzind92}.  In particular,  when emission regions move along the helical path of a jet with a  large bulk Lorentz factor, $\Gamma$,  relativistic effects become dominant such that  periodic flux modulations can appear due to the periodic changes  in the viewing angle. In such scenario, The rest frame flux (${F}'_{{\nu'}}$) and observed flux ($F_{\nu}$) are related through the relations
  \begin{equation}
F_{\nu}(\nu)=\delta(t)^{3+\alpha}{F}'_{{\nu}'} (\nu)   \quad\text{and}\quad      \delta(t)=1/\Gamma \left ( 1-\beta cos\theta(t)  \right ).
\label{flux}
\end{equation}

If we let intrinsic flux of the emission region unchanged but change viewing angle,  the corresponding ratio of the observed  flux to the intrinsic  flux for a given change in the angle $\Delta \theta$ can be expressed as
 \begin{equation}
 \Delta logF=-\left ( 3+\alpha  \right )\delta \Gamma \beta sin\theta \Delta \theta ,
 \end{equation} 
 For illustrative purpose, for blazars emission with typical \gama-ray spectral index ($\alpha=1.5$) and viewing angles in the $1-5 ^o$ range, a slight change in the viewing angle e.g., $\sim 1^o$, is sufficient to produce an apparent flux twice as bright (see \citealt{Bhatta2018d} Figure 4). Similarly,  QPOs can originate in blazars jets owing to recurring boosts of turbulent cells behind a propagating shock. If a biggest dominant structure stands out, it will exhibit enhanced Doppler boosting contributing to QPO component. However, it is possible that,  due to the stochastic nature of turbulence, the cell would gradually decay causing the amplitude of the QPO to diminish accordingly over a period of time \citep[see][]{Wiita2011}.

 \end{itemize}

It should be stressed that the dominance of red-noise in blazar light curves often gives rise to a general skepticism towards the actual presence of the QPOs in blazar, particularly QPOs at the low-frequency (LF) ends frequently reported in the literature. Consequently, many authors tend to  adopt a conservative measure for the significance,  such as $\gtrapprox$ 99.99\% over the PSD, required to establish their existence. However, we argue that if we take such a strict approach towards the significance,  there could be a risk of overlooking many interesting features in AGN  and thereby we may miss exciting physics. To illustrate our point we present periodogram of a pure power-law of spectral index 1.5 on which purely sinusoidal waves of the periods 10, 100, 500, and 2000 days but of the same intensity are superimposed as shown in Figure \ref{fig:8}. The simulated light curve is evenly spaced so that the 99\% significance is computed using Equation 16 in \cite{Vaughan2005}. The figure shows that for the same amplitude of the periodic modulation, the significance of the peaks gradually decreases as we move from HF to LF such that even in a relatively ideal situation of purely sinusoidal modulations present in the evenly spaced observations, the corresponding spectral peaks can get drowned into the strong power-law trend which is ever-rising, and consequently fail to pass the 99 \% significance test. Similar situation might arise when LF QPOs are unable to maintain phase coherence over more a few oscillations. In such cases,  performing statistical analysis using multiple methods e.g. carry out both frequency and time-based analysis \citep[see][]{Bhatta2019} would be more useful.  Furthermore, a year-scale QPOs could arise in various scenarios discussed above \citep[see also][]{Bhatta2019, Bhatta2018d,Bhatta2017,Ackermann15}; now it is a challenging task to break the apparent degeneracy in the models to single out the actual process behind the detection. The task would require an in-depth analysis of multi-frequency light curves applying multiple approaches to the time series analysis.
\end{itemize}

\section{Conclusion \label{sec:5}}
We performed an in-depth time domain analysis of decade long (2008-2018) Fermi/LAT light curves of a sample of 20 bright blazars.  We found that  \gama-ray emission from blazars is highly pronounced and variable over diverse timescale. As one of characteristic features, a steep linear trend was observed in the correlation between fractional variability and the \gama-ray spectral index suggesting that the variability is highly sensitive to its spectral slope. The \gama-ray flux of the blazars is found to be distributed in a way that is closely approximated as lognormal  PDF.  Statistical analysis of flux rising and decay rates in the \gama-ray light curves show that both the distribution are very similar and therefore no significant asymmetry between the flux rising and decay profiles was detected. Moreover, most of the sources appear to exhibit a linear RMS-flux relation indicating higher flux states are often more variable. Furthermore, to constrain the statistical nature of such variability over a wide range of temporal frequencies, extensive MC simulations were performed to estimate the PSDs which best represent the blazar \gama-ray periodogram. The study shows that the PSDs are consistent with a single power-law, $P(\nu)\propto1/\nu$,  with spectral indexes centered around 1.0 indicating the nature of variability as flicker-noise and, therefore, might be driven by long-memory processes. Additionally, a closer inspection of the Lomb-Scargle and WWZ periodograms of some of the sources in the sample, including S5 0716+714, Mrk 421, ON +325, PKS 1424-418 and PKS 2155-304, reveal spectral features that signify presence of year timescales QPOs that are highly significant over the possible artifacts usually found in blazar light curves.

\acknowledgments

GB acknowledges the financial support by the Narodowe Centrum Nauki (NCN) grant UMO-2017/26/D/ST9/01178. We would like to express   
our gratitude to Prof. Staszek Zola for kindly allowing us to use their computational facility for this research. We would also like to thank Prof. Michal Ostrowski and Prof. Alan Marscher for fruitful discussion on the \gama-ray variability of blazars. The authors are grateful to the anonymous referee for his/her comments that significantly improved the quality of the work.

%

\vspace{5mm}
\facilities{Fermi/LAT}


\software{HEAsoft (HEASARC 2014),  FTOOLS (Blackburn 1995),  and\  fitdistrplus(Delignette-Muller \& Dutang 2015) }



\appendix

\section{Decade-long Fermi/LAT light curves of blazars \label{apndx1}}

\begin{figure*}[!b]
\begin{center}
\plotone{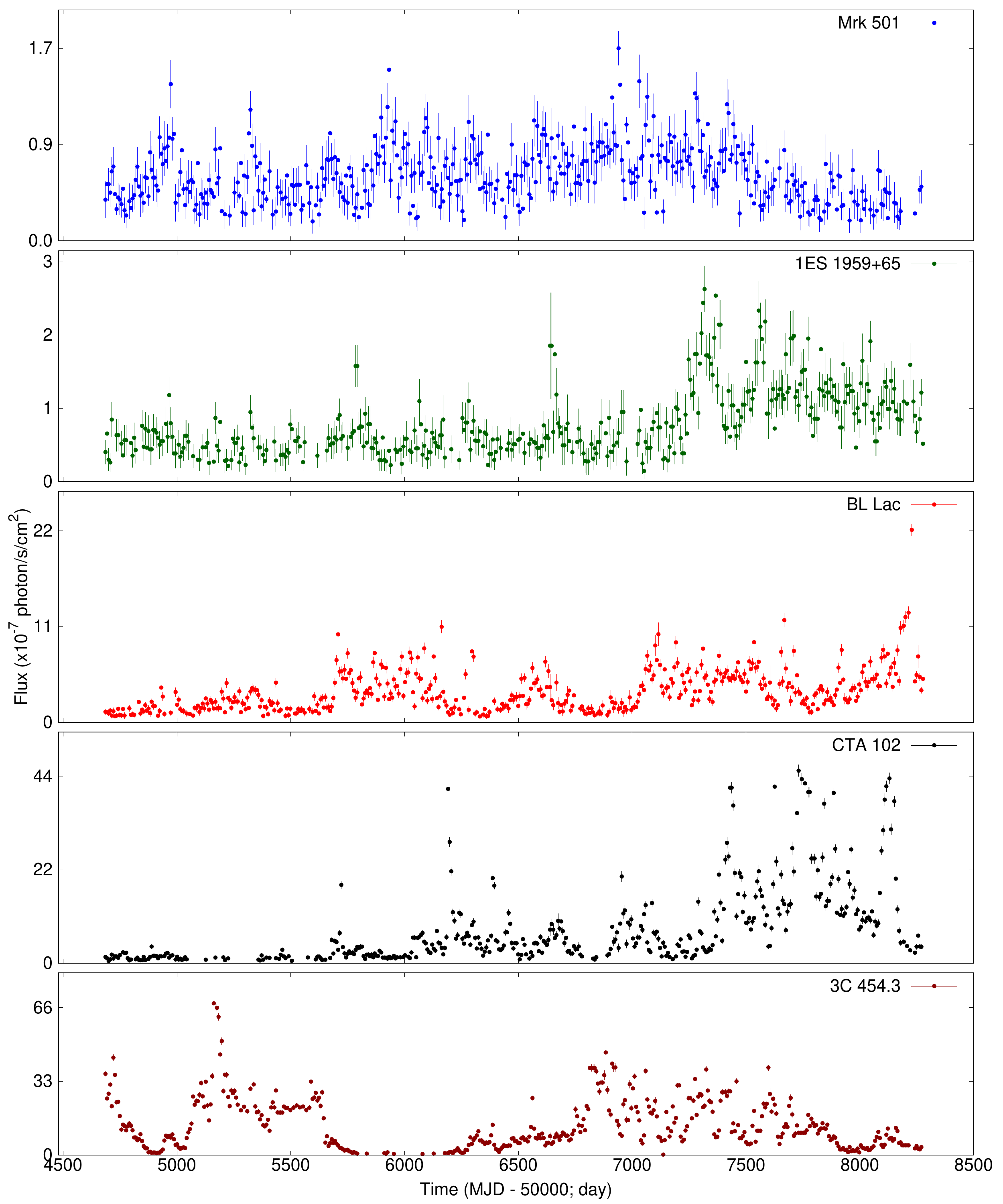}
\caption{Weekly binned Fermi/LAT light curves of  the sample blazar listed in Table \ref{table:1} \label{Fig:9}}
\end{center}
\end{figure*}

\renewcommand{\thefigure}{\arabic{figure} (Cont.)}
\addtocounter{figure}{-1}

\begin{figure*}[!ht]
\begin{center}
\plotone{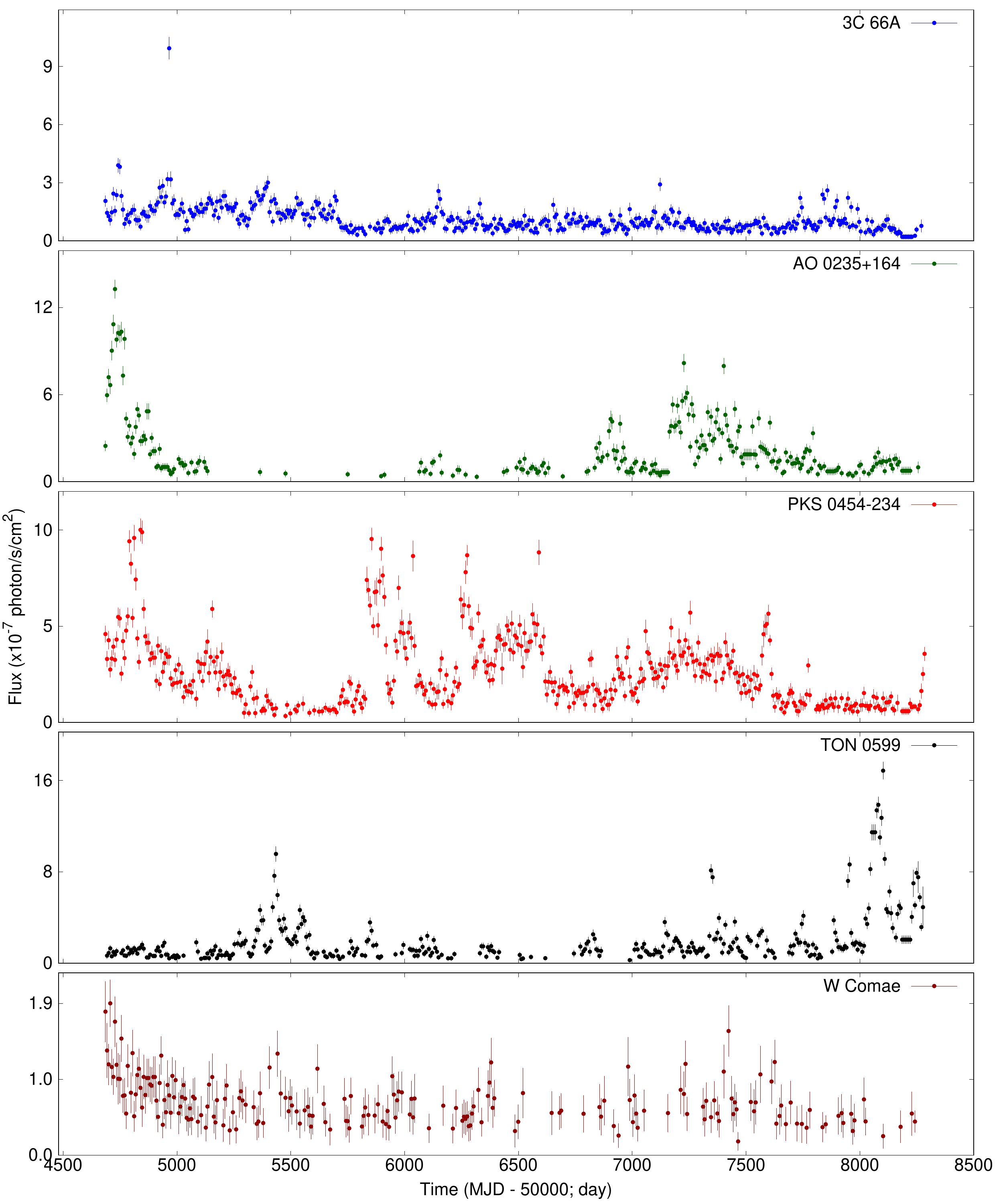}
\caption{\label{Fig:10}}
\end{center}
\end{figure*}

\clearpage

\section{Fitting of blazar flux distribution using weighted least square method \label{apndx2}}

\begin{deluxetable*}{llll|lll}[!b]
\tablecaption{Lognormal and normal distribution fit statistics for the \gama-ray flux distribution of the Fermi/LAT sources using weighted least square method.\label{table:6}}
\tablewidth{500pt}
\tabletypesize{\scriptsize}
\tablehead{
\colhead{} & 
\multicolumn{3}{c}{Lognormal fit} & \multicolumn{3}{c}{Normal fit} \\
\hline
\colhead{Source name} &
\colhead{$m$} & \colhead{$s$} & 
\colhead{$\chi^{2}/dof$}  &  
\colhead{$\mu$} & \colhead{$\sigma$} &
\colhead{$\chi^{2}/dof$} \\
} 
\colnumbers
\startdata
  3C 66A & -0.03 $\pm$ 0.02 & 0.52 $\pm$ 0.02 & 4.19/6 & 0.81 $\pm$ 0.07 & 0.75 $\pm$ 0.07 & 24.58/6 \\ 
  AO 0235+164 & 0.34 $\pm$ 0.08 & 0.90 $\pm$ 0.09 & 7.63/10 & -24.37 $\pm$ 2.07 & 6.73 $\pm$ 0.53 & 11.66/10 \\ 
  PKS 0454-234 & 0.72 $\pm$ 0.04 & 0.75 $\pm$ 0.03 & 17.04/17 & 0.82 $\pm$ 0.51 & 2.54 $\pm$ 0.27 & 15.21/17 \\ 
  S5 0716+714 & 0.63 $\pm$ 0.04 & 0.72 $\pm$ 0.03 & 20.02/16 & 1.63 $\pm$ 0.14 & 1.63 $\pm$ 0.12 & 34.72/16 \\ 
  Mrk 421 & 0.62 $\pm$ 0.02 & 0.35 $\pm$ 0.01 & 36.22/16 & 1.94 $\pm$ 0.04 & 0.71 $\pm$ 0.03 & 56.54/16 \\ 
  TON 0599 & 0.25 $\pm$ 0.07 & 0.79 $\pm$ 0.09 & 19.27/12 & -19.07 $\pm$ 1.42 & 5.33 $\pm$ 0.36 & 27.05/12 \\ 
  ON +325 & -0.32 $\pm$ 0.03 & 0.62 $\pm$ 0.03 & 25.45/9 & 0.78 $\pm$ 0.03 & 0.48 $\pm$ 0.03 & 23.88/9 \\ 
  W Comae & -0.51 $\pm$ 0.03 & 0.40 $\pm$ 0.02 & 11.25/10 & 0.61 $\pm$ 0.02 & 0.21 $\pm$ 0.02 & 41.99/10 \\ 
  4C +21.35 & 0.83 $\pm$ 0.09 & 1.11 $\pm$ 0.10 & 13.12/13 & -47.73 $\pm$ 3.67 & 12.97 $\pm$ 0.93 & 23.43/13 \\ 
  3C 273 & 0.84 $\pm$ 0.04 & 0.70 $\pm$ 0.06 & 25.98/12 & -34.07 $\pm$ 2.64 & 9.52 $\pm$ 0.65 & 39.21/12 \\ 
  3C 279 & 1.30 $\pm$ 0.04 & 0.87 $\pm$ 0.06 & 39.84/16 & -63.96 $\pm$ 4.19 & 17.56 $\pm$ 1.06 & 57.24/16 \\ 
  PKS 1424-418 & 1.52 $\pm$ 0.03 & 0.66 $\pm$ 0.02 & 8.54/15 & 4.16 $\pm$ 0.26 & 3.51 $\pm$ 0.52 & 73.60/15 \\ 
  PKS 1502+106 & 1.08 $\pm$ 0.05 & 0.85 $\pm$ 0.05 & 8.26/10 & -51.32 $\pm$ 5.73 & 14.17 $\pm$ 0.99 & 13.33/10 \\ 
  4C +38.41 & 0.64 $\pm$ 0.07 & 1.01 $\pm$ 0.07 & 24.30/13 & -37.57 $\pm$ 2.98 & 10.25 $\pm$ 0.75 & 38.12/13 \\ 
  Mrk 501 & -0.63 $\pm$ 0.02 & 0.45 $\pm$ 0.02 & 20.36/15 & 0.54 $\pm$ 0.02 & 0.27 $\pm$ 0.01 & 15.55/15 \\ 
  1ES 1959+65 & -0.39 $\pm$ 0.03 & 0.53 $\pm$ 0.02 & 29.06/15 & 0.71 $\pm$ 0.03 & 0.35 $\pm$ 0.03 & 98.54/15 \\ 
  PKS 2155-304 & 0.09 $\pm$ 0.02 & 0.48 $\pm$ 0.02 & 16.45/9 & 1.10 $\pm$ 0.03 & 0.48 $\pm$ 0.03 & 42.73/9 \\ 
  BL Lac & 1.03 $\pm$ 0.04 & 0.71 $\pm$ 0.03 & 35.63/17 & 1.05 $\pm$ 0.69 & 3.41 $\pm$ 0.32 & 23.65/17 \\ 
  CTA 102 & 1.49 $\pm$ 0.09 & 1.22 $\pm$ 0.11 & 18.90/15 & -97.12 $\pm$ 11.90 & 26.18 $\pm$ 2.98 & 40.00/15 \\ 
  3C 454.3 & 2.34 $\pm$ 0.06 & 1.00 $\pm$ 0.05 & 52.79/12 & -33.57 $\pm$ 28.02 & 27.80 $\pm$ 7.17 & 46.30/12 \\
\enddata
\tablecomments{For the normal fit $\mu$ and $\sigma$ are presented in the unit of flux in 10$^{-7}$ $\times$  counts/sec/cm$^2$, whereas for the lognormal fit $m$ is in the unit of natural log of flux.}
\end{deluxetable*}

\clearpage
\renewcommand{\thefigure}{\arabic{figure}}
\section{Probability Density Function for Blazar Flux Distribution \label{apndx3}}
\begin{figure*}[!b]
\gridline{
          \fig{{"Plots2/Flux_Dist/3C66A"}.pdf}{0.3\textwidth}{}\hspace{-0.7cm}
          \fig{{"Plots2/Flux_Dist/AO0235+164"}.pdf}{0.3\textwidth}{}\hspace{-0.7cm}
           \fig{{"Plots2/Flux_Dist/PKS0454-234"}.pdf}{0.3\textwidth}{}\hspace{-0.7cm}
}
 \vspace{-0.5cm}
\gridline{
          \fig{{"Plots2/Flux_Dist/S50716+714"}.pdf}{0.3\textwidth}{}\hspace{-0.7cm}
          \fig{{"Plots2/Flux_Dist/Mrk421"}.pdf}{0.3\textwidth}{}\hspace{-0.7cm}
          \fig{{"Plots2/Flux_Dist/TON0599"}.pdf}{0.3\textwidth}{}\hspace{-0.7cm}
          }
         \vspace{-0.5cm}
\gridline{
         \fig{{"Plots2/Flux_Dist/ON+325"}.pdf}{0.3\textwidth}{}\hspace{-0.7cm}
         \fig{{"Plots2/Flux_Dist/WComae"}.pdf}{0.3\textwidth}{}\hspace{-0.7cm}
          \fig{{"Plots2/Flux_Dist/4C+21.35"}.pdf}{0.3\textwidth}{}\hspace{-0.7cm}
}
\vspace{-0.5cm}
\caption{Lognormal and normal distribution fit to the \gama-ray flux distribution of the Fermi/LAT sources listed in Table \ref{table:1} using MLE method. \label{fig:LogN}}
\end{figure*}

\renewcommand{\thefigure}{\arabic{figure} (Cont.)}
\addtocounter{figure}{-1}
\begin{figure*}

         \gridline{          
\fig{{"Plots2/Flux_Dist/3C273"}.pdf}{0.3\textwidth}{}\hspace{-0.7cm}
\fig{{"Plots2/Flux_Dist/3C279"}.pdf}{0.3\textwidth}{}\hspace{-0.7cm}
\fig{{"Plots2/Flux_Dist/PKS1424-418"}.pdf}{0.3\textwidth}{}\hspace{-0.7cm}
}
\vspace{-0.5cm}

\gridline{
\fig{{"Plots2/Flux_Dist/PKS1502+106"}.pdf}{0.3\textwidth}{}\hspace{-0.7cm}
\fig{{"Plots2/Flux_Dist/4C+38.41"}.pdf}{0.3\textwidth}{}\hspace{-0.7cm}
\fig{{"Plots2/Flux_Dist/Mrk501"}.pdf}{0.3\textwidth}{}\hspace{-0.7cm}
}
\vspace{-0.2cm}

\gridline{
		\fig{{"Plots2/Flux_Dist/1ES1959+65"}.pdf}{0.3\textwidth}{}\hspace{-0.7cm}
\fig{{"Plots2/Flux_Dist/PKS2155-304"}.pdf}{0.3\textwidth}{}\hspace{-0.7cm}
\fig{{"Plots2/Flux_Dist/BLLac"}.pdf}{0.3\textwidth}{}\hspace{-0.7cm}
}
\vspace{-0.2cm}
\gridline{
\fig{{"Plots2/Flux_Dist/CTA102"}.pdf}{0.3\textwidth}{}\hspace{-0.7cm}
		\fig{{"Plots2/Flux_Dist/3C454.3"}.pdf}{0.3\textwidth}{}\hspace{-0.7cm}
}
\vspace{-0.2cm}

\caption{ \label{fig:LogN2}}
\end{figure*}

\clearpage

\renewcommand{\thefigure}{\arabic{figure}}

\section{RMS-flux relation in blazar \label{apndx4}}

\begin{figure*}[!b]
\gridline{\fig{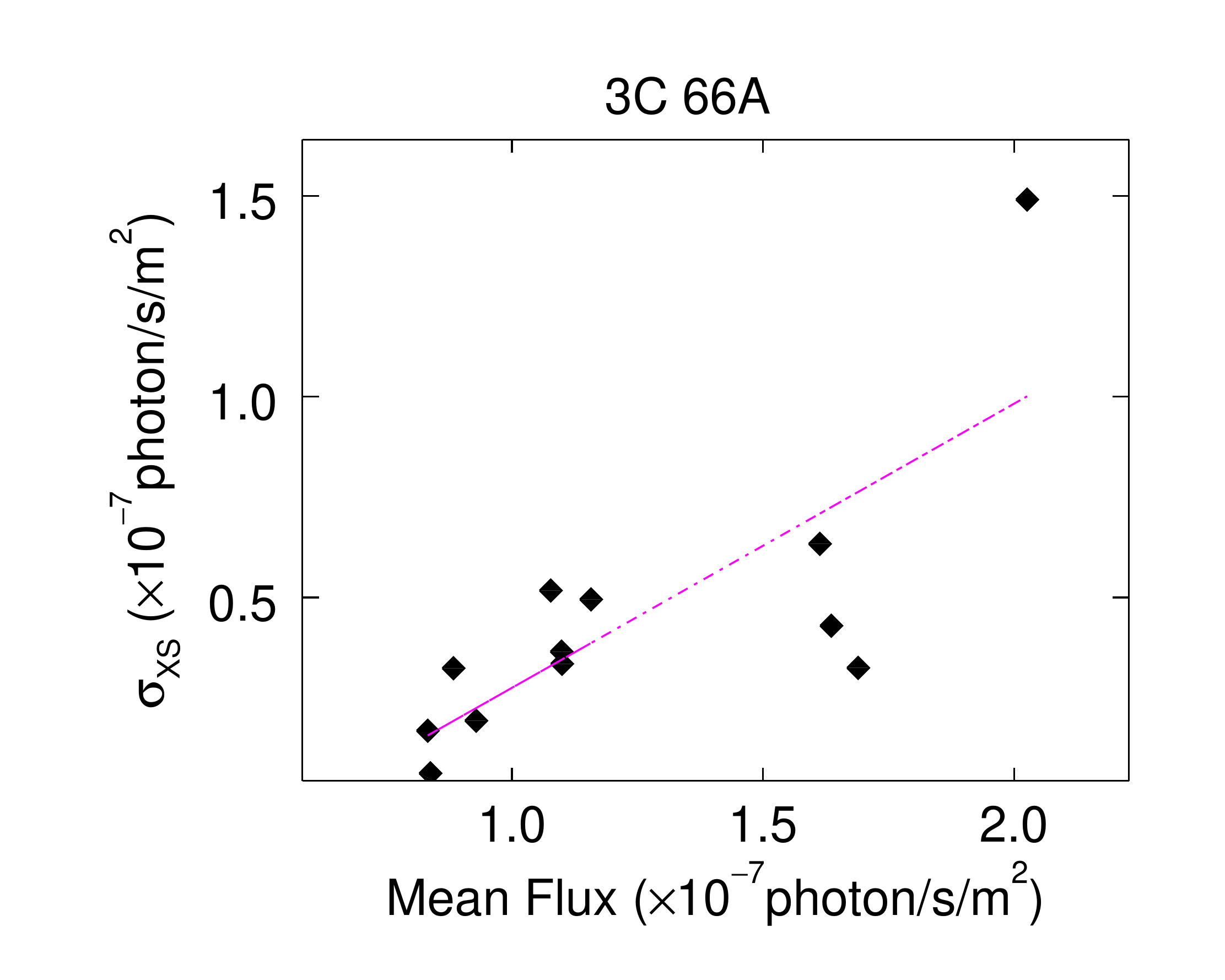}{0.25\textwidth}{}\hspace{-0.7cm}
\fig{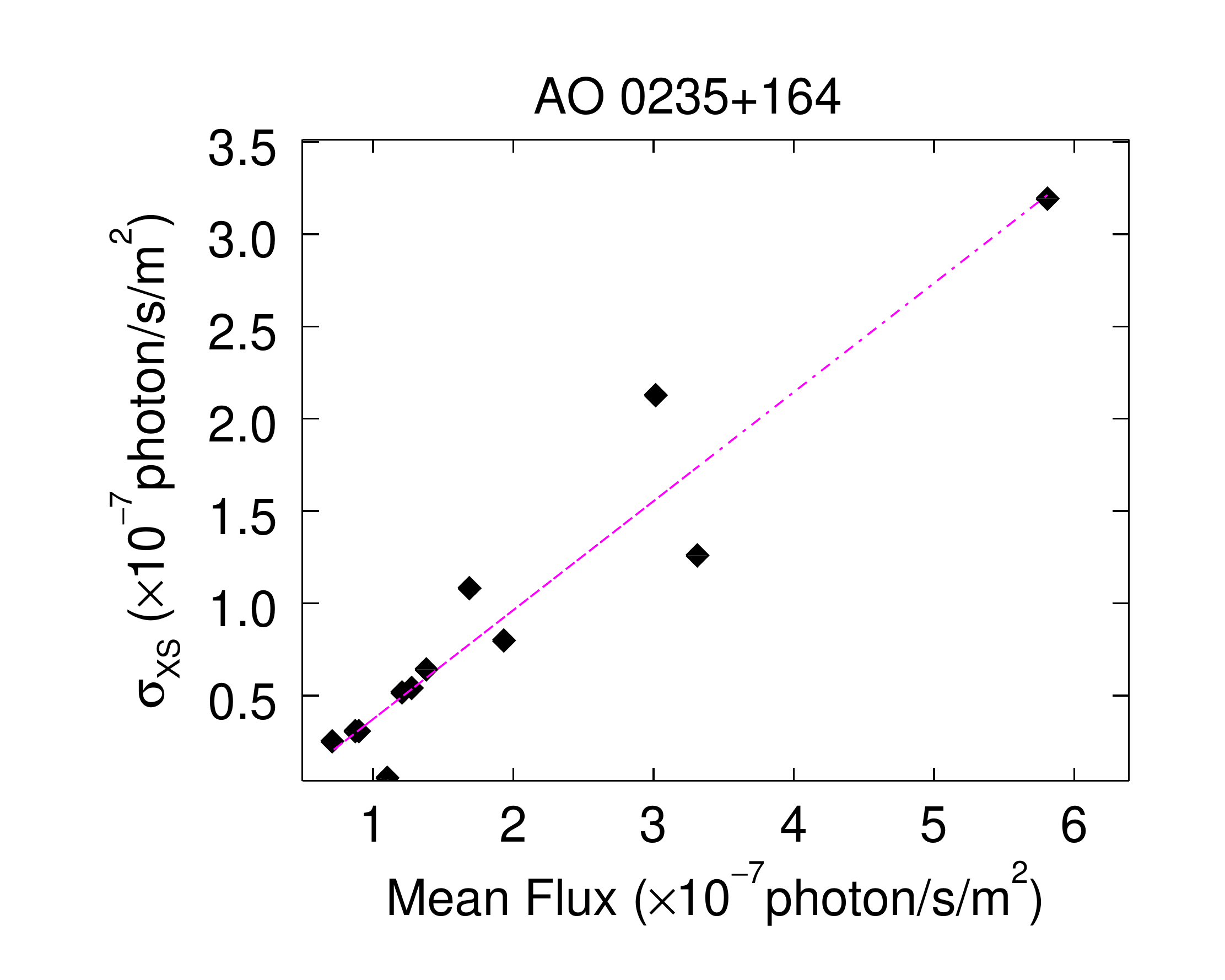}{0.25\textwidth}{}\hspace{-0.7cm}
\fig{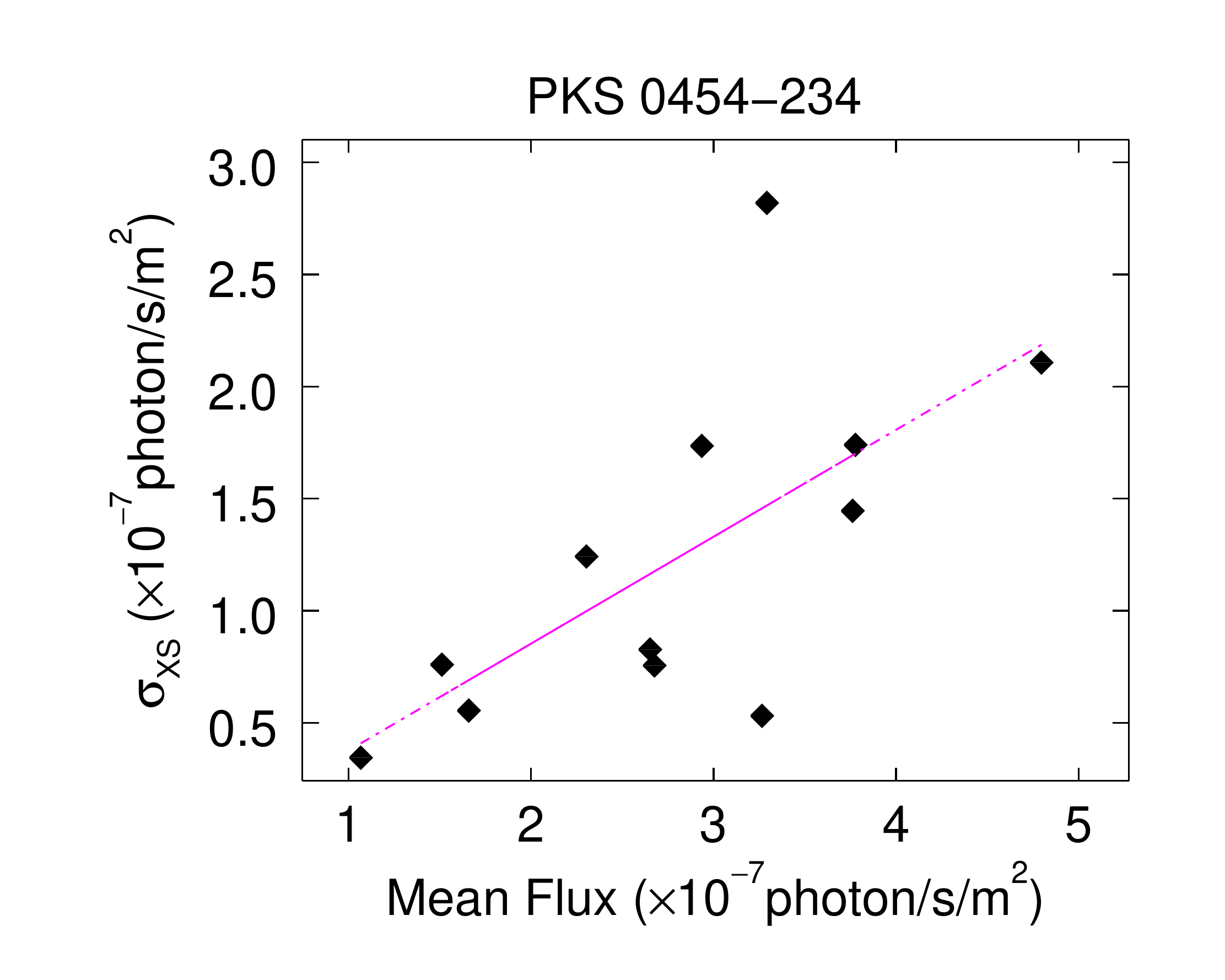}{0.25\textwidth}{}\hspace{-0.7cm}
\fig{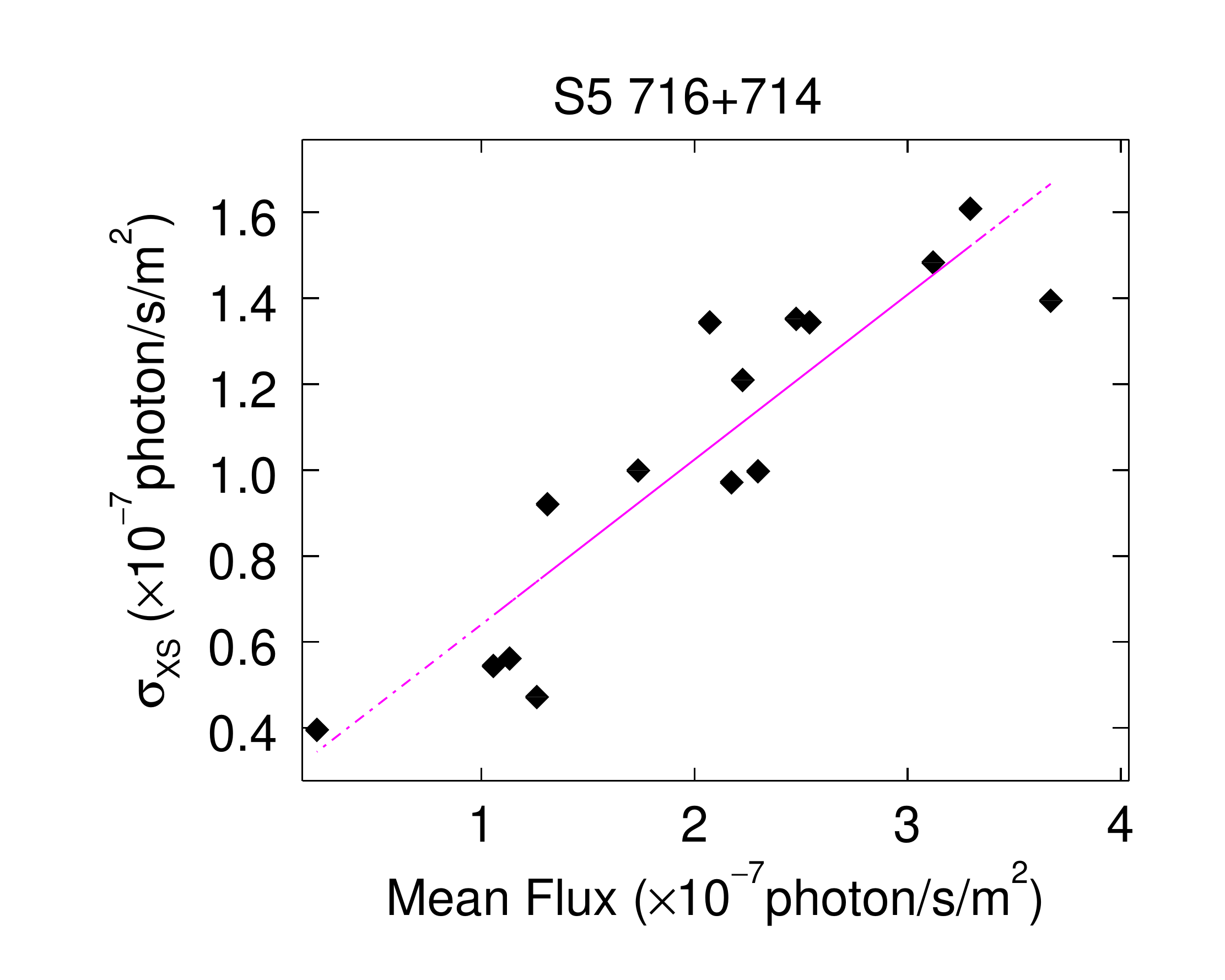}{0.25\textwidth}{}\hspace{-0.7cm}
          }
          \vspace{-0.5cm}
\gridline{\fig{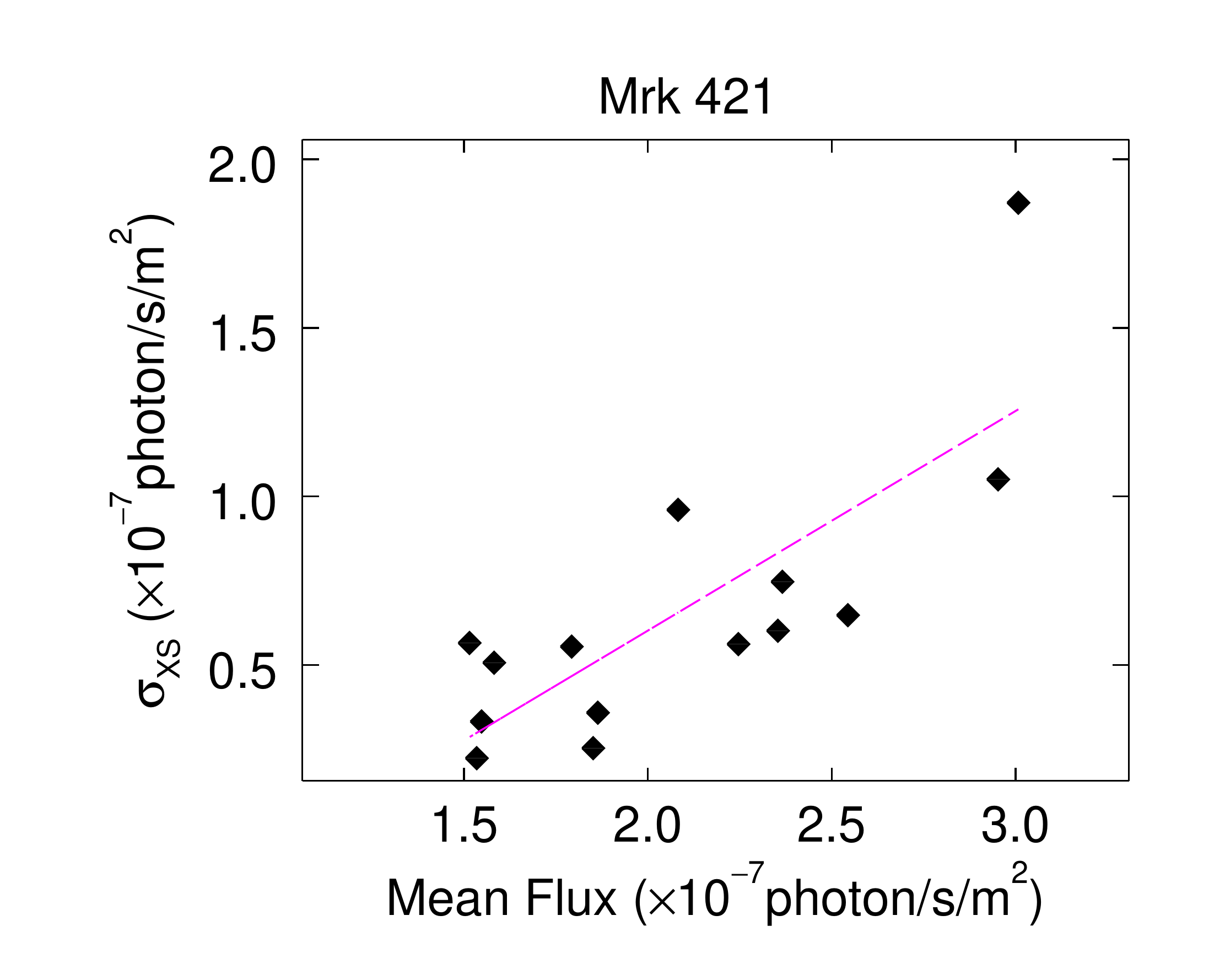}{0.25\textwidth}{}\hspace{-0.7cm}
\fig{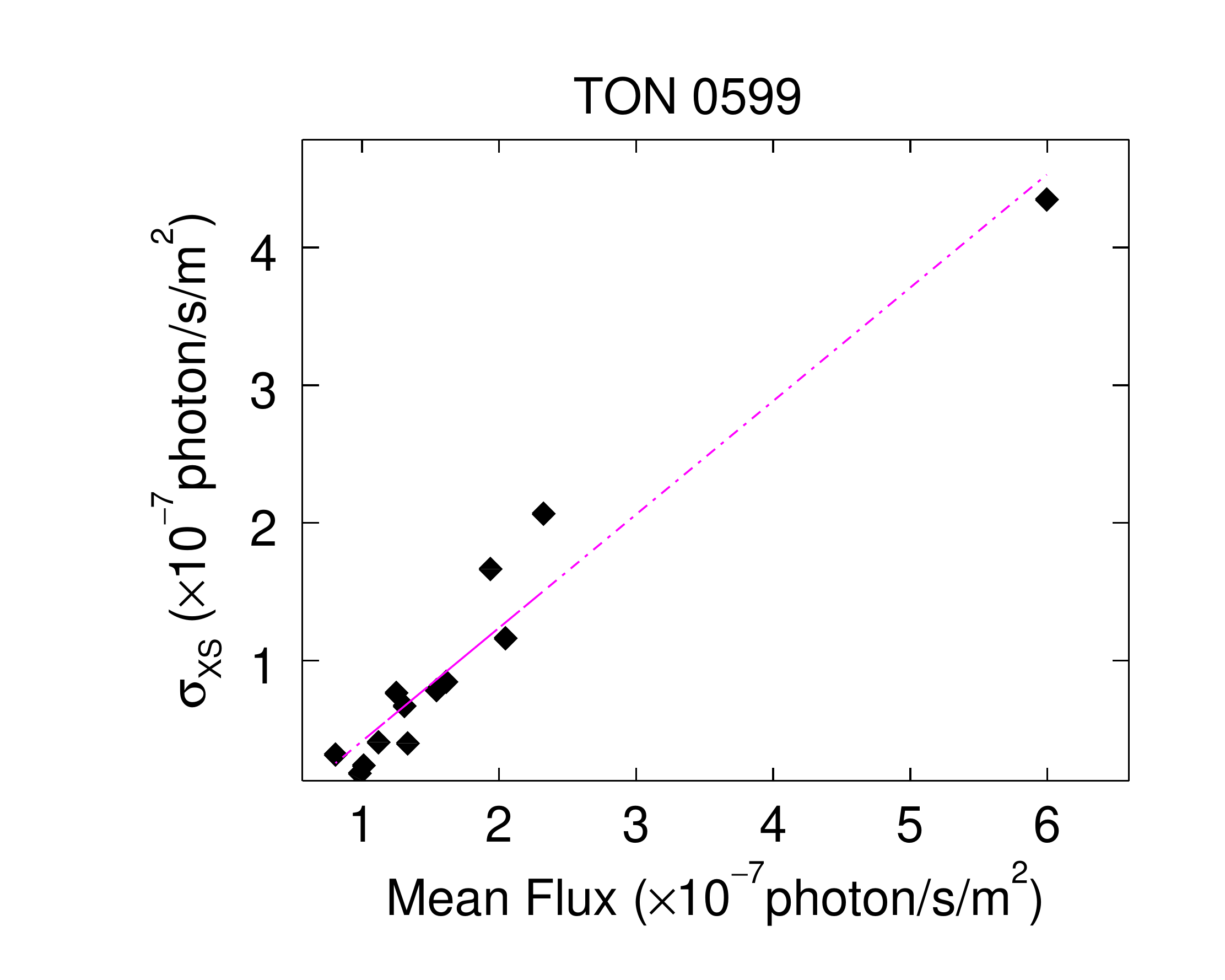}{0.25\textwidth}{}\hspace{-0.7cm}
\fig{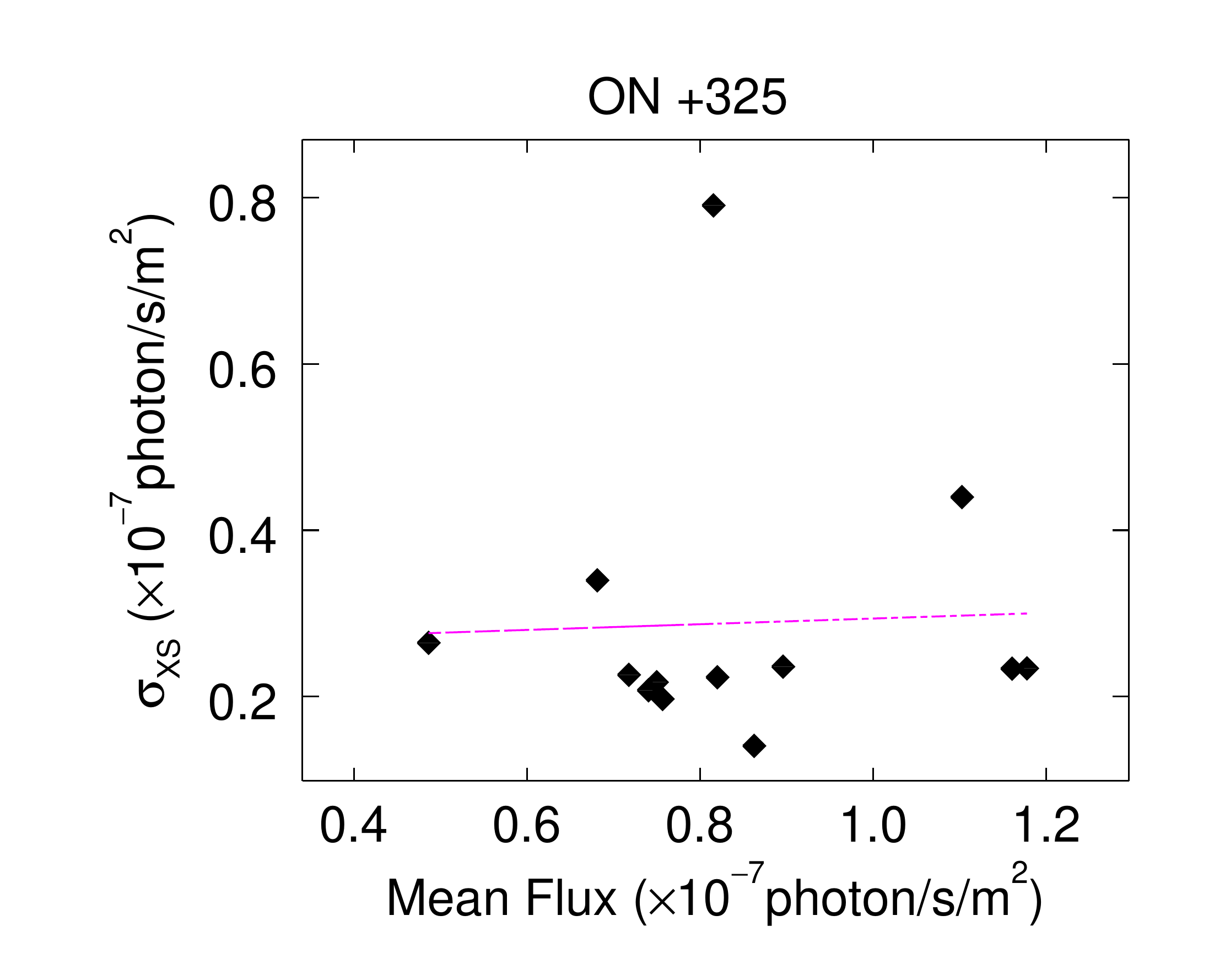}{0.25\textwidth}{}\hspace{-0.7cm}
\fig{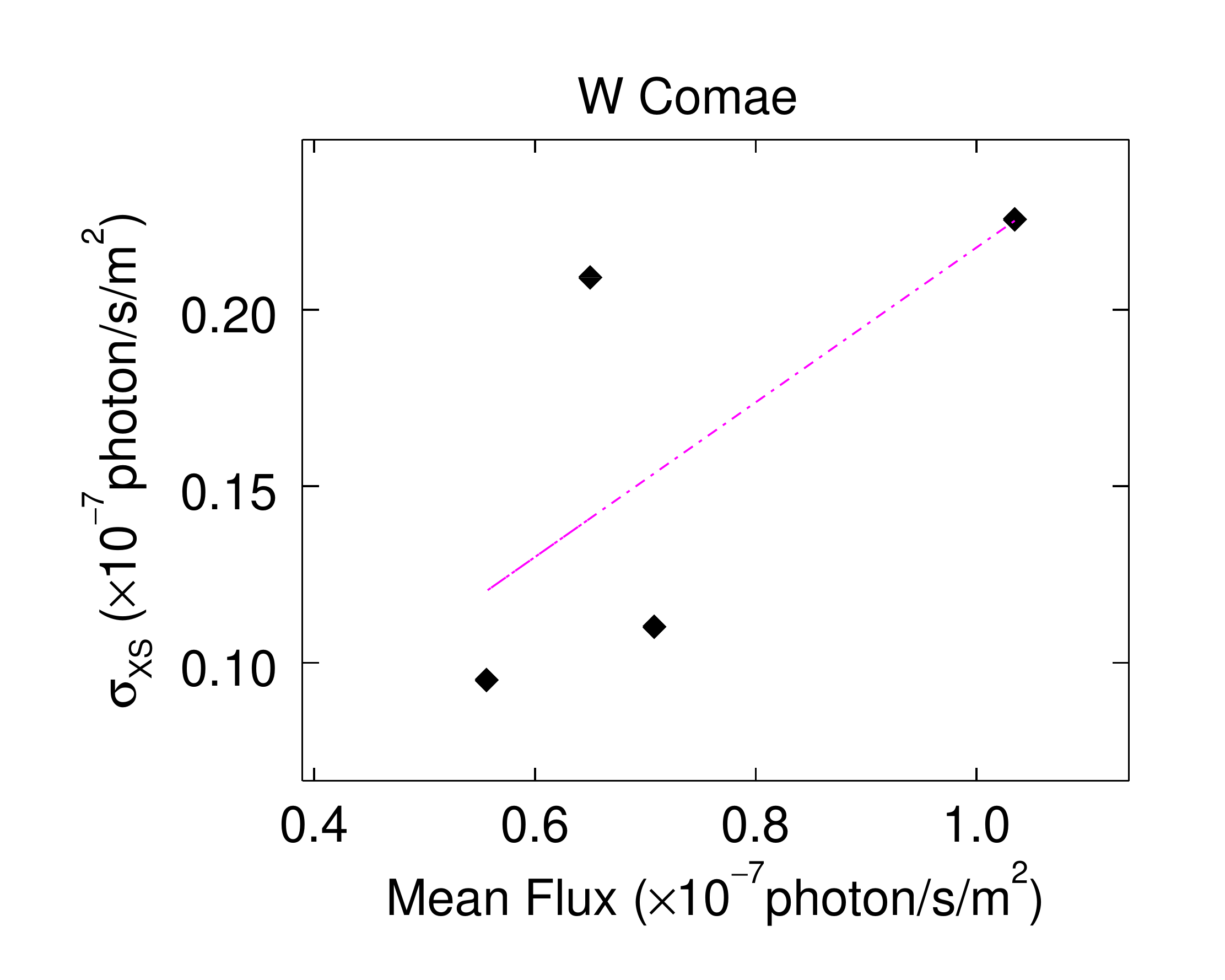}{0.25\textwidth}{}\hspace{-0.7cm}
          }
          \vspace{-0.5cm}
\gridline{\fig{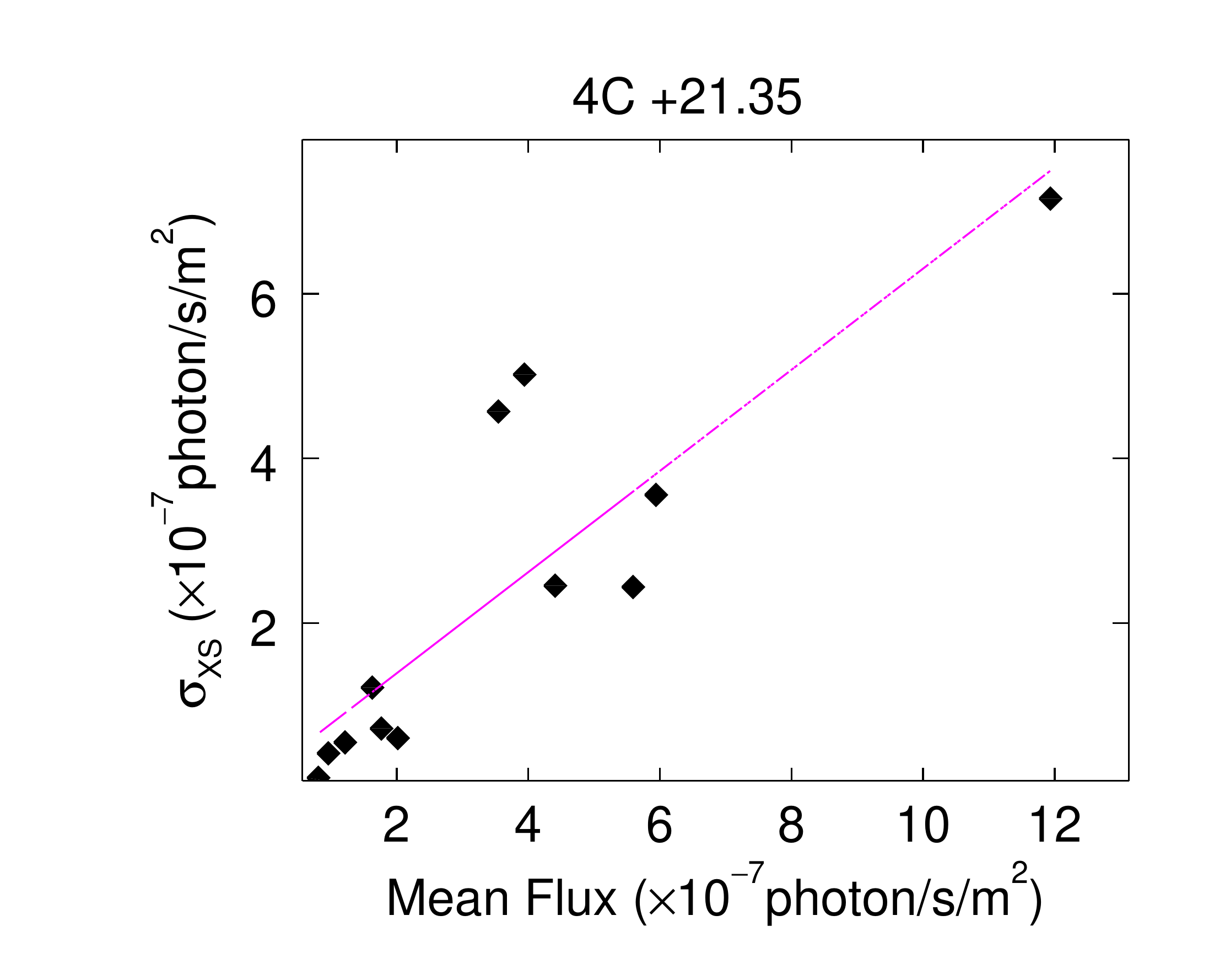}{0.25\textwidth}{}\hspace{-0.7cm}
\fig{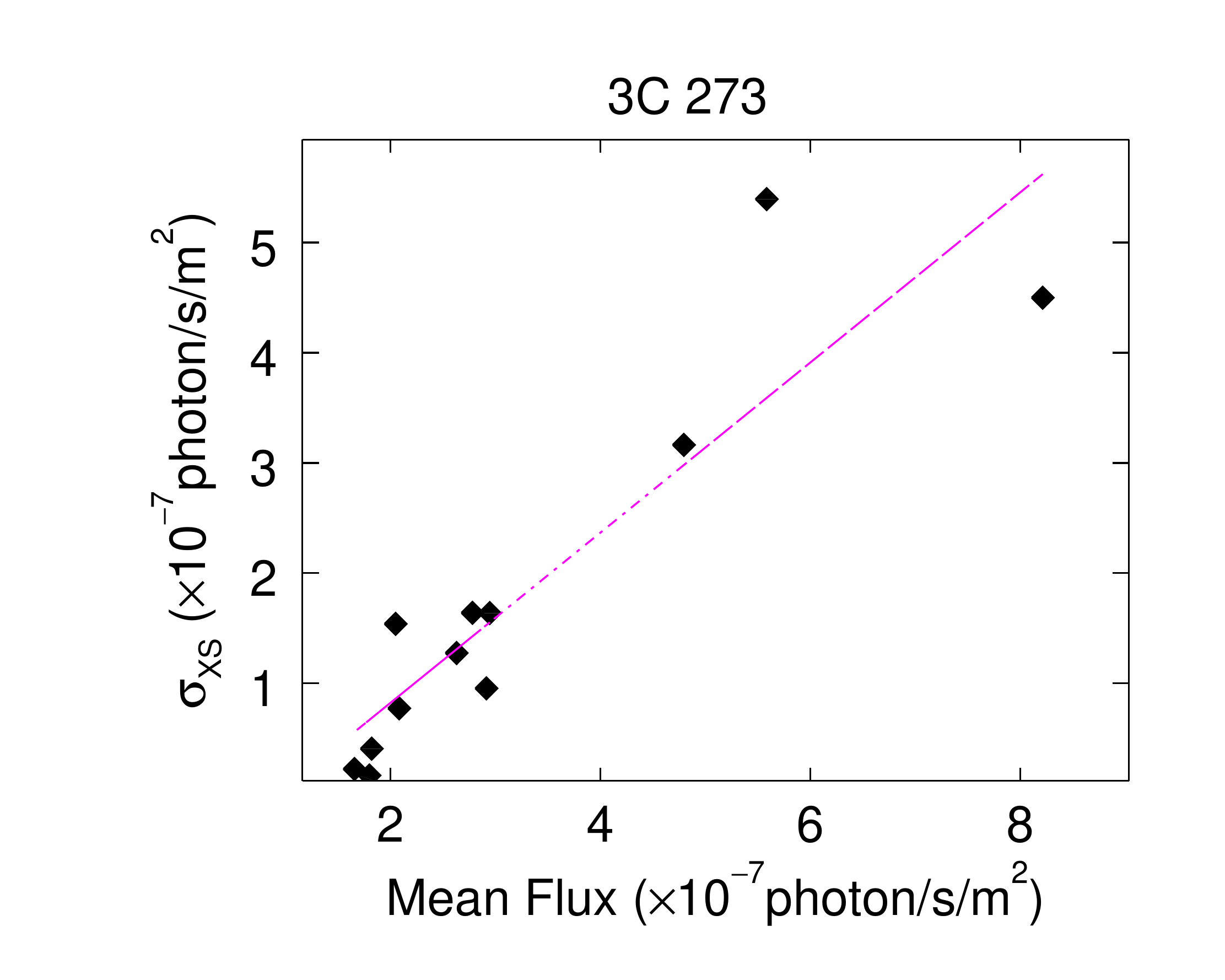}{0.25\textwidth}{}\hspace{-0.7cm}
\fig{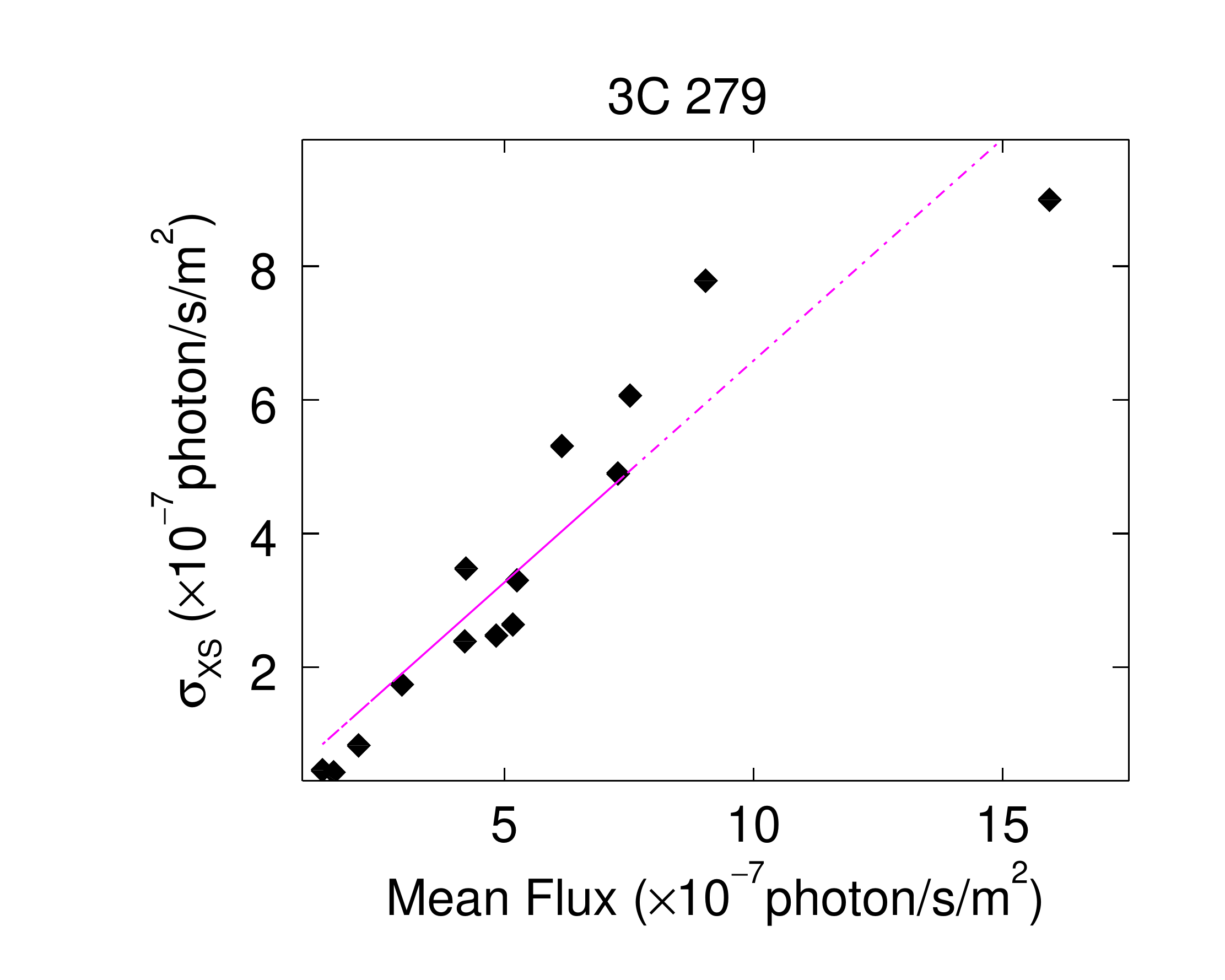}{0.25\textwidth}{}\hspace{-0.7cm}
\fig{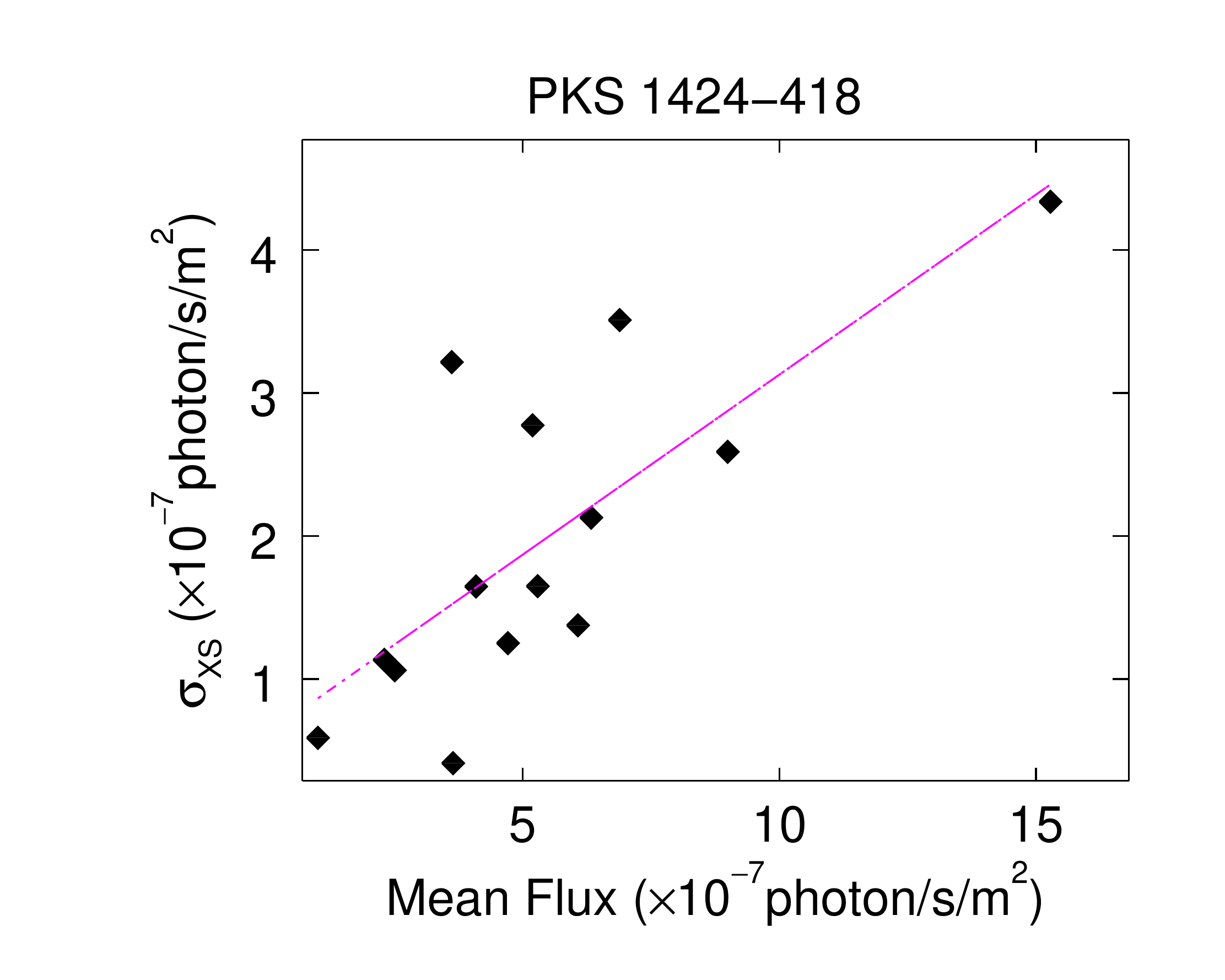}{0.25\textwidth}{}\hspace{-0.7cm}
          }
                    \vspace{-0.5cm}
\gridline{\fig{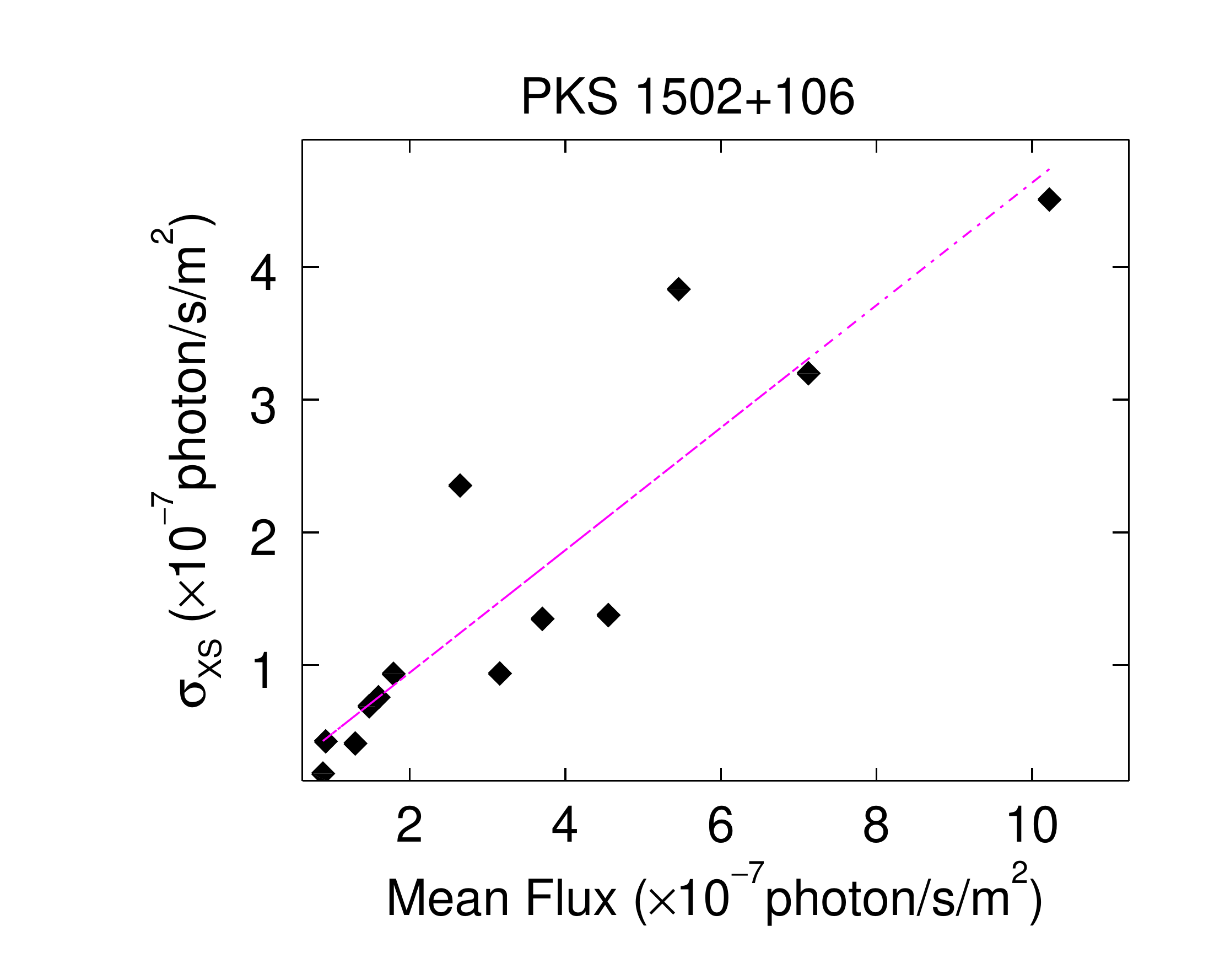}{0.25\textwidth}{}\hspace{-0.7cm}
\fig{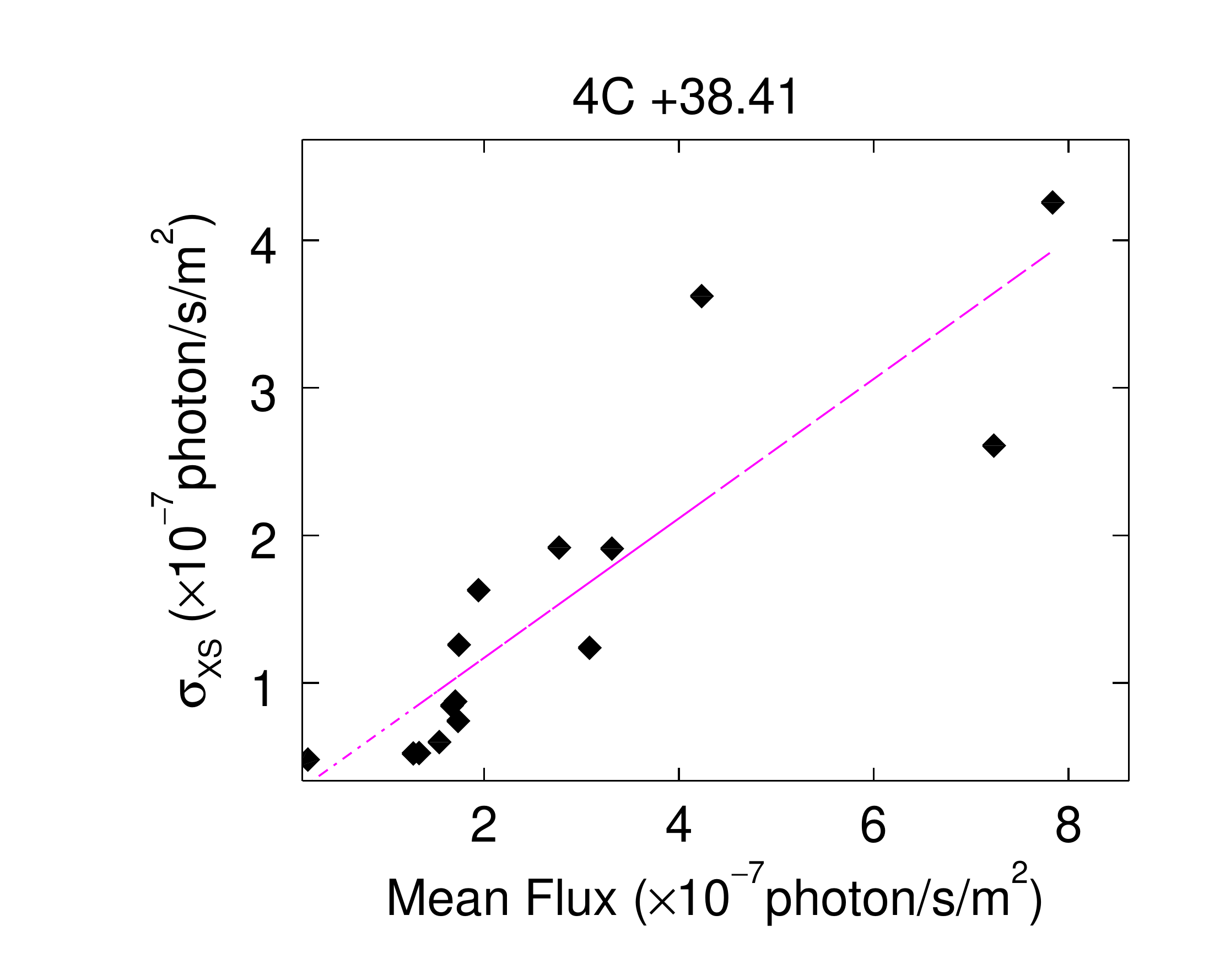}{0.25\textwidth}{}\hspace{-0.7cm}
\fig{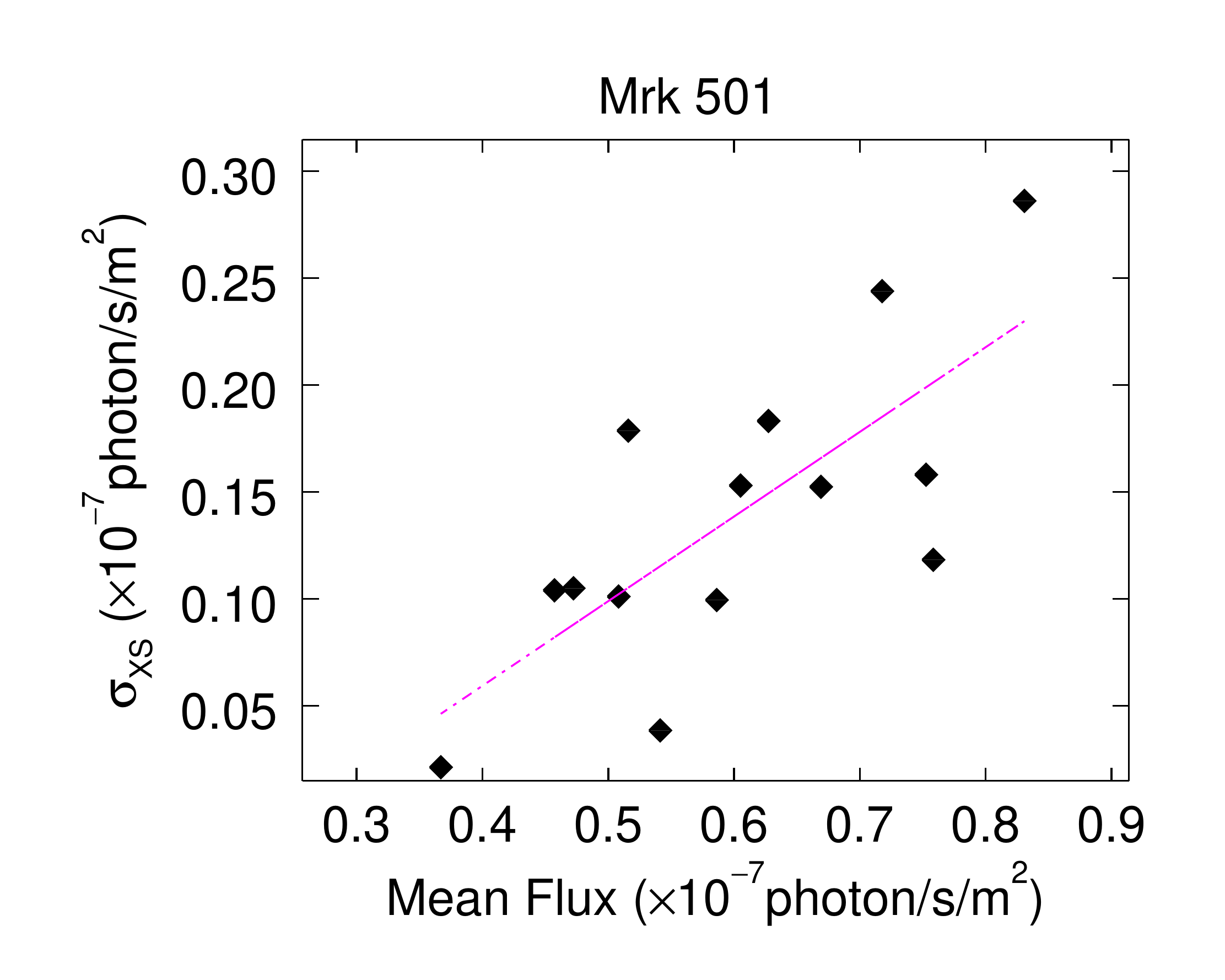}{0.25\textwidth}{}\hspace{-0.7cm}
\fig{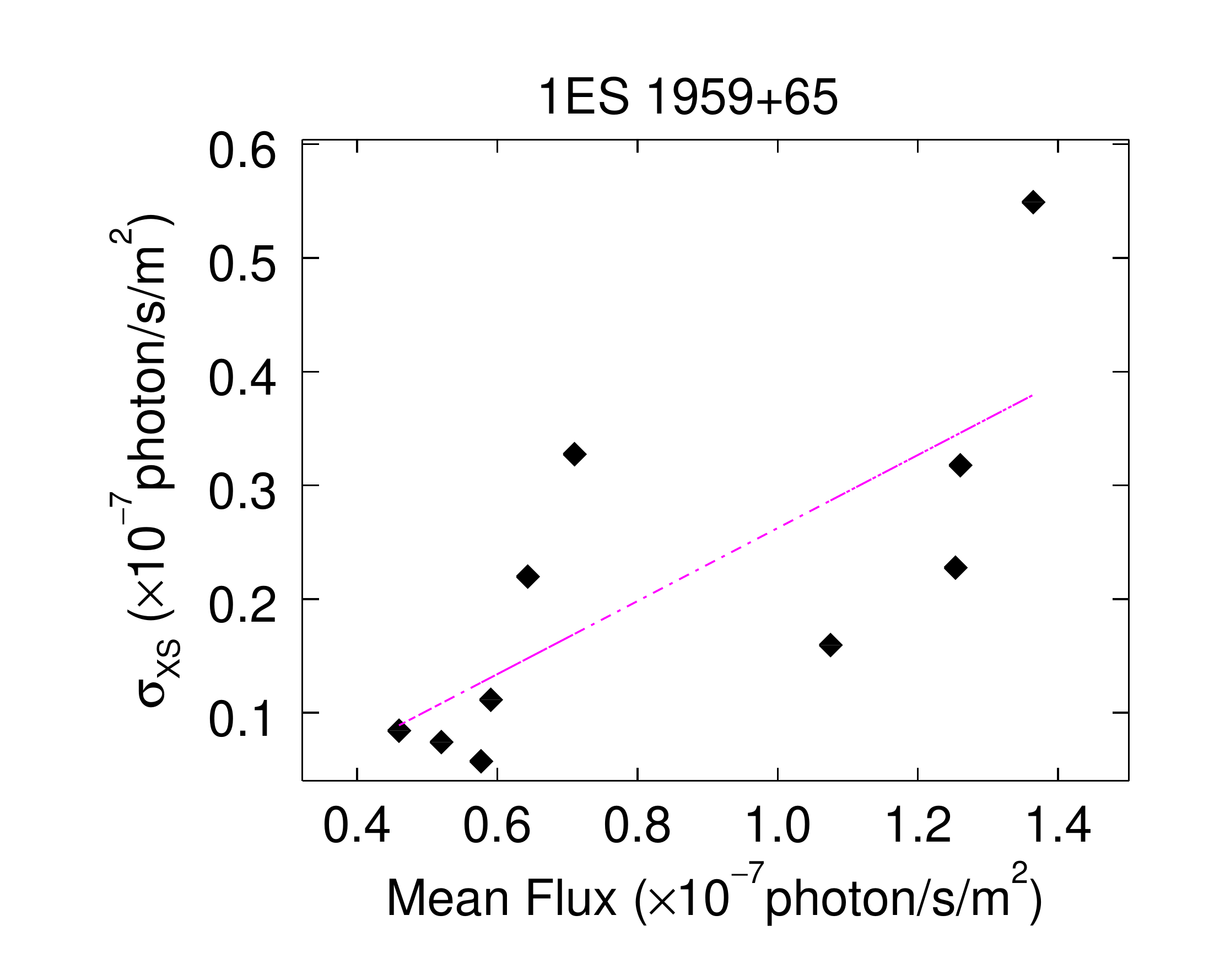}{0.25\textwidth}{}\hspace{-0.7cm}
              }
                    \vspace{-0.5cm}
\gridline{\fig{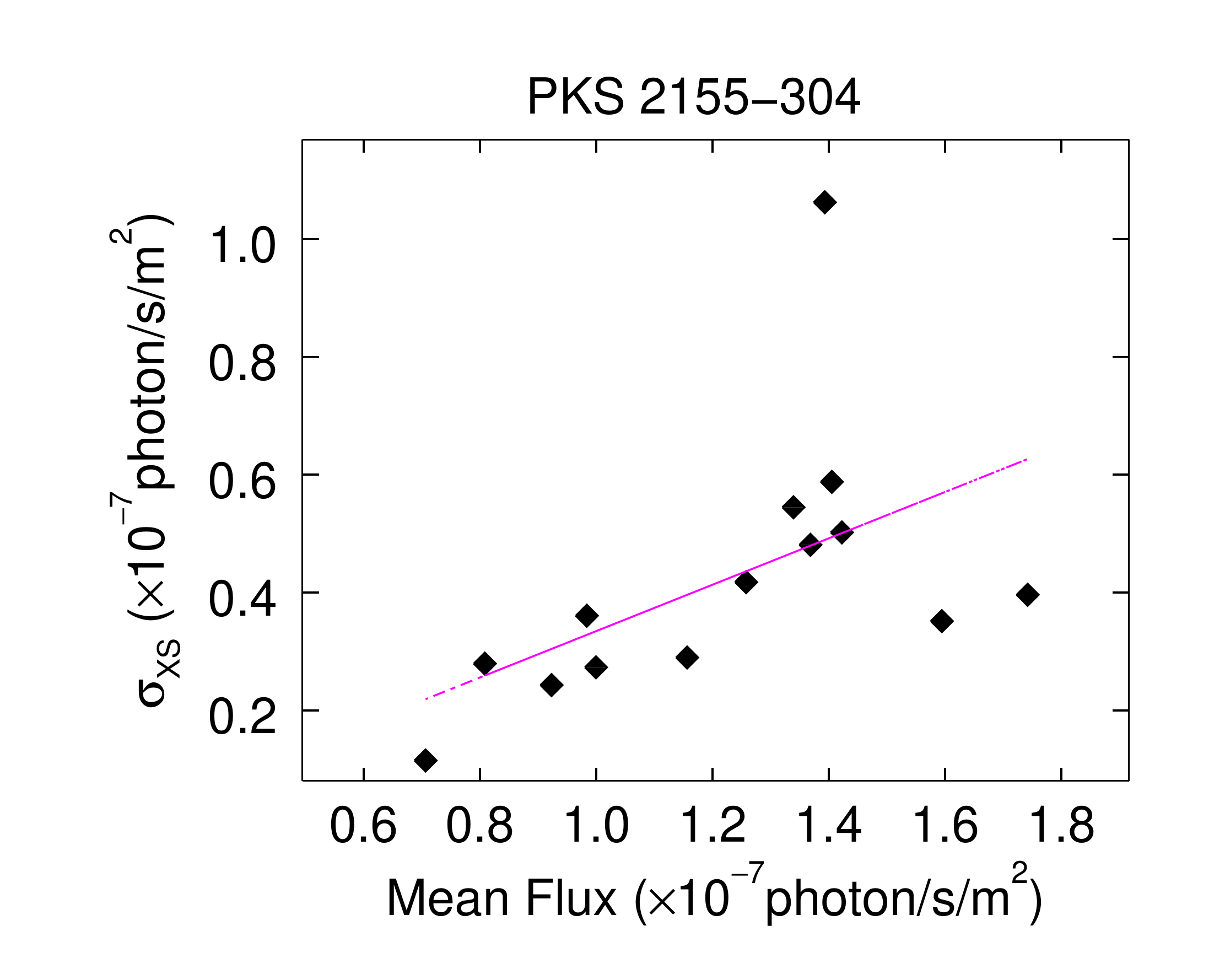}{0.25\textwidth}{}\hspace{-0.7cm}
\fig{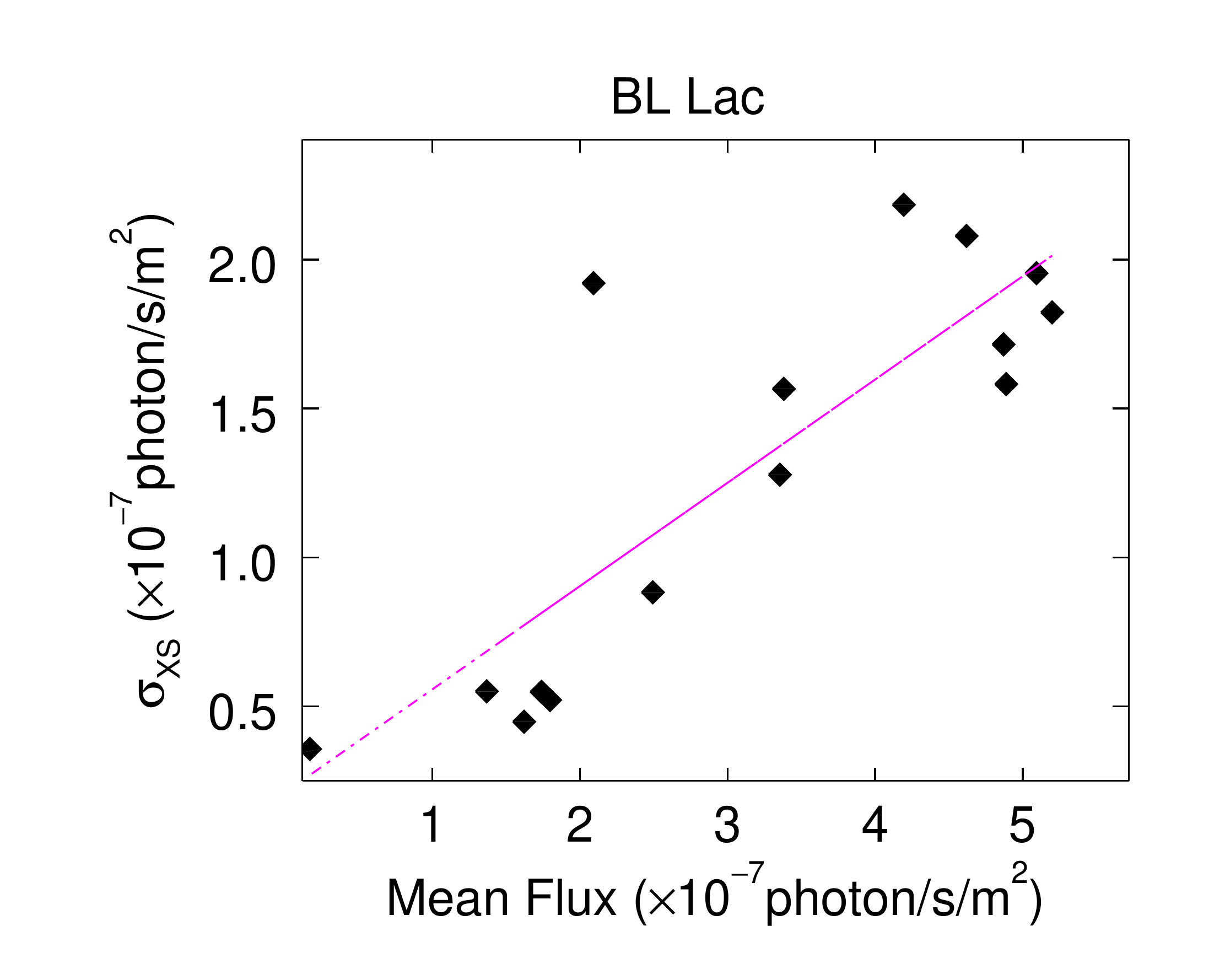}{0.25\textwidth}{}\hspace{-0.7cm}
\fig{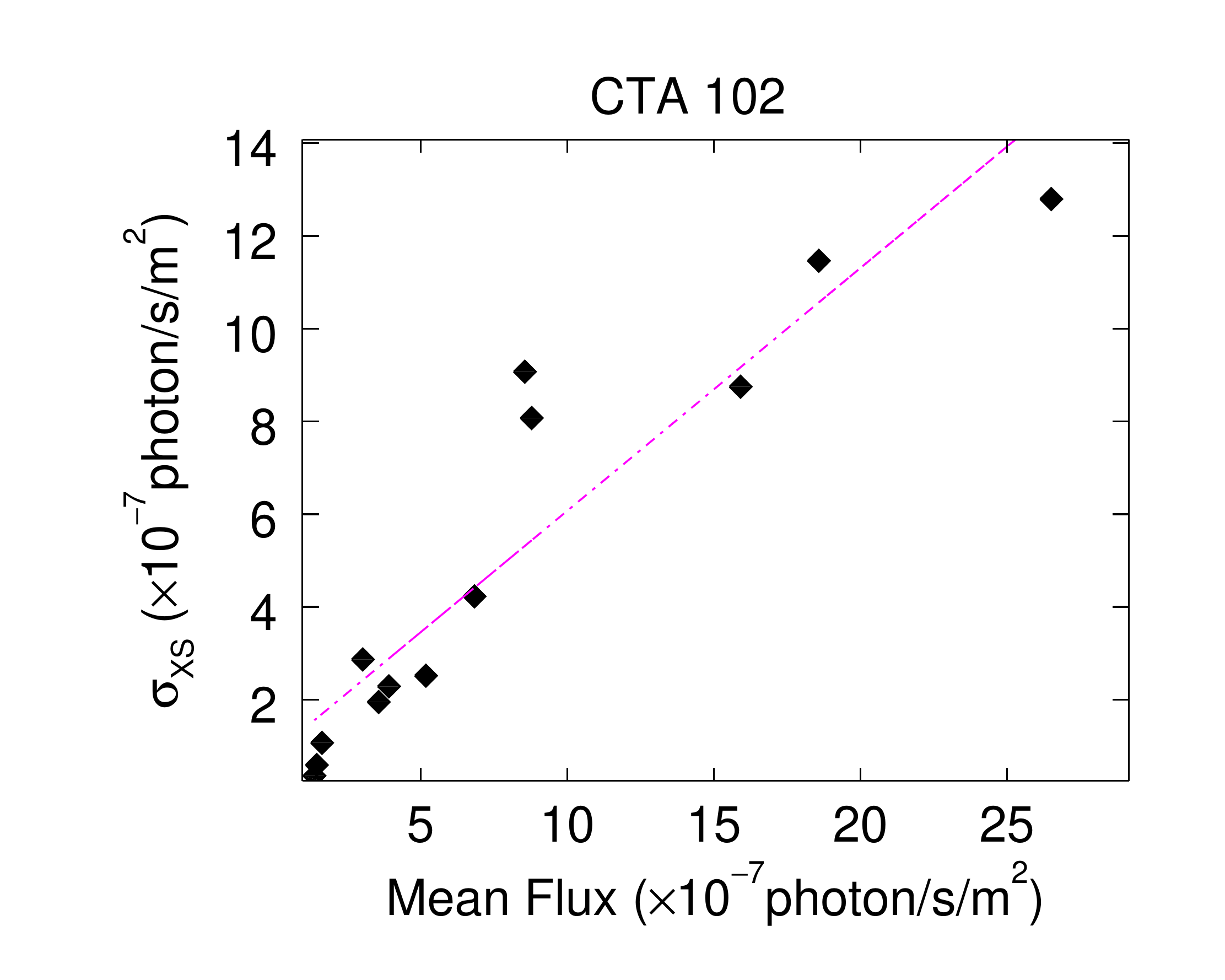}{0.25\textwidth}{}\hspace{-0.7cm}
\fig{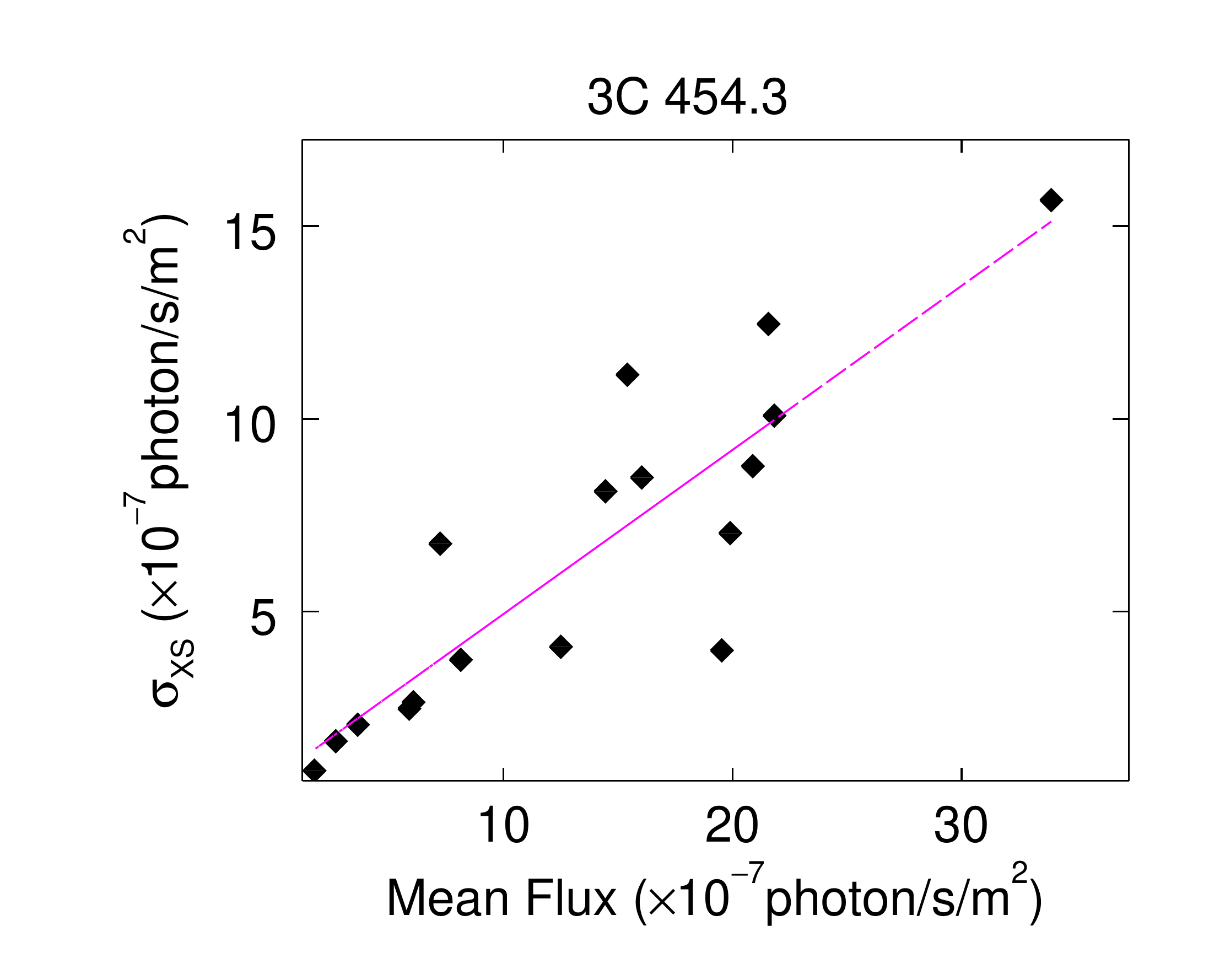}{0.25\textwidth}{}\hspace{-0.7cm}              
          }
\caption{RMS-Flux relation in the gamma-ray light curves of the sample  blazars. The magenta line represents the linear fit to the observations. \label{fig:PSD}}
\end{figure*}

\clearpage

\section{Power spectral density of blazar \label{apndx5}}

\begin{figure*}[!b]
\gridline{\fig{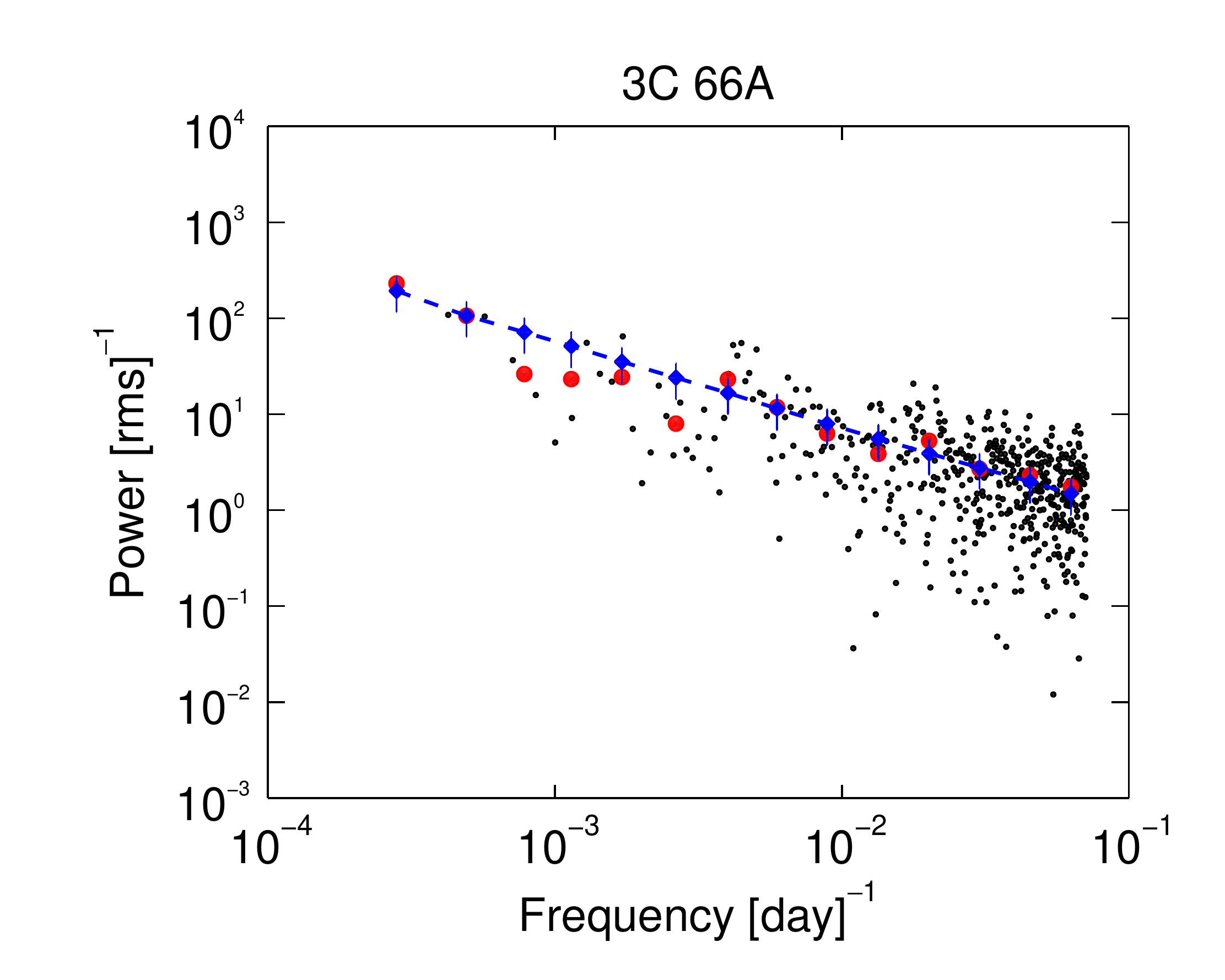}{0.25\textwidth}{}\hspace{-0.7cm}
          \fig{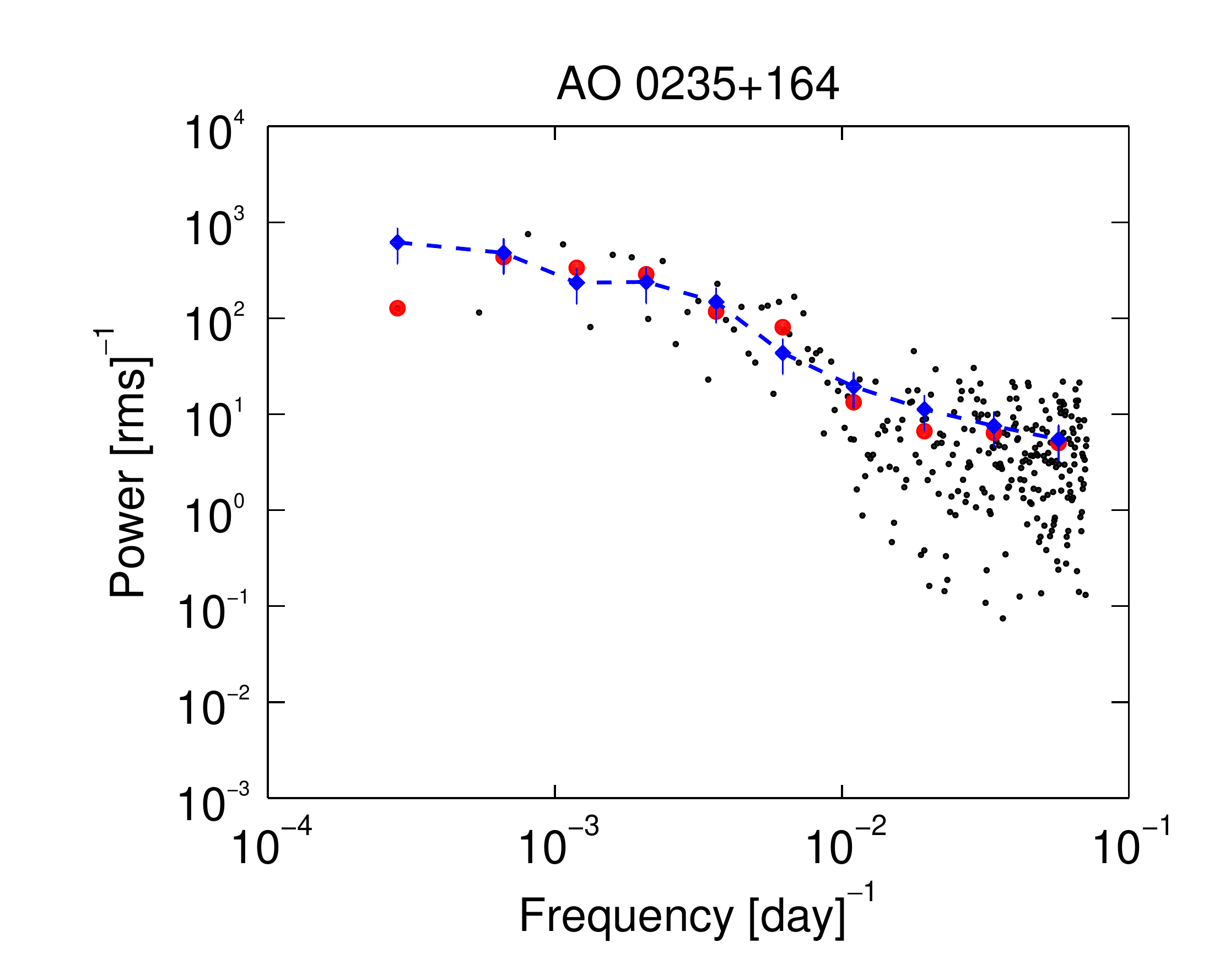}{0.25\textwidth}{}\hspace{-0.7cm}
          \fig{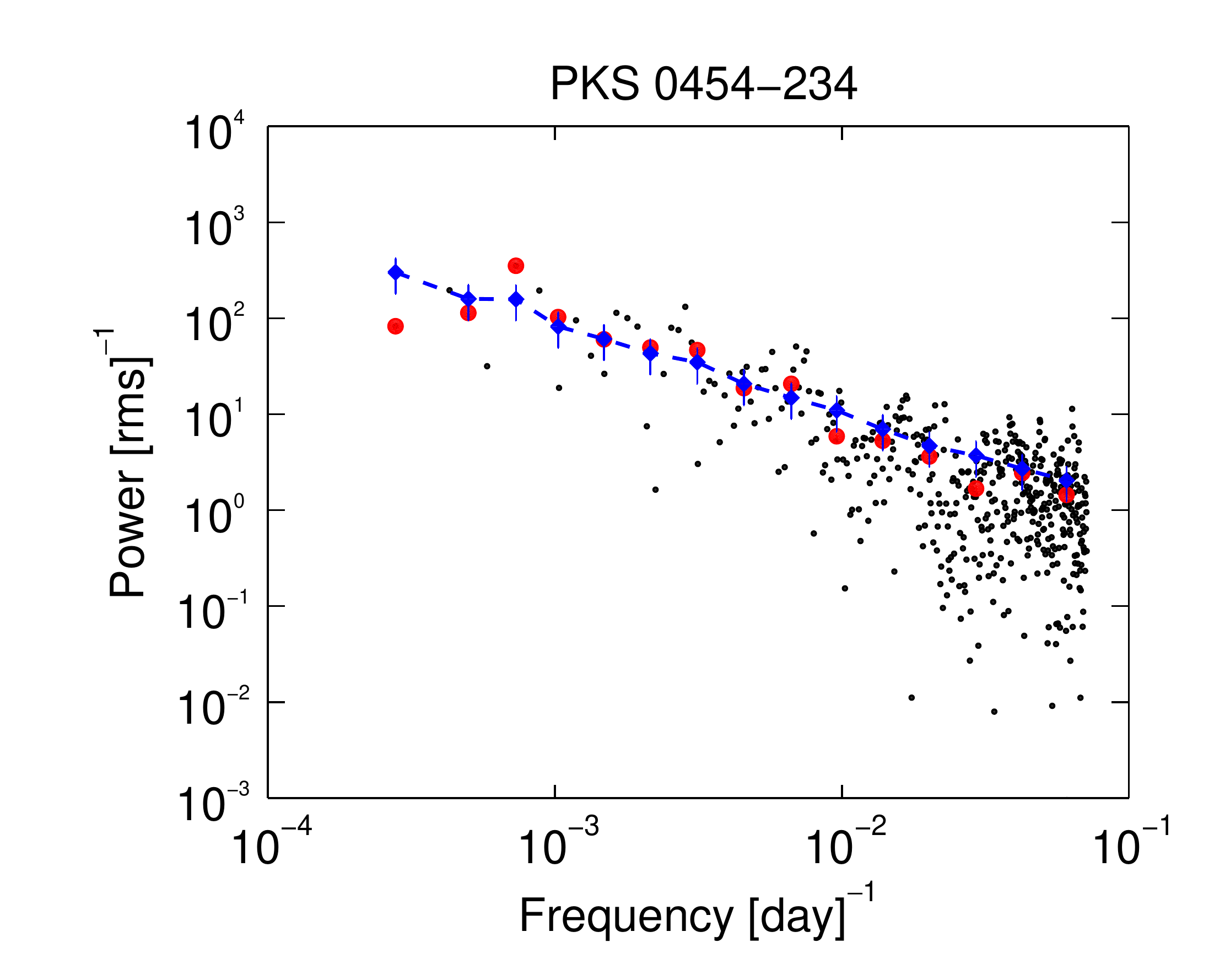}{0.25\textwidth}{}\hspace{-0.7cm}
           \fig{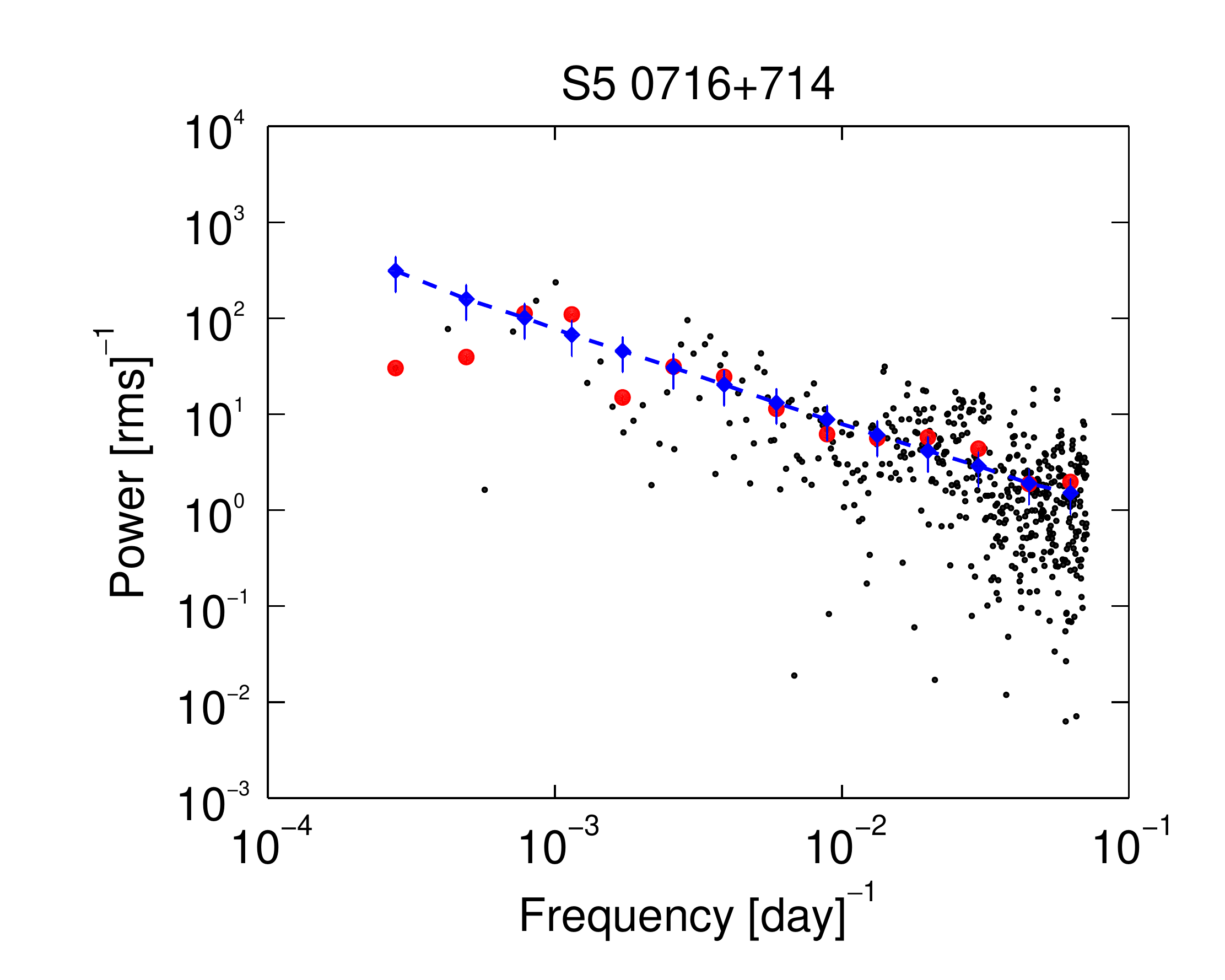}{0.25\textwidth}{}\hspace{-0.7cm}
          }
          \vspace{-0.5cm}
\gridline{\fig{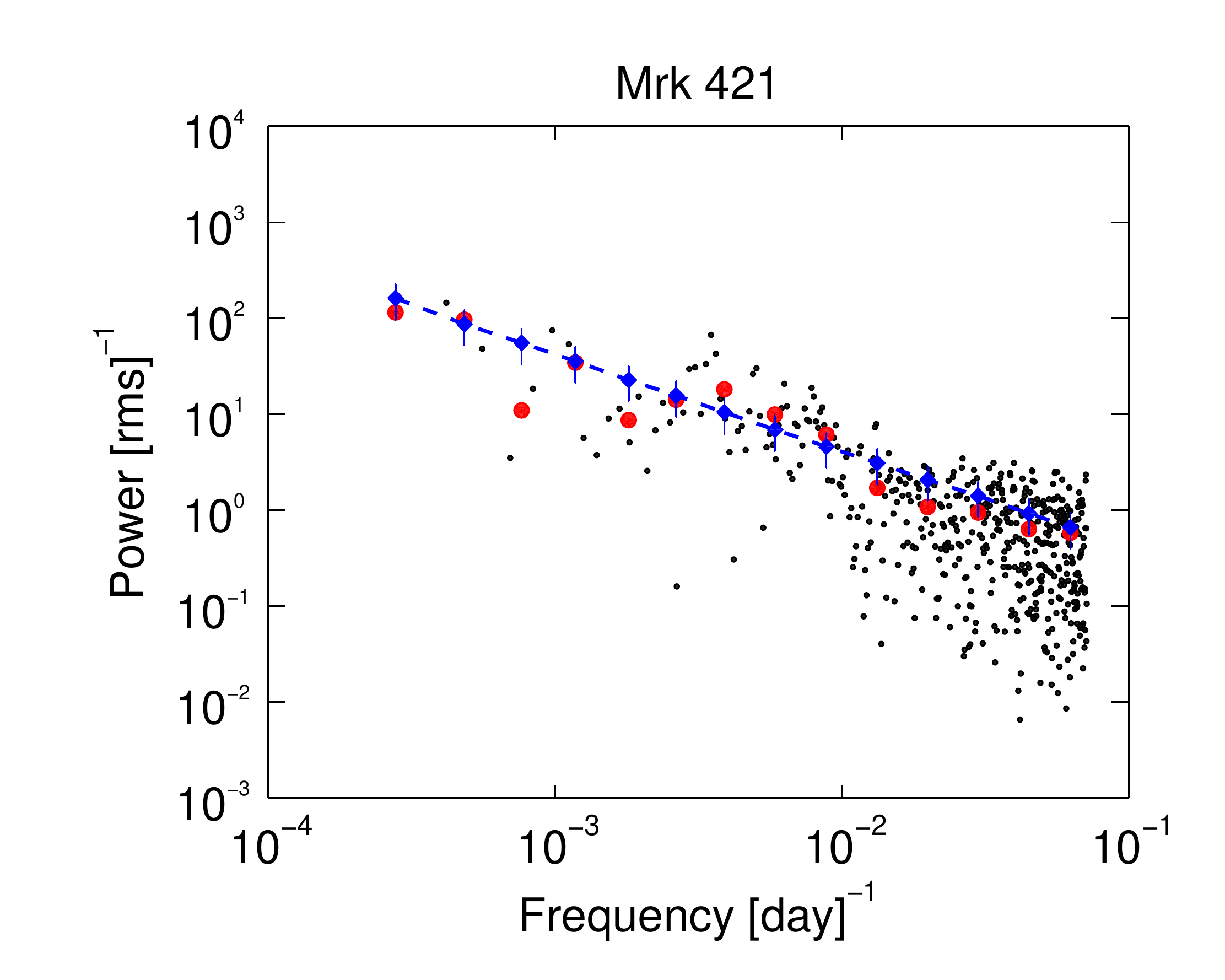}{0.25\textwidth}{}\hspace{-0.7cm}
          \fig{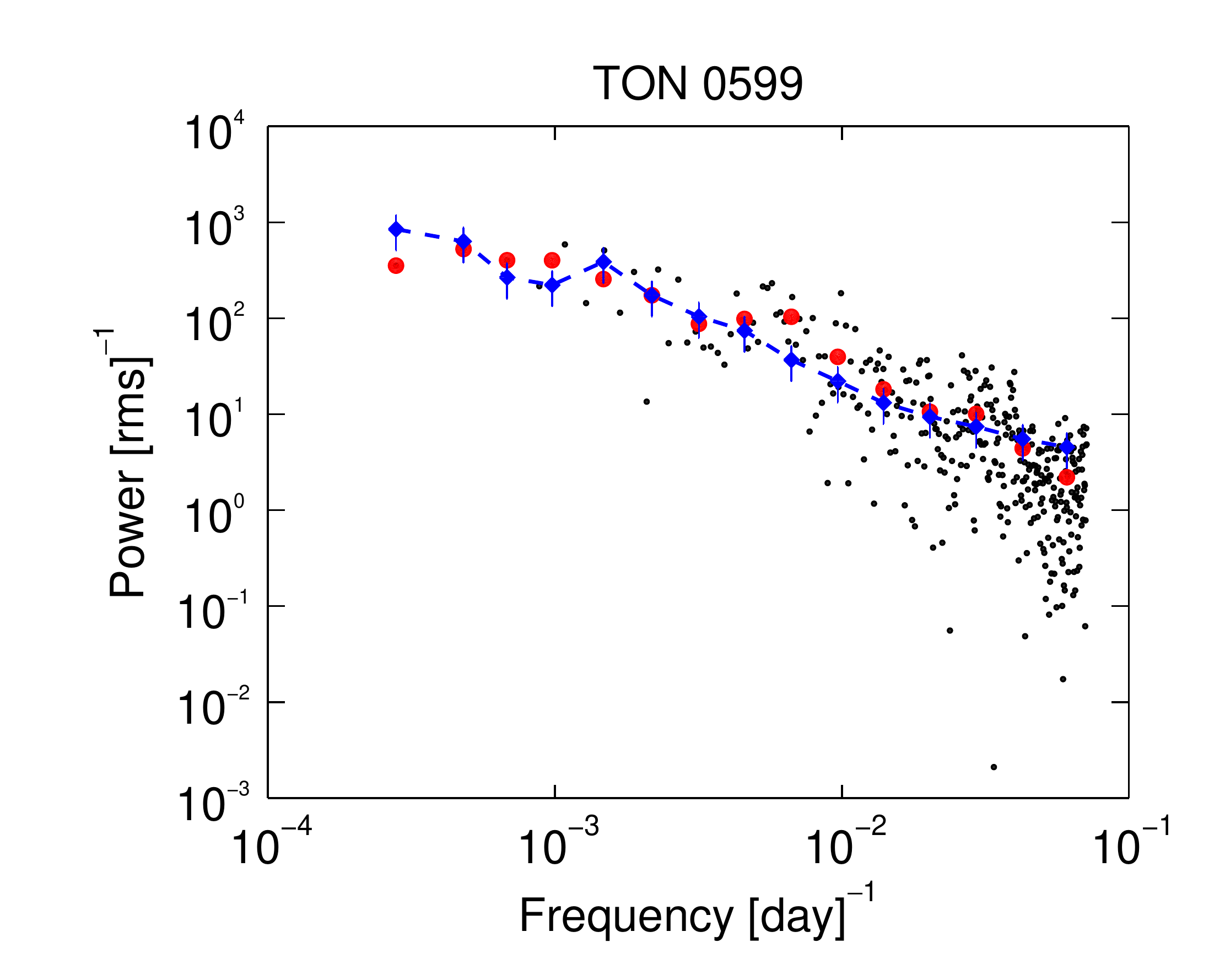}{0.25\textwidth}{}\hspace{-0.7cm}
          \fig{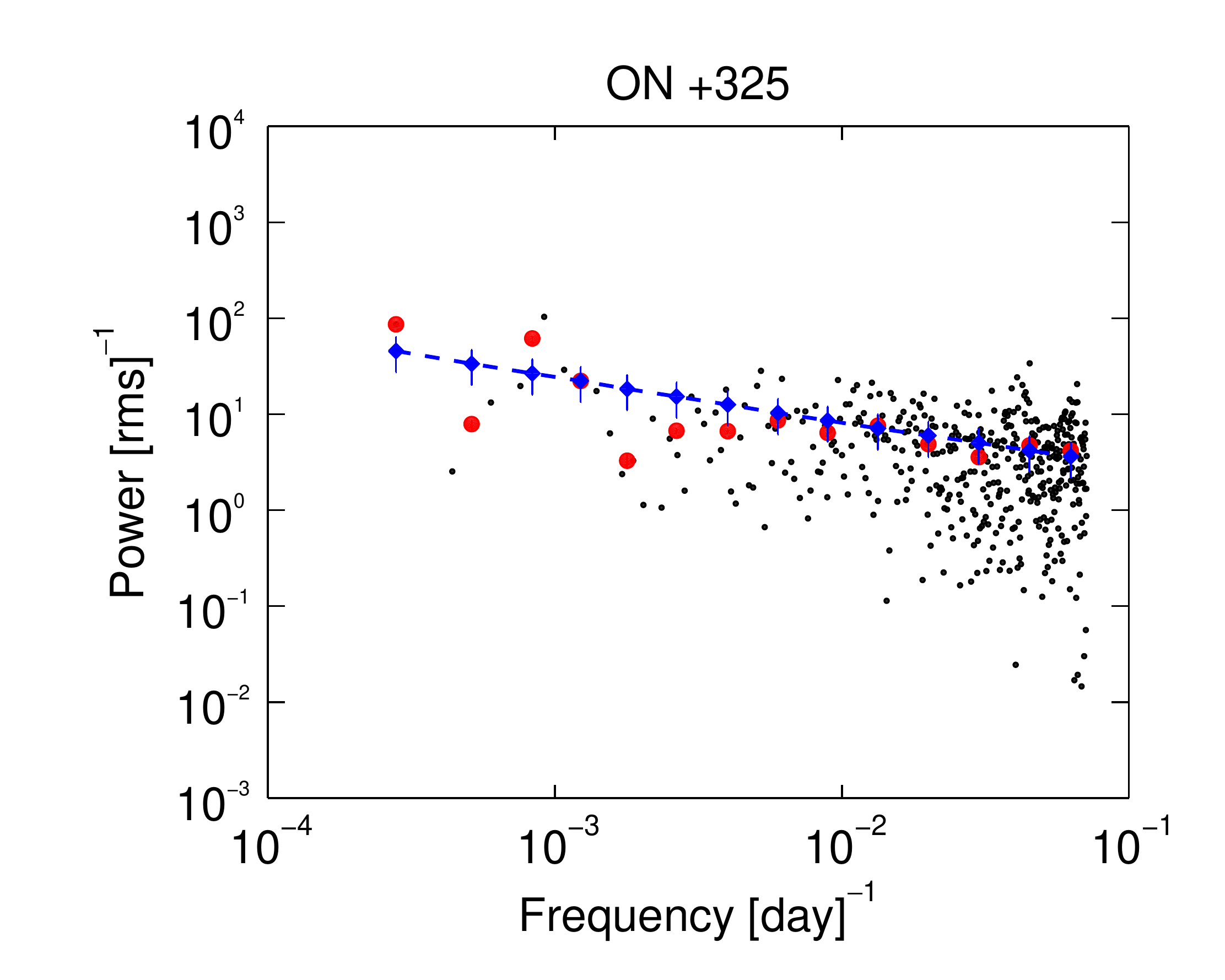}{0.25\textwidth}{}\hspace{-0.7cm}
           \fig{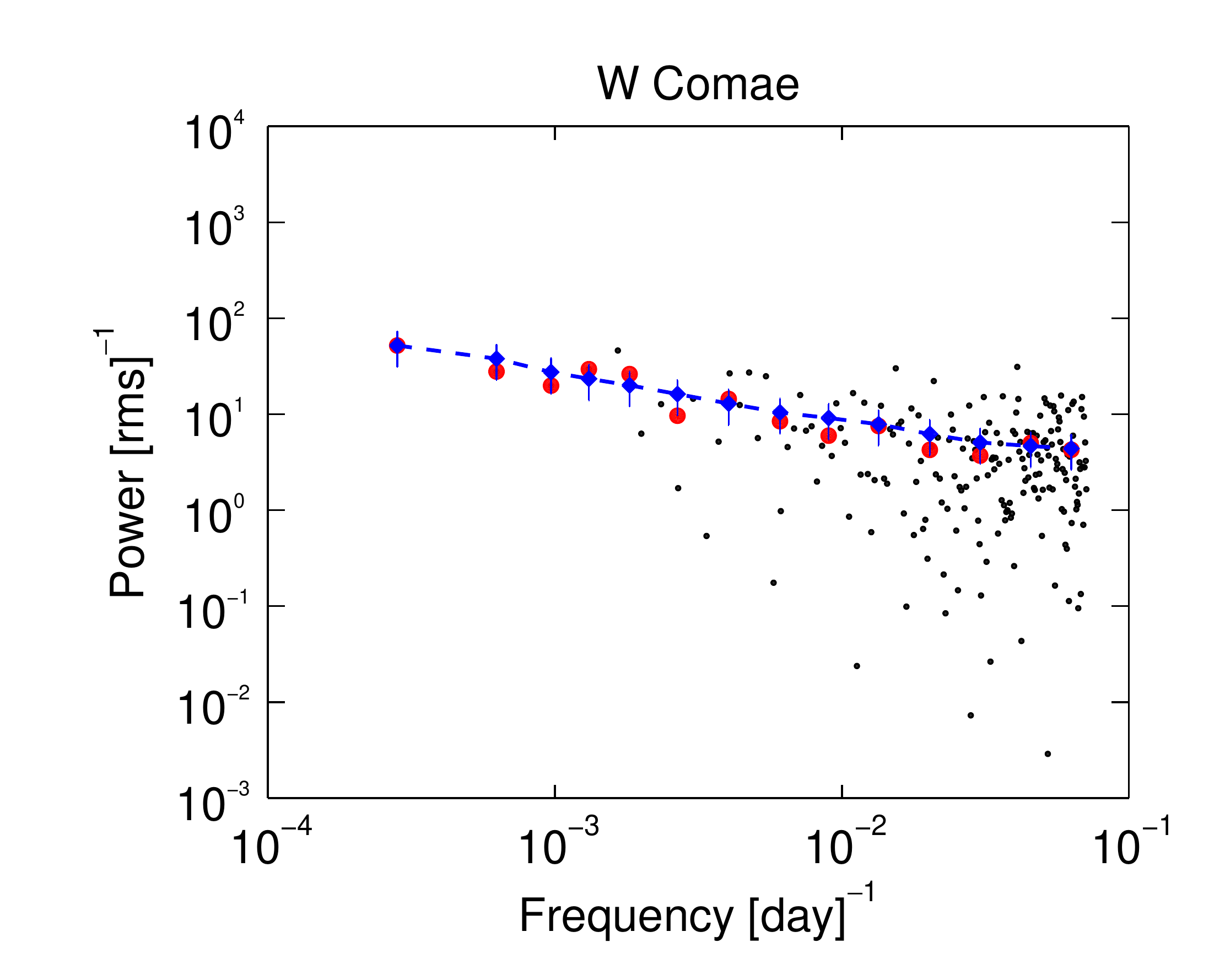}{0.25\textwidth}{}\hspace{-0.7cm}
          }
          \vspace{-0.5cm}
\gridline{\fig{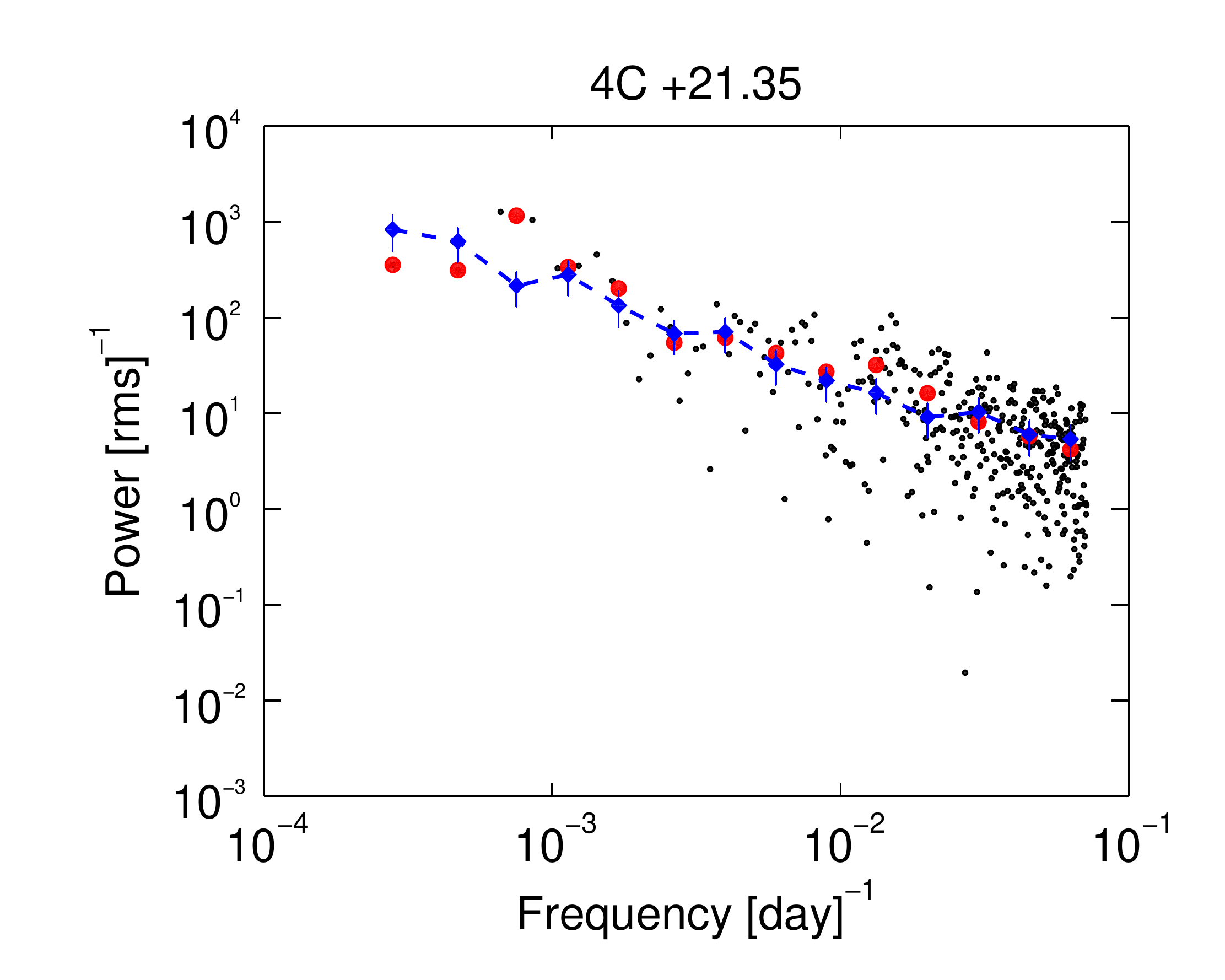}{0.25\textwidth}{}\hspace{-0.7cm}
          \fig{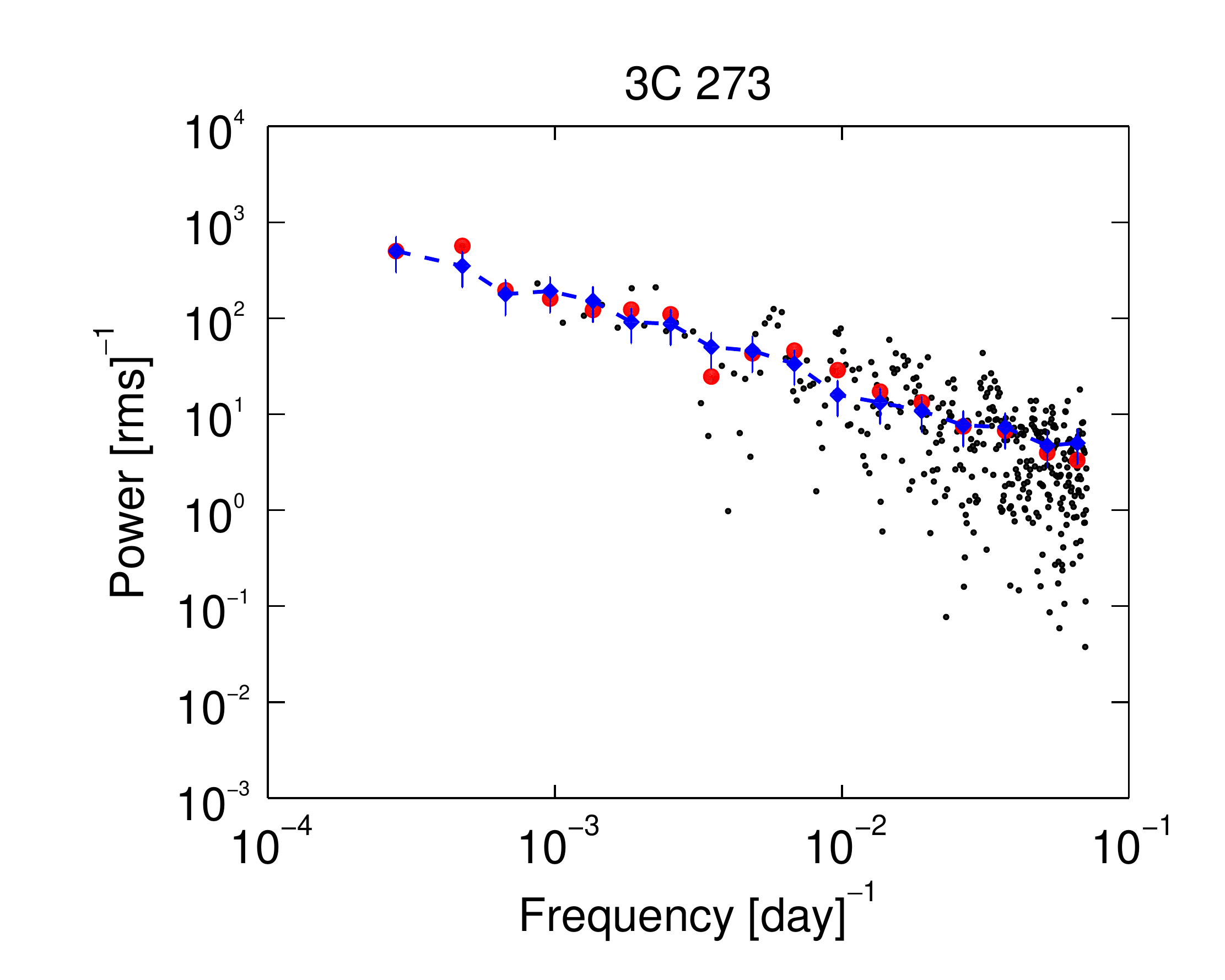}{0.25\textwidth}{}\hspace{-0.7cm}
          \fig{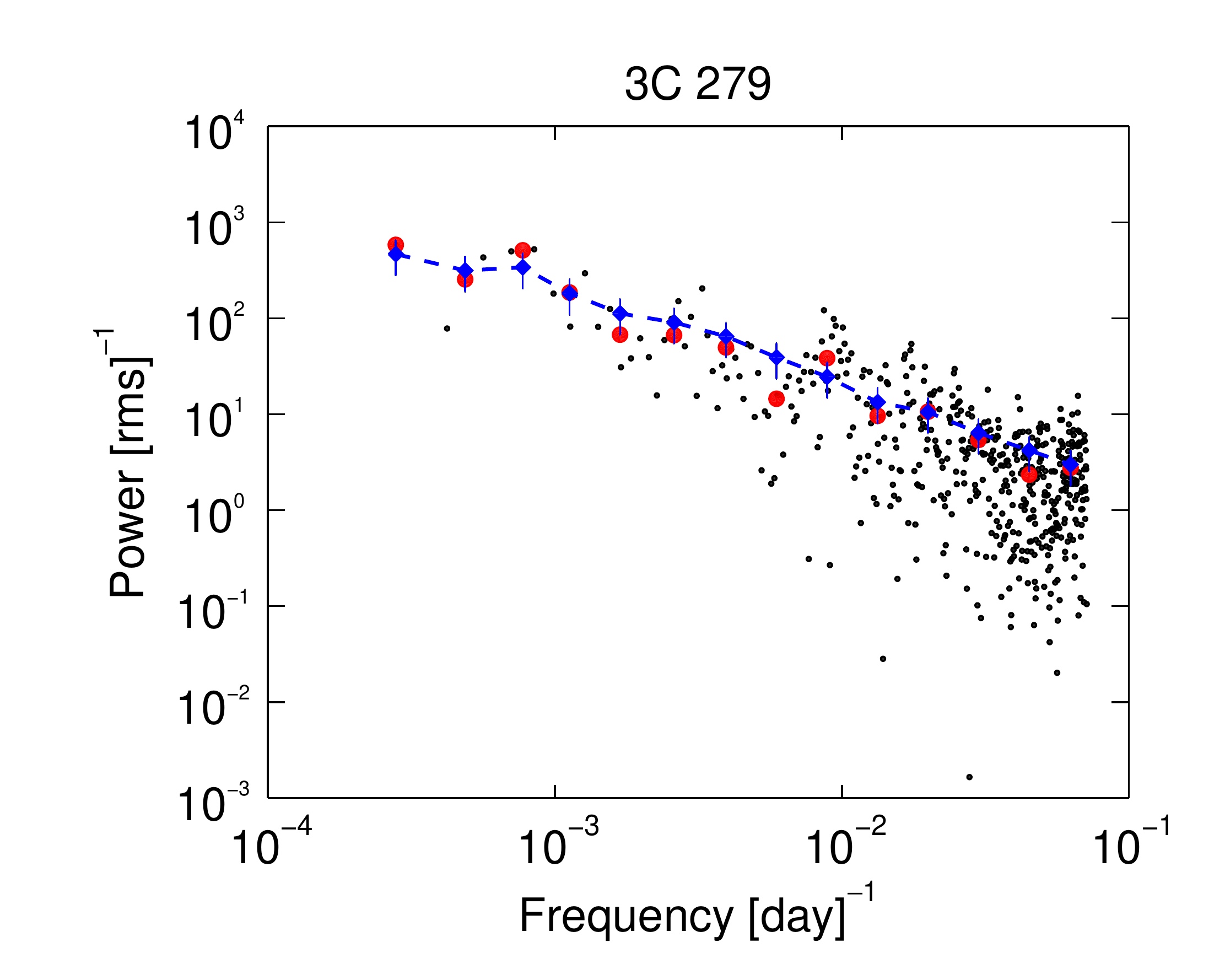}{0.25\textwidth}{}\hspace{-0.7cm}
           \fig{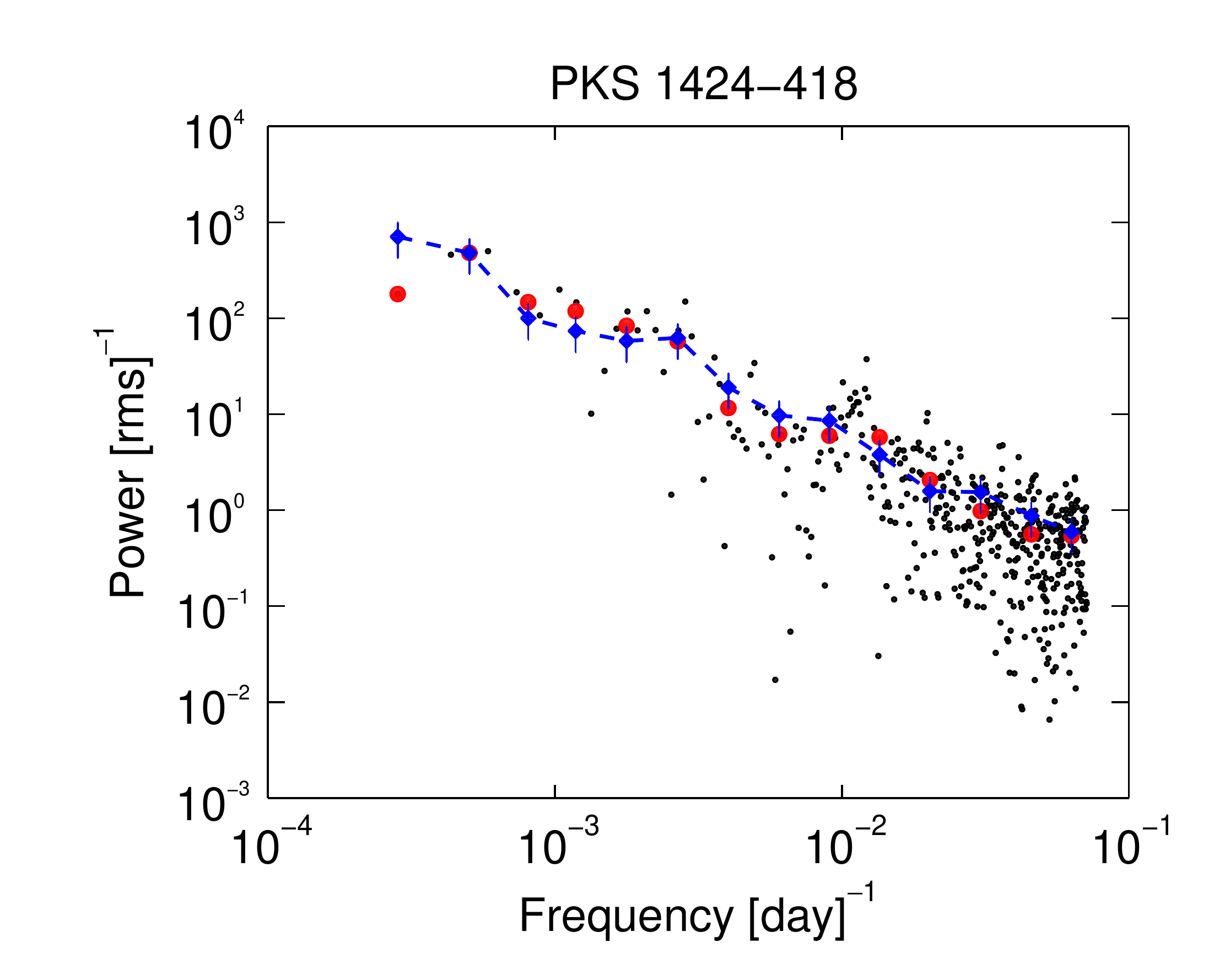}{0.25\textwidth}{}\hspace{-0.7cm}
          }
                    \vspace{-0.5cm}
\gridline{\fig{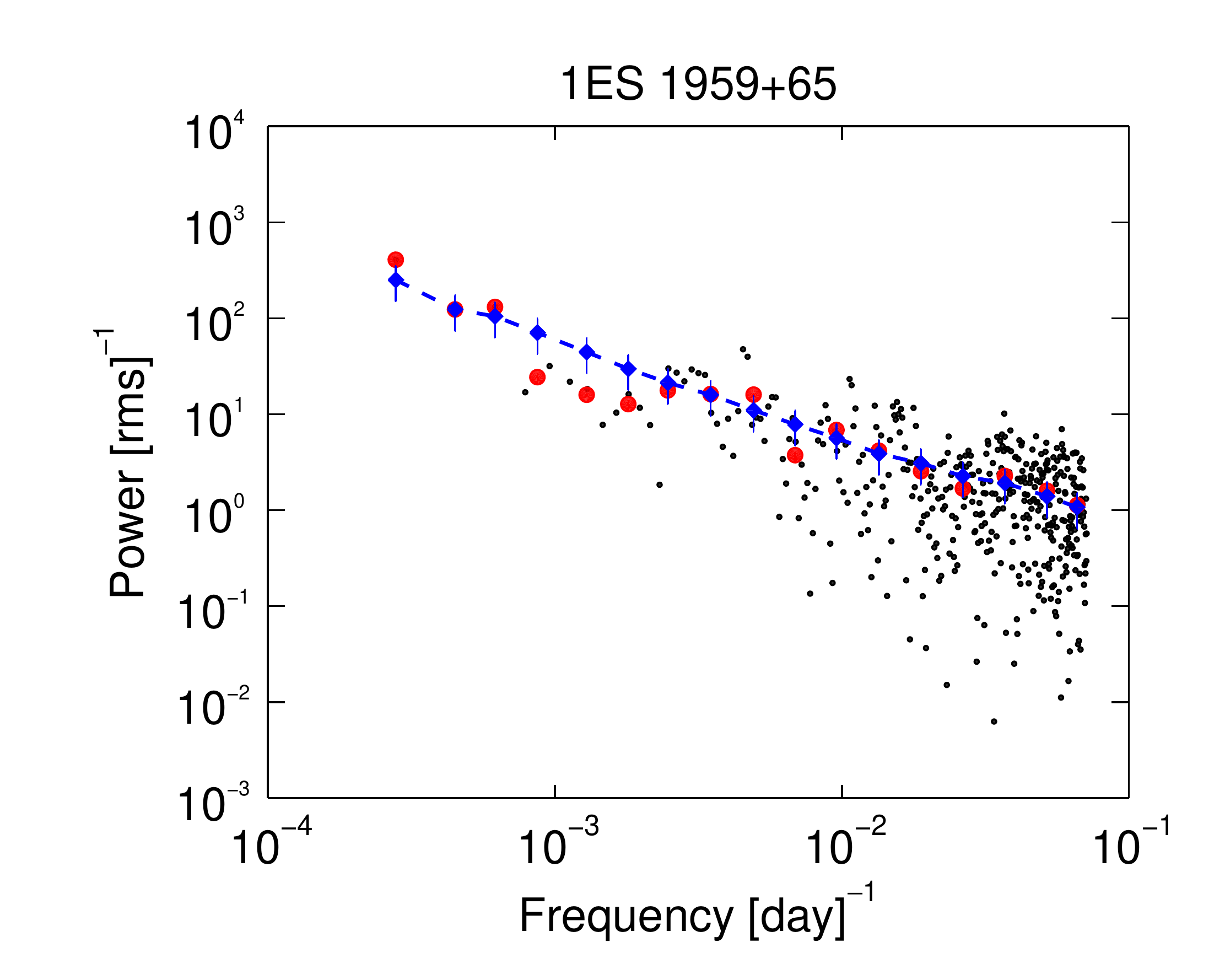}{0.25\textwidth}{}\hspace{-0.7cm}
          \fig{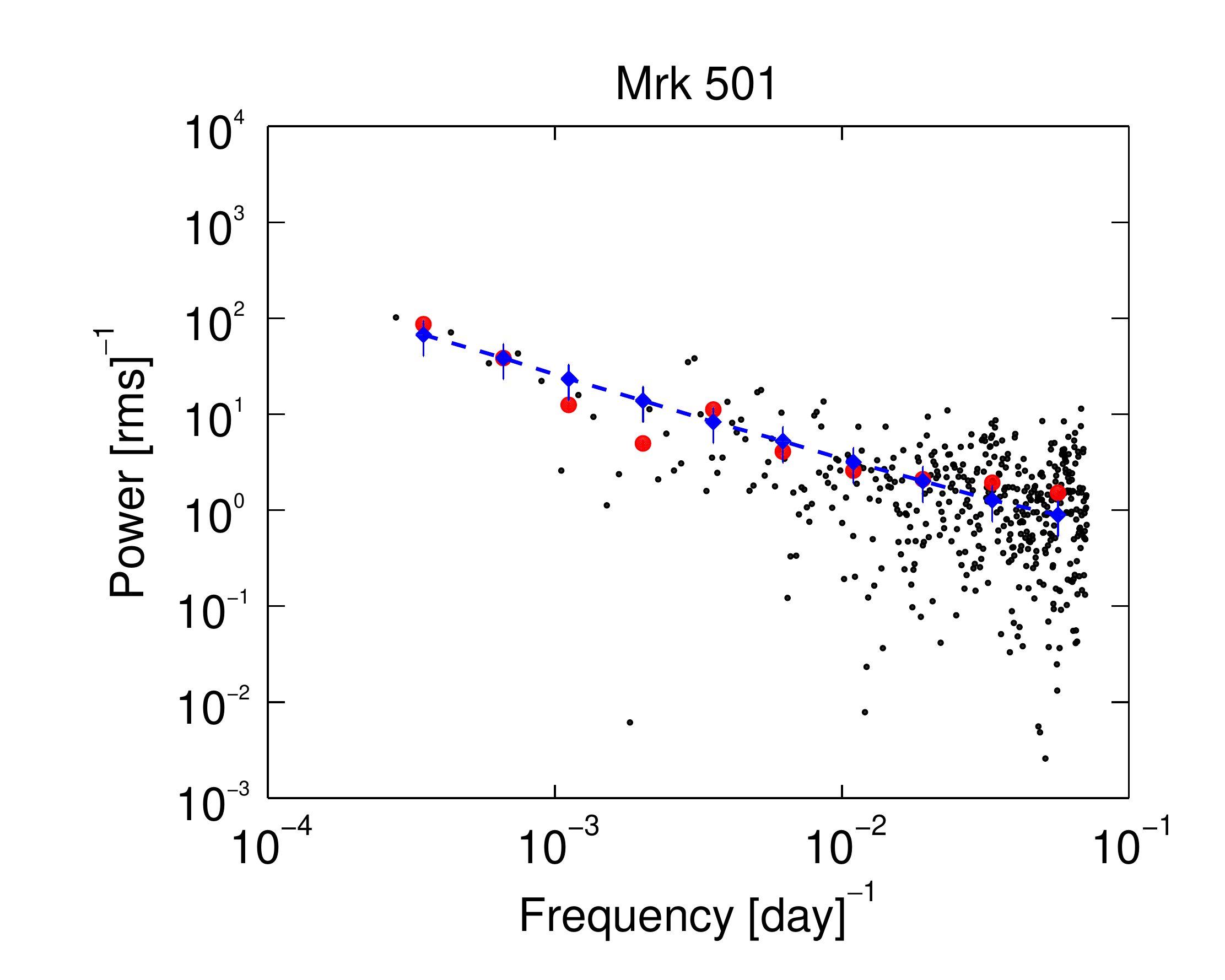}{0.25\textwidth}{}\hspace{-0.7cm}
          \fig{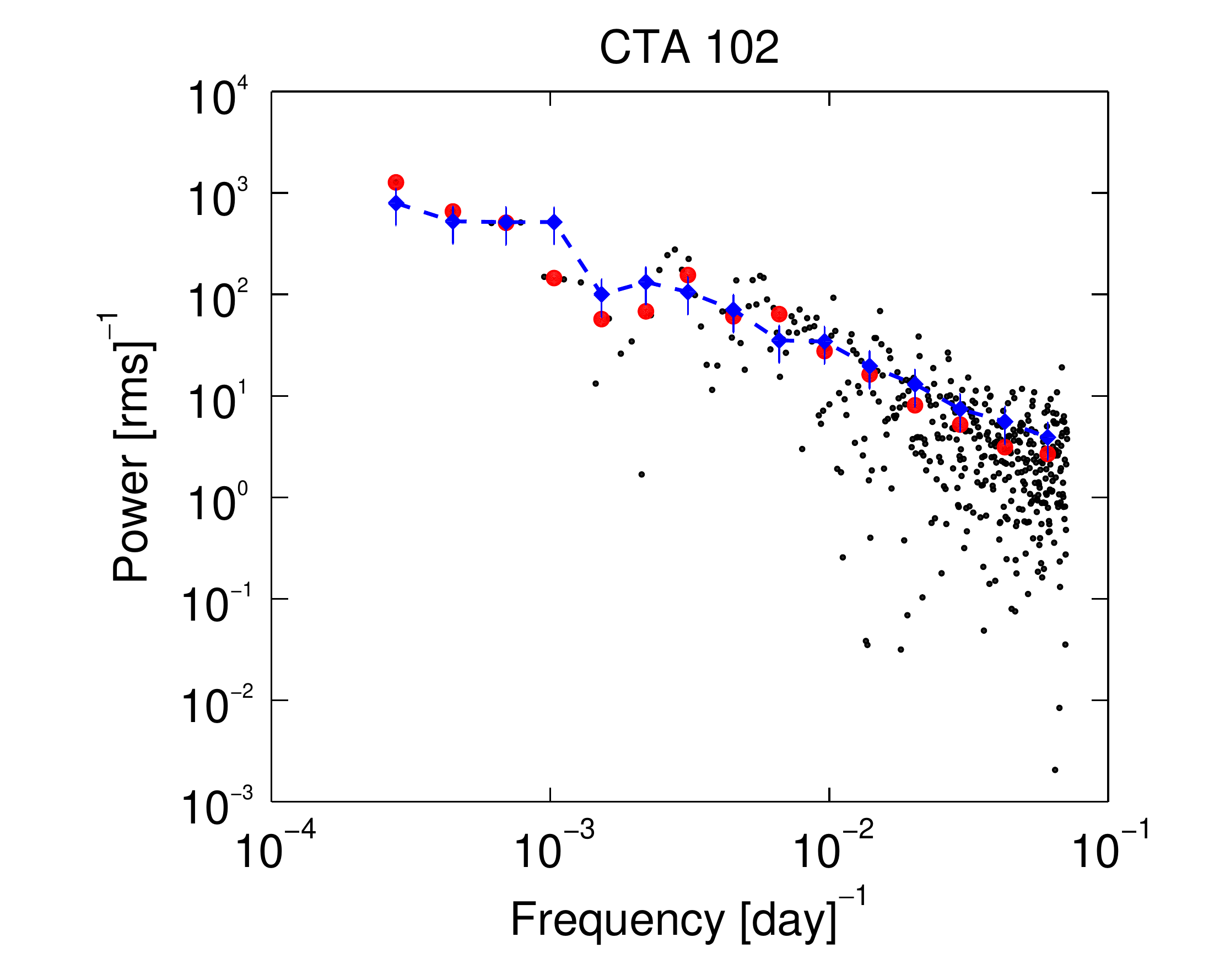}{0.25\textwidth}{}\hspace{-0.7cm}
           \fig{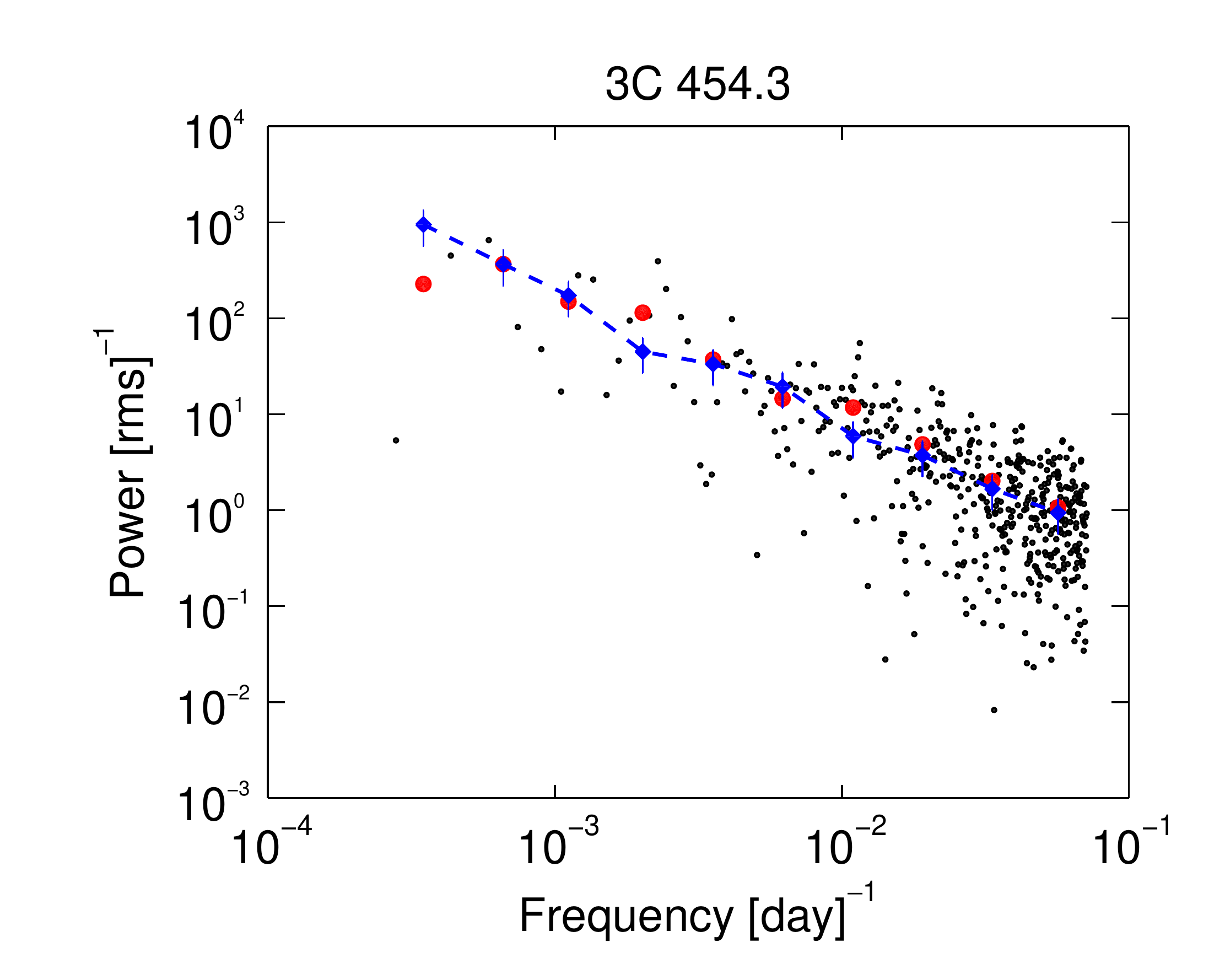}{0.25\textwidth}{}\hspace{-0.7cm}
          }
                    \vspace{-0.5cm}
\gridline{\fig{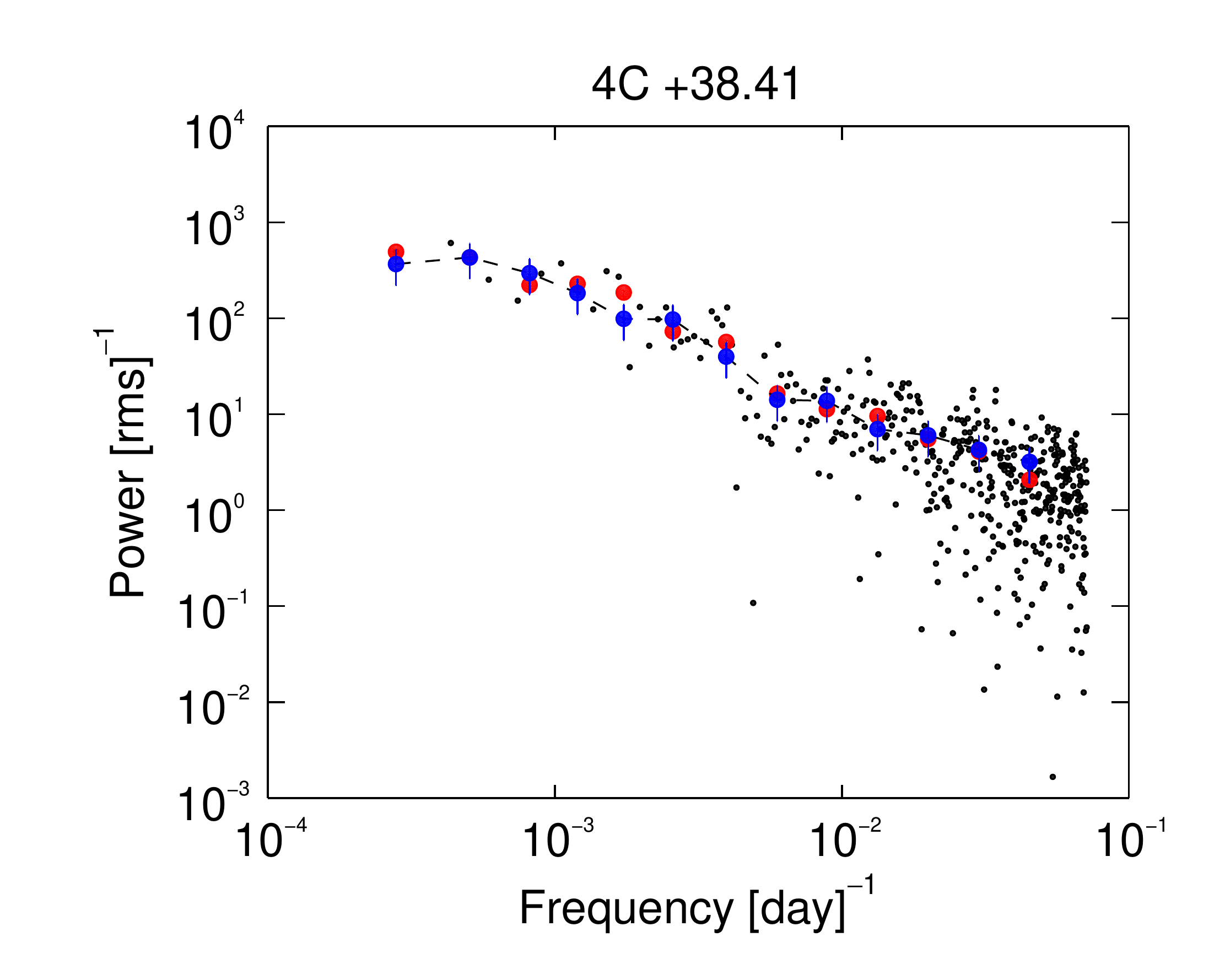}{0.25\textwidth}{}\hspace{-0.7cm}
          \fig{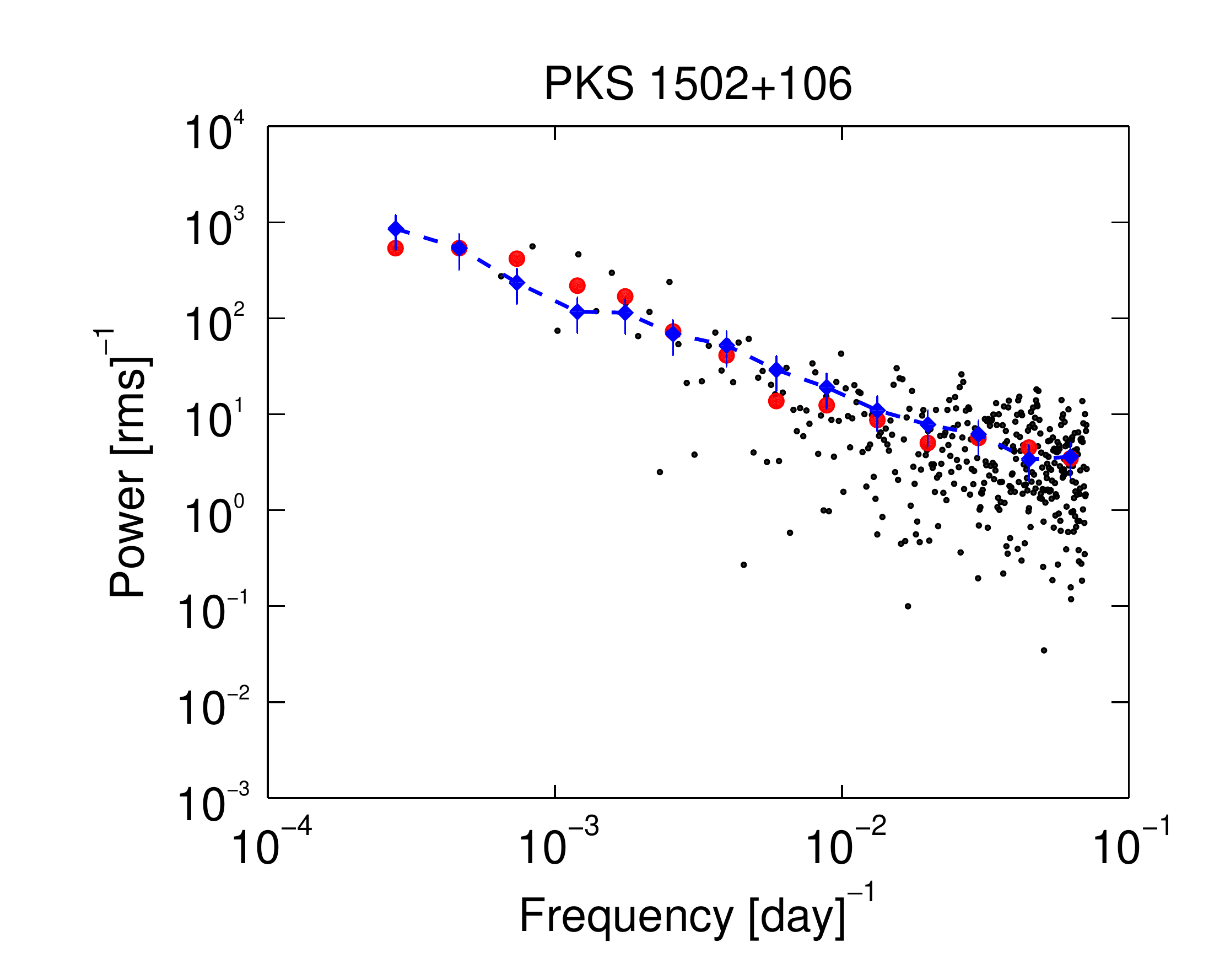}{0.25\textwidth}{}\hspace{-0.7cm}
          \fig{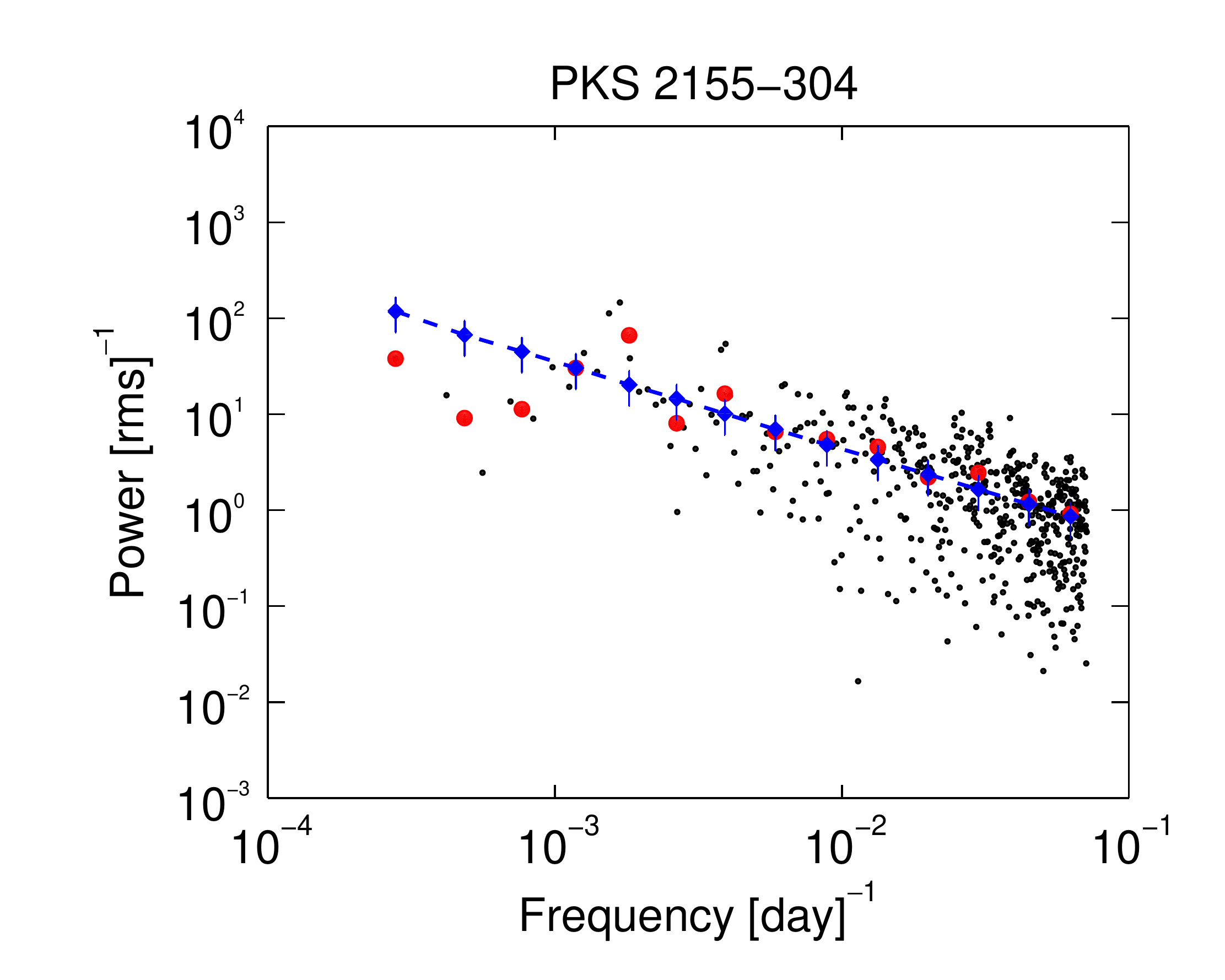}{0.25\textwidth}{}\hspace{-0.7cm}
           \fig{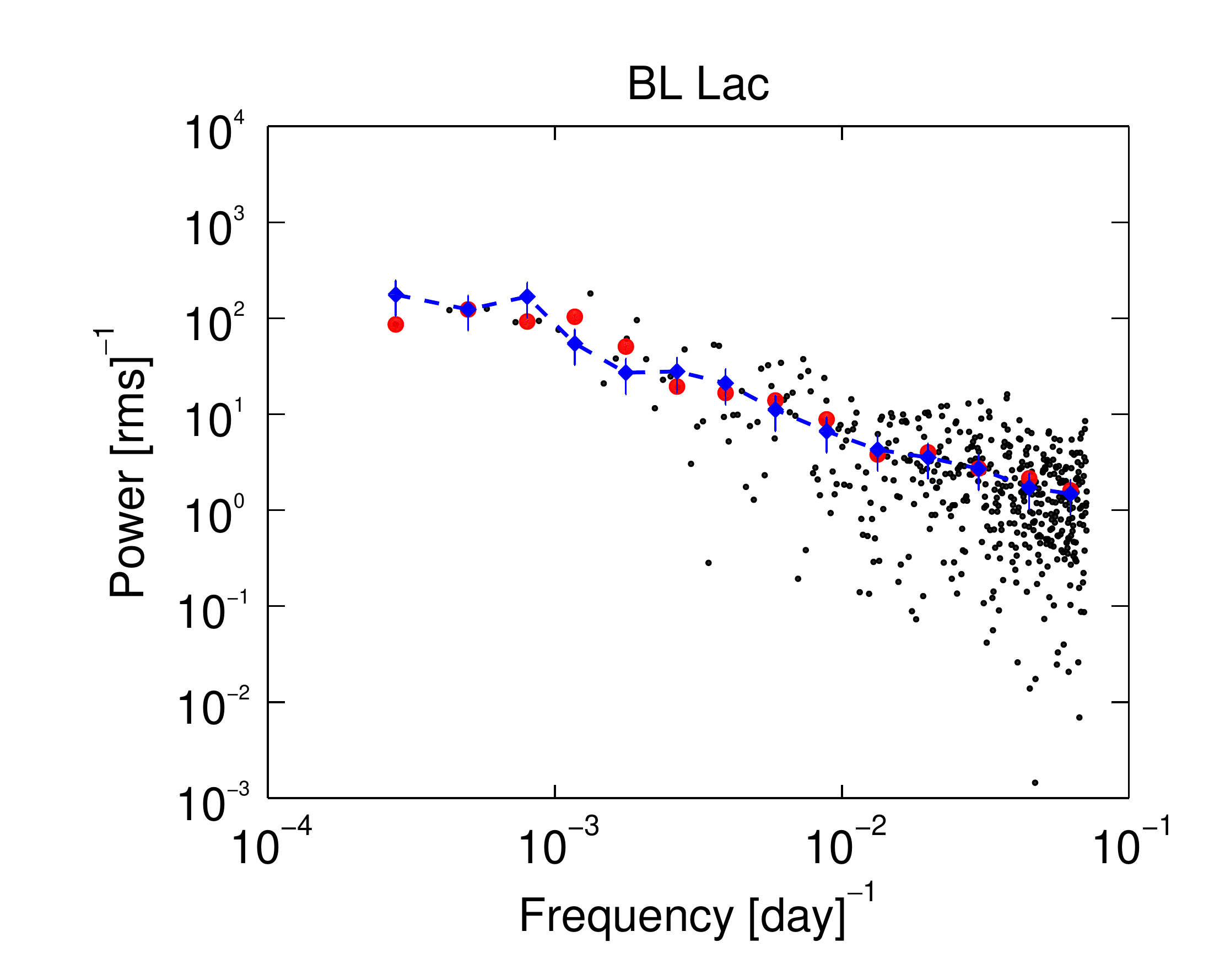}{0.25\textwidth}{}\hspace{-0.7cm}
          }
\caption{Power spectral density of the gamma-ray light curves of the blazars. Discrete Fourier periodogram  (black), binned periodogram (red), and the best fit PSD (blue) \label{fig:PSD}}
\end{figure*}

\end{document}